\documentclass[%
reprint,
superscriptaddress, 
amsmath,amssymb,
aps,
prb,
prstab,
floatfix,
longbibliography
]{revtex4-2}

\usepackage[caption=false]{subfig}
\usepackage{multirow}
\usepackage{graphicx}
\usepackage{dcolumn}
\usepackage{bm}
\usepackage{threeparttable}
\usepackage{color}
\usepackage[dvipsnames]{xcolor}
\usepackage{chemformula}
\usepackage{xspace}
\usepackage{orcidlink}
\usepackage{hyperref}
\usepackage{booktabs}
\hypersetup{colorlinks=true,linkcolor=blue,filecolor=blue,urlcolor=blue,citecolor=blue,pdftitle = {Polar phonons and magnetic excitations in the antiferromagnet $\mathbf{CoF}_{2}$}, pdfauthor = {R.~M.~Dubrovin}}

\newcommand{\CoF}{$\mathrm{CoF}_{2}$}
\newcommand{\MnF}{$\mathrm{MnF}_{2}$}
\newcommand{\TF}{$\mathrm{TF}_{2}$}

\definecolor{myblue}{RGB}{0,0,0}
\definecolor{newtext}{RGB}{0,0,0} 
\definecolor{newtext2}{RGB}{0,0,0} 

\bibpunct{[}{]}{;}{n}{}{}

\begin{document}
\title{Polar phonons and magnetic excitations in the antiferromagnet $\mathbf{CoF}_{2}$}

\author{R.~M.~Dubrovin\,\orcidlink{0000-0002-7235-7805}}
\email{dubrovin@mail.ioffe.ru}
\altaffiliation[Also at ]{University of Nizhny Novgorod, 603022 Nizhny Novgorod, Russia}
\affiliation{Ioffe Institute, Russian Academy of Sciences, 194021 St.\,Petersburg, Russia}
\author{A.~Tellez-Mora\,\orcidlink{0000-0002-3080-5868}}
\affiliation{Department of Physics and Astronomy, West Virginia University, WV-26506-6315 Morgantown, West Virginia, USA}
\author{A.~C.~Garcia-Castro\,\orcidlink{0000-0003-3379-4495}}
\affiliation{School of Physics, Universidad Industrial de Santander, 680002 Bucaramanga, Colombia}
\author{N.~V.~Siverin\,\orcidlink{0000-0002-4643-845X}}
\altaffiliation[Now at ]{Experimental Physics 2, Department of Physics, TU Dortmund, 44227, Dortmund, Germany}
\affiliation{Ioffe Institute, Russian Academy of Sciences, 194021 St.\,Petersburg, Russia}
\author{N.~N.~Novikova\,\orcidlink{0000-0003-2428-6114}}
\affiliation{Institute of Spectroscopy, Russian Academy of Sciences, 108840 Moscow, Troitsk, Russia}
\author{K.~N.~Boldyrev\,\orcidlink{0000-0002-3784-7294}}
\affiliation{Institute of Spectroscopy, Russian Academy of Sciences, 108840 Moscow, Troitsk, Russia}
\author{E.~A.~Mashkovich\,\orcidlink{0000-0002-0825-1408}}
\affiliation{Institute of Physics II, University of Cologne, 50937 Cologne, Germany}
\author{Aldo~H.~Romero\,\orcidlink{0000-0001-5968-0571}}
\affiliation{Department of Physics and Astronomy, West Virginia University, WV-26506-6315 Morgantown, West Virginia, USA}
\author{R.~V.~Pisarev\,\orcidlink{0000-0002-2008-9335}}
\affiliation{Ioffe Institute, Russian Academy of Sciences, 194021 St.\,Petersburg, Russia}

\date{\today}

\begin{abstract}
The coupling between antiferromagnetic spins and infrared-active phonons in solids is responsible for many intriguing phenomena and is a field of intense research with extensive potential applications in the modern devices based on antiferromagnetic spintronics and phononics.
Insulating rutile antiferromagnetic crystal \CoF{} is one of the model materials for studying nonlinear magneto-phononics due to the strong spin-lattice coupling as a result of the orbitally degenerate ground state of $\mathrm{Co}^{2+}$ ions manifested in the plethora of static and induced piezomagnetic effects. 
Here, we report results on the complete infrared spectroscopy study of lattice and magnetic dynamics in \CoF{} in a wide temperature range and their careful analysis.
We observed that infrared-active phonons demonstrate frequency shifts at the antiferromagnetic ordering.  
Furthermore, using first-principles calculations, we examined the lattice dynamics and disclosed that these frequency shifts are rather due to the spin-phonon coupling than geometrical lattice effects.
Next we found that the low-frequency dielectric permittivity demonstrates distinct changes at the antiferromagnetic ordering due to the spontaneous magnetodielectric effect caused by the behavior of infrared-active phonons.
In addition, we have observed magnetic excitations in the infrared spectra and identified their magnetodipole origin.
To strengthen our conclusions, we analyze the theoretical phonon-magnon coupling overall phonons at the $\Gamma$ point.
We conclude that the largest effect comes from the $A_{1g}$ and $B_{2g}$ Raman-active modes.
As such, our results establish a solid basis for further investigations and more deeper understanding of the coupling of phonons with spins and magnetic excitations in antiferromagnets.
\end{abstract}

\maketitle

\section{Introduction}

Antiferromagnets are promising materials for next-generation spintronic devices because of their \textcolor{newtext}{several} considerable advantages over ferromagnets.
These advantages cover, for example, orders of magnitude higher frequencies of spin dynamics (typically in the terahertz range), insensitivity to external magnetic perturbations, and the absence of stray field~\cite{jungwirth2016antiferromagnetic,baltz2018antiferromagnetic,jungwirth2018multiple,nemec2018antiferromagnetic,liu2019antiferromagnetic,fukami2020antiferromagnetic,brataas2020spin,florez2022lattice,xiong2022antiferromagnetic,han2023coherent,meer2023antiferromagnetic}.
In turn, the coupling between the optical phonons with spins in antiferromagnets has tremendous interest and it is, nowadays, a hot topic area of modern condensed matter physics.
The latter interest lies at the intersection of spintronics and phononics because it provides the attractive route to control of the magnetic and structural properties of materials~\cite{nova2017effective,streib2019magnon,juraschek2020phono,juraschek2021sum,stupakiewicz2021ultrafast,afanasiev2021ultrafast,disa2021engineering,ueda2023non,davies2024phononis,basini2024terahertz}.
Nevertheless, despite the intensive research in this area in recent years, the microscopic mechanisms for the spin-phonon coupling in insulator crystals remain poorly understood~\cite{gu2022spin,guo2023spin}.


Among magnetic crystals, $3d$ transition metal fluorides \TF{}, with a rutile structure, are attractive materials for antiferromagnetic spintronic applications.
Thus, the spin Seebeck effect was experimentally observed in thin films of $\mathrm{MnF}_{2}$~\cite{wu2016antiferromagnetic} and $\mathrm{FeF}_{2}$~\cite{li2019spin} that can be used for generation of nonequilibrium magnons for spin caloritronic devices.
The controllable generation of coherent spin currents at terahertz frequencies, due to the spin-pumping effect, has been demonstrated in thin film of~$\mathrm{MnF}_{2}$~\cite{vaidya2020subterahertz}.
It has been shown that phonon frequencies~\cite{lockwood1988spin,schleck2010elastic} and dielectric permittivity in the $\mathrm{MnF}_{2}$~\cite{seehra1984anomalous,seehra1986effect} are changed at the antiferromagnetic ordering that indicates a coupling of phonons with spins in the \TF{} materials.

The cobalt fluoride \CoF{} is one of the most fascinating \TF{} compounds because the $\mathrm{Co}^{2+}$ cation exhibits a not fully frozen orbital momentum with a strong spin-orbit interaction leading to an exceptionally high-magnetic anisotropy~\cite{moriya1956magnetic,correa2018electronic} and large magnetostriction~\cite{jauch2004gamma,chatterji2010magnetoelasticCoF2}.
Thus, the strong static~\cite{borovik1960piezomagnetism,moriya1959piezomagnetism,phillips1967piezomagnetism}, phonon-driven~\cite{disa2020polarizing} and laser-induced THz~\cite{formisano2022laser_JPCM,formisano2022laser_AP} piezomagnetic effects have been observed.
Furthermore, \CoF{} has been a model material for nonlinear phononics and magneto-phononics in which it was discovered that infrared-active phonons and magnons pumped by terahertz pulses efficiently excite coherent Raman-active phonons~\cite{disa2020polarizing,mashkovich2021terahertz,metzger2023impulsive}.
It is worth mentioning that most of basic physical properties of this, as well as of some other \TF{} crystals, are fairly known. These studies include the experimental exploration of the magnetic and optical properties even considering experiments in the high magnetic field~\cite{ozhogin1964antiferromagnets,ozhogin1968behavior,gufan1974dependence,kharchenko1982magnetooptical,gurtovoi1982noncolinear}.  
The latter studies partially cover the spin excitations and Raman-active phonons at the center of the Brillouin zone have been investigated experimentally in a wide temperature range~\cite{macflane1970raman,ishikawa1971magnons,meloche2014one,cipriani1971raman,natoli1973two,meloche2007two,cowley1973magnetic,macfarlane1970light}.
Nevertheless, there is a lack of understanding of the infrared-active phonons and magnetic-dipole active excitations in a wide range of temperatures and some studies only cover a few temperature points~\cite{barker1965infrared,balkanski1966infrared,haussler1981infrared}. Moreover, the temperature behavior of dielectric properties of \CoF{} remains generally unexplored.

In this paper, we report the results of a detailed experimental study of the lattice and spin dynamics of \CoF{} in a wide temperature range by using far-infrared and dielectric spectroscopy techniques.
The latter findings are supported by appropriate first-principles calculations and their careful analysis. 
We show that frequencies of the polar phonons present a clear shift at the antiferromagnetic ordering. 
Lattice dynamics calculations have shown that these frequency shifts are due to the spin-phonon coupling.
Moreover, we have revealed that observed changes in the low-frequency dielectric permittivity at the antiferromagnetic ordering, due to the spontaneous magnetodielectric effect, are caused by the coupling of spins with polar phonons. 

\section{Methods}
\subsection{Crystal and magnetic structures:}

\begin{figure}
\centering
\includegraphics[width=\columnwidth]{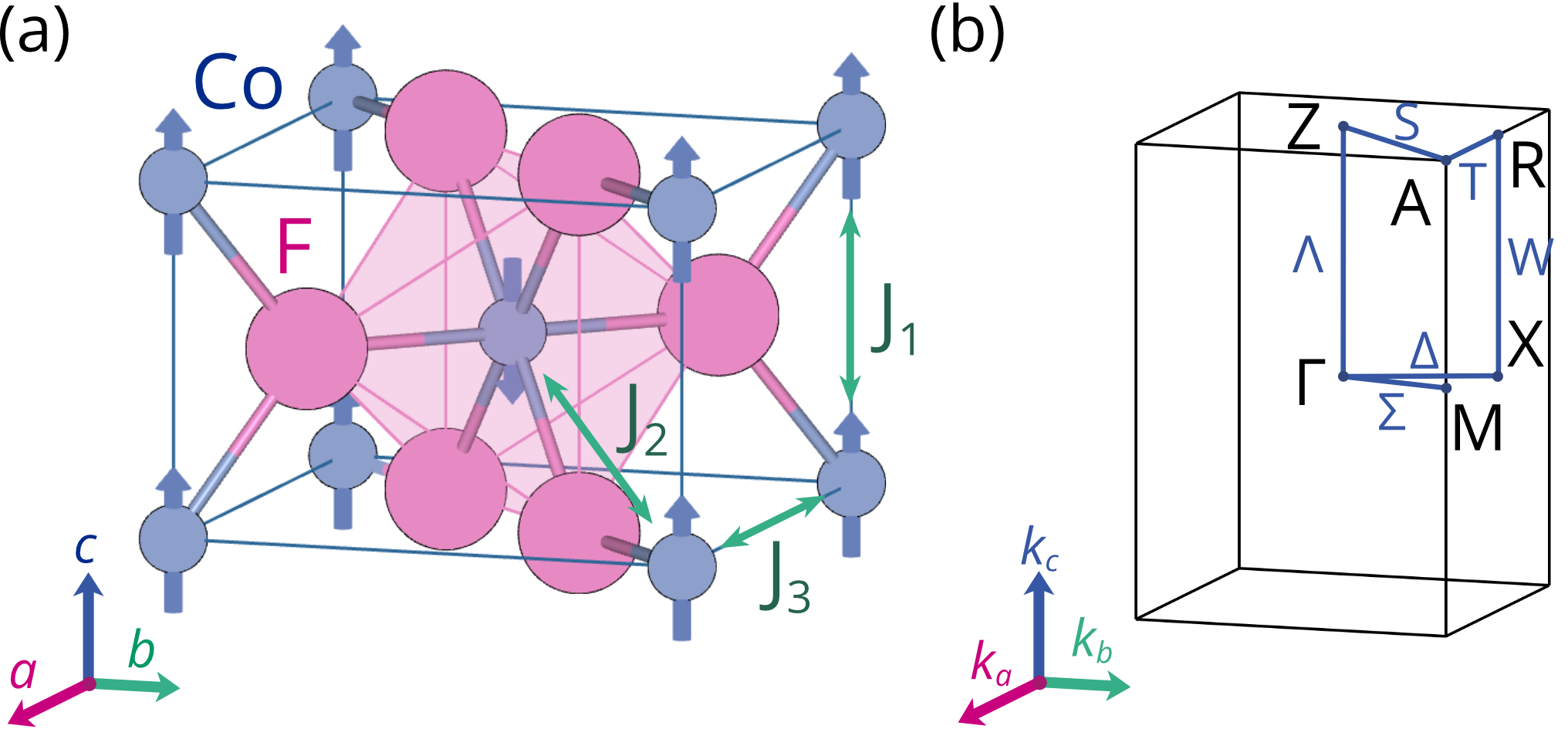}
\caption{\label{fig:structure}
(a)~Crystal and magnetic structures of \CoF{} with tetragonal space group $P4_{2}/mnm$.
The spins of \ch{Co^{2+}} ions are antiferromagnetically ordered along the $c$ axis as marked by arrows.
The \ch{Co^{2+}} ions in the unit cell are numbered.
The dominant exchange interactions $J_{1}$, $J_{2}$, and $J_{3}$ are also marked.
(b)~The Brillouin zone of rutile indicating the high-symmetry $\mathrm{M}$--$\Gamma$--$\mathrm{X}$--$\mathrm{R}$--$\mathrm{A}$--$\mathrm{Z}$--$\Gamma$ path used in the lattice dynamic and magnon calculations.
}
\end{figure}

Cobalt fluoride \CoF{} crystallizes in a centrosymmetric tetragonal rutile crystal structure with the space group $P4_{2}/mnm$ ($D_{4h}^{14}$, \#136, $Z=2$)~\cite{stout1954crystal} as shown in Fig.~\ref{fig:structure}(a).
The lattice parameters measured at room temperature are $a=b=4.695$\,\AA{} and $c=3.1817$\,\AA{}~\cite{costa1993charge}.
The rutile structure consists of alternating neighboring $\mathrm{CoF}_{6}$ octahedra which share edges and corners.
The $\mathrm{Co}^{2+}$ and $\mathrm{F}^{1-}$ ions are located at the Wyckoff positions 2a (0, 0, 0) and 4f (0.30346, 0.30346, 0.0000) with site symmetries $D_{2h}$ and $C_{2v}$, respectively~\cite{costa1993charge}.
The antiferromagnetic phase transition in \CoF{} takes place at $T_{N}=39$\,K and the magnetic space group $P4'_{2}/mnm'$ is realized~\cite{thomson2014cof2}.
At low temperatures in the antiferromagnetic phase the spins $S=3/2$ of \ch{Co^2+} ($3d^{7}$) ions are aligned along the $c$ axis with opposite directions at the center and corners~\cite{stout1953magnetic,erickson1953neutron,strempfer2004magnetic} as shown by purple arrows in Fig.~\ref{fig:structure}(a).
The experimental magnetic structure of \CoF{} can be adequately described by the three exchange integrals between the nearest-neighboring \ch{Co^2+} ions with values $J_{1}=-0.15$\,meV, $J_{2}=1.6$\,meV, and $J_{3}\sim{0}$\,meV~\cite{meloche2007two} which are indicated in Fig.~\ref{fig:structure}(a). 

\subsection{Experimental Details:}
The single crystal of \CoF{} was grown by the Bridgman method in platinum crucibles in a helium atmosphere as described in Ref.~\cite{eremenko1982rearrangement}.
Samples with the $ab$ and $ac$ planes were cut from the x-ray oriented single crystal and polished to optical quality.
The surface size of the samples for far-infrared experiments were about $8\times{8}$\,mm$^{2}$ whereas the thicknesses were about 5\,mm.
Samples for dielectric experiments were prepared in a form of plane-parallel plates with a thickness of about 1.5\,mm and an area of about 50\,mm$^{2}$.

The far-infrared (IR) reflectivity \textcolor{newtext2}{and transmission} measurements were carried out in the spectral range 50--650\,cm$^{-1}$ \textcolor{newtext2}{and 30--300\,cm$^{-1}$, respectively,} using Bruker~IFS~125HR spectrometer with a liquid helium-cooled bolometer.
The $ac$ plane sample was mounted on a cold finger of a closed-cycle helium cryostat Cryomech~ST403.
\textcolor{newtext2}{The polarization of electric field $\bm E$ was switched between two positions of the linear thin film polarizer along the $a$ and $c$ axes of the crystal.}
\textcolor{newtext2}{In reflectivity measurements,} the incident light beam was at 10$^{\circ}$ from the normal to the sample plane.
The relative reflectivity spectra from the sample with respect to a reference reflectivity of a gold mirror at room temperature were measured at slow continuous cooling from 300 to 5\,K.
No corrections on the surface quality and shape of the sample, as well as change of positions due to cold finger thermal contraction, were done.
The absolute reflectivity spectra were obtained by means of the Bruker~IFS~66v/S spectrometer in the range of 50--7500\,cm$^{-1}$ in order to determine the values of high-frequency dielectric permittivity $\varepsilon_{\infty}$.

The dielectric permittivity $\varepsilon^{\mathrm{lf}}$ in the frequency range from 20\,Hz to 1\,MHz was measured using precision RLC meter AKTAKOM AM-3028.
The $ab$ and $ac$ plane samples were used to measure the dielectric permittivity $\varepsilon^{\mathrm{lf}}$ along the $a$ ($a=b$) and $c$ axes, respectively.
Electric contacts were deposited on the sample surfaces using silver paint to form a capacitor.
Samples were placed in a helium flow cryostat Cryo CRC-102 and capacitance measurements were performed at continuous heating from 5 to 300\,K.
Experimental data are presented only for the frequency of 100\,kHz because no noticeable dispersion was observed in the studied frequency range.
The dielectric losses are very small, on the order of 10$^{-5}$, and no noticeable temperature changes were detected.

\subsection{Computational Details:}
The obtained experimental results were supplemented by the lattice dynamic calculations within the density-functional theory (DFT) framework~\cite{hohenberg1964inhomogeneous,kohn1965self} and the projector-augmented wave method~\cite{blochl1994projector} as implemented in the Vienna \textit{ab initio} simulation package (\textsc{vasp})~\cite{kresse1996efficient,kresse1999from}.
The configurations considered in the pseudopotentials for the valence electrons were $\mathrm{Co}$ ($3p^{6}3d^{7}4s^{2}$, version 23Apr2009) and $\mathrm{F}$ ($2s^{2}2p^{5}$, version 08Apr2002).
The Monkhorst-Pack scheme with a mesh 6$\times$6$\times$8 has been employed for the Brillouin zone integration~\cite{monkhorst1976special}.
As such, an $E_{\mathrm{cut}} = 600$\,eV energy cutoff was used to give forces convergence of less than 0.001\,eV$\cdot$\AA$^{-1}$.
The exchange-correlation energy was represented within the generalized gradient approximation (GGA) in the Perdew-Burke-Ernzerhof revised for solids PBEsol form~\cite{perdew2008restoring}.
The $\mathrm{GGA}+U+J$ approximation within the Dudarev's formalism~\cite{dudarev1998electron} with the on-site Coulomb interaction $U=6.0$\,eV and an exchange correction $J=1.0$\,eV were used to account for the strong correlation between electrons in the $3d^{7}$ shell of the \ch{Co^{2+}} ions~\cite{bousquet2010dependence}.
The dielectric constants, phonon dispersions, eigenvectors, and Born effective charges were calculated within the density-functional perturbation theory~\cite{gonze1997dynamical} and analyzed through the \textsc{Phonopy} code~\cite{togo2015first}.
The longitudinal-transverse optical (LO-TO) phonon splitting near the $\Gamma$-point was included using nonanalytical corrections to the dynamical matrix~\cite{wang2010mixed}.
Finally, the correlation between eigendisplacements of the LO and TO modes was evaluated using \textsc{LADYtools}~\cite{ladyteam}.
Moreover, we use the LKAG Green's function method~\cite{liechtenstein1987local} as implemented in the TB2J code~\cite{he2021tb2j} to calculate the exchange interaction constants of \CoF{} following the process described in Ref.~\cite{tellez2024systematic}.
Particularly, we interface TB2J with the Siesta package~\cite{soler2002siesta}, where we use a 400\,Ry mesh-cutoff energy and a double-zeta polarized basis in addition to the previously mentioned DFT parameters.
We compare the LKAG results with those obtained from \textsc{vasp} with an energy mapping procedure. 
Finally, we use linear spin-wave theory~\cite{toth2015linear} to calculate the magnon frequencies. 

\section{Results and Discussion}
\subsection{Far-infrared spectroscopy:}
The group-theoretical analysis for the rutile \CoF{} predicts 13 phonon modes at the $\Gamma$ point of the Brillouin zone~\cite{kroumova2003bilbao}:
\begin{equation}
\label{eq:group_irrep_total}
\begin{split} 
\Gamma_{\mathrm{total}} = \underbrace{A_{2u} \oplus E_{u}}_{\Gamma_{\mathrm{acoustic}}} \oplus \underbrace{A_{1g} \oplus B_{1g} \oplus B_{2g} \oplus E_{g}}_{\Gamma_{\mathrm{Raman}}} \oplus \\ \oplus \underbrace{A_{2u} \oplus 3 E_{u}}_{\Gamma_{\mathrm{IR}}} \oplus \underbrace{A_{2g} \oplus 2 B_{1u}}_{\Gamma_{\mathrm{Silent}}},
\end{split}
\end{equation}
among which there are 2 acoustic $\Gamma_{\mathrm{acoustic}}  =  A_{2u} \oplus E_{u}$, three ``silent'' $\Gamma_{\mathrm{silent}}  =  A_{2g} \oplus 2 B_{1u}$, four Raman-active $\Gamma_{\mathrm{Raman}}  = A_{1g} \oplus B_{1g} \oplus B_{2g} \oplus E_{g}$, and four infrared-active (polar) $\Gamma_{\mathrm{IR}}  = A_{2u} \oplus 3 E_{u}$ phonons.
The $A$ and $B$ modes are nondegenerated, whereas the $E$ modes are doubly degenerated.
The polar $A_{2u}$ phonon is polarized along the $c$ axis whereas $3 E_{u}$ phonons are active in the $ab$ plane.

\begin{figure}
\centering
\includegraphics[width=\columnwidth]{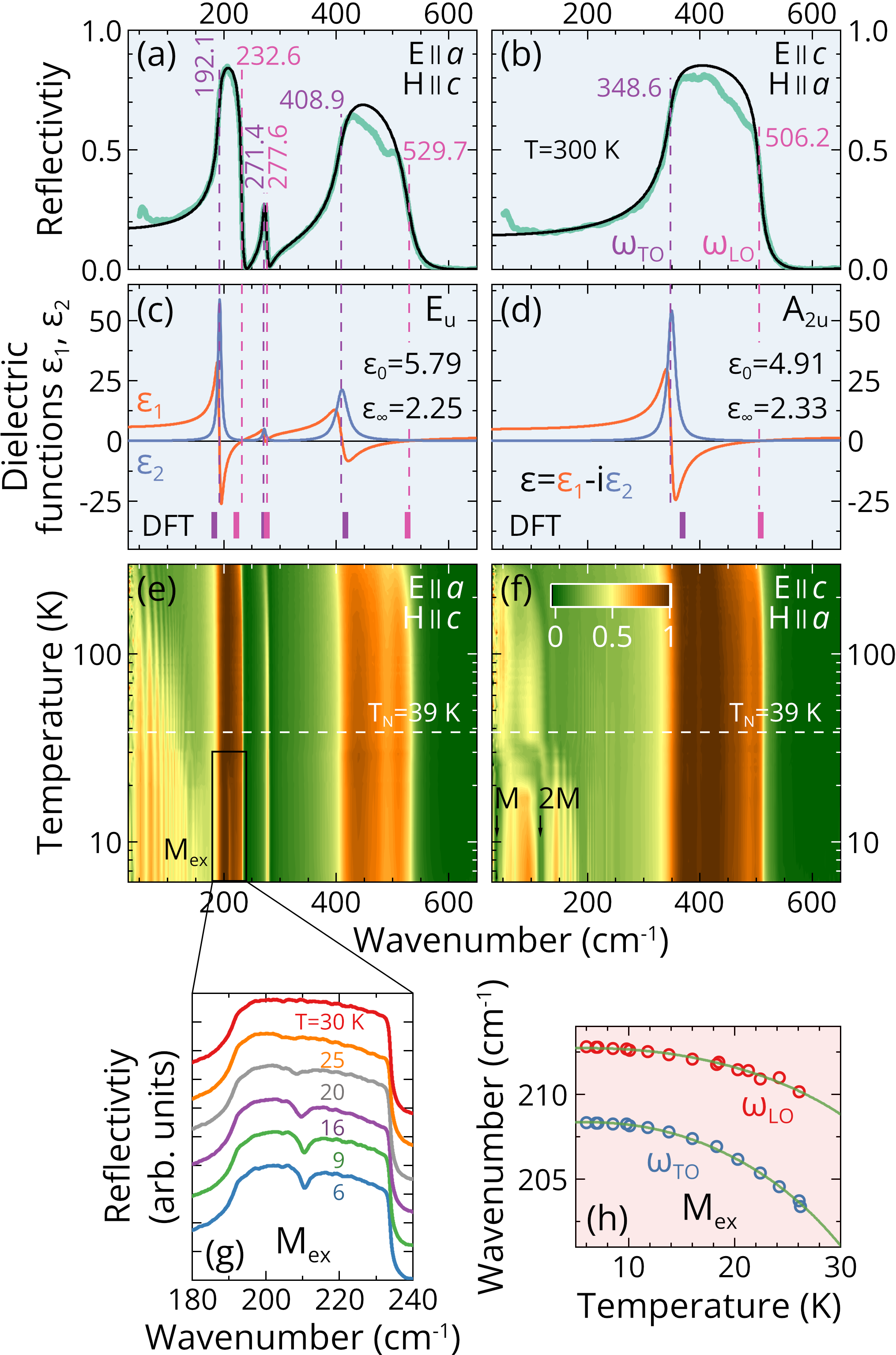}
\caption{\label{fig:reflectivity}
IR reflectivity spectra with the electric field of light $\bm E$ polarized along (a)~$a$ and (b)~$c$ axes in the $ac$ plane for \CoF{} at room temperature.
The black lines are fits based on a generalized oscillator model according to Eq.~\eqref{eq:epsilon_TOLO}.
Real $\varepsilon_{1}$ and imaginary $\varepsilon_{2}$ parts of the dielectric functions corresponding to the (c)~$E_{u}$ and (d)~$A_{2u}$ polar phonons.
Colored dashed lines indicate the experimental $\omega_{\mathrm{TO}}$ and $\omega_{\mathrm{LO}}$ phonon frequencies which are indicated.
Colored sticks on the (c) and (d) panels present the calculated frequencies of polar phonons.
Temperature maps of IR reflectivity spectra for the light polarizations (e)~$\mathbf{E} \parallel a$ and (f)~$\mathbf{E} \parallel c$.
Magnon ($\mathrm{M}$), two-magnon ($\mathrm{2M}$) and magnetic $\mathrm{M}_{\mathrm{ex}}$ excitations are marked. 
(g)~Effect of magnetic excitation $\mathrm{M}_{\mathrm{ex}}$ on the reflectivity spectra at the indicated temperatures.
(h)~Temperature dependencies of frequencies of the magnetic excitation $\mathrm{M}_{\mathrm{ex}}$.}
\end{figure}

The IR reflectivity spectra at room temperature with the polarization $\mathbf{E}$ of light parallel to the $a$ and $c$ axis are shown in Figs.~\ref{fig:reflectivity}(a) and~\ref{fig:reflectivity}(b), respectively.
Three and one reflection bands are observed in the spectra for nonequivalent polarizations, which correspond to the $E_{u}$ ($\mathbf{E}\parallel{a}$) and $A_{2u}$ ($\mathbf{E}\parallel{c}$) polar phonons in full agreement with group theory predictions.  
To obtain the phonon parameters the reflectivity spectra $R(\omega)$ were fitted using the Fresnel equation~\cite{born2013principles}
\begin{equation}
\label{eq:reflectivity}
R(\omega) = \Bigl|\frac{\sqrt{\varepsilon(\omega)} - \sqrt{\mu(\omega)}}{\sqrt{\varepsilon(\omega)} + \sqrt{\mu(\omega)}}\Bigr|^2,
\end{equation}
with the factorized form of the complex dielectric permittivity~\cite{gervais1974anharmonicity}
\begin{equation}
\label{eq:epsilon_TOLO}
\varepsilon(\omega) = \varepsilon_{1}(\omega) - i\varepsilon_{2}(\omega) = \varepsilon_{\infty}\prod\limits_{j}^{N}\frac{{\omega^{2}_{j\mathrm{LO}}} - {\omega}^2 + i\gamma_{j\mathrm{LO}}\omega}{{\omega^{2}_{j\mathrm{TO}}} - {\omega}^2 + i\gamma_{j\mathrm{TO}}\omega},
\end{equation}
where $\varepsilon_{\infty}$ is the high-frequency dielectric permittivity, $\omega_{j\mathrm{LO}}$, $\omega_{j\mathrm{TO}}$, $\gamma_{j\mathrm{LO}}$ and $\gamma_{j\mathrm{TO}}$ correspond to $\mathrm{LO}$ and $\mathrm{TO}$ frequencies ($\omega_{j}$) and dampings ($\gamma_{j}$) of the $j$th polar phonon, respectively.
Multiplication occurs over all $N$ polar phonons which are active for the relevant polarization of light.
Far from the magnetic excitations, it is assumed that magnetic permeability $\mu \equiv 1$. 
There is, in general, a good agreement between experimental green and fit black lines, as shown in Figs.~\ref{fig:reflectivity}(a) and~\ref{fig:reflectivity}(b).
Slight deviations appear at the highest-frequency phonons presumably due to multiphonon absorption~\cite{benoit1988dynamical}.
The real $\varepsilon_{1}$ and imaginary $\varepsilon_{2}$ parts of the complex dielectric functions $\varepsilon = \varepsilon_{1} - i\varepsilon_{2}$ which correspond to the fits are shown in Figs.~\ref{fig:reflectivity}(c) and~\ref{fig:reflectivity}(d).
The obtained parameters of the polar phonons from fits using Eqs.~\eqref{eq:reflectivity} and~\eqref{eq:epsilon_TOLO} of the experimental data at room temperature are listed in Table~\ref{tab:phonon_parameters}.
These parameters are in fair agreement with the literature data for \CoF~\cite{barker1965infrared,balkanski1966infrared}.
\textcolor{newtext}{Note that LO-TO splitting does not depend on the degeneracy of polar phonons~\cite{berker1964transverse}}.

\begin{table}[b]
    \caption{\label{tab:phonon_parameters} Experimental TO and LO frequencies $\omega_{j}$ (cm$^{-1}$), dampings $\gamma_{j}$ (cm$^{-1}$) and dielectric strengths $\Delta\varepsilon_{j}$ of the polar phonons, static $\varepsilon_{0}$ and high-frequency $\varepsilon_{\infty}$ dielectric permittivities in \CoF{} at room temperature.
    For comparison, the calculated frequencies for the nonmagnetic state are also given in parentheses.}
    \begin{ruledtabular}
            \begin{tabular}{ccccccc}
             Sym. & $j$ & $\omega_{j\mathrm{TO}}$ & $\gamma_{j\mathrm{TO}}$ & $\omega_{j\mathrm{LO}}$ & $\gamma_{j\mathrm{LO}}$ & $\Delta\varepsilon_{j}$\\
             \hline
              \multirow{3}{*}{$E_{u}$} & 1 & 192.1 (182.7) & 7.0  & 232.6 (222.3) &  4.2 & 2.14 (2.26)\\
                                       & 2 & 271.4 (271.8) & 6.7  & 277.6 (276.2) &  5.3 & 0.12 (0.103)\\
                                       & 3 & 408.9 (415.7) & 24.5 & 529.7 (526.9) & 28.1 & 1.28 (1.33)\\\cmidrule{2-7} 
                                       & \multicolumn{6}{c}{$\varepsilon_{\infty}=2.25$ (2.54) \qquad $\varepsilon_{0} = 5.79$ (6.24)}\\
             \hline
             \multirow{1}{*}{$A_{2u}$} & 1 & 348.6 (369.8) & 16.6 & 506.2 (507.1) & 13.0 & 2.58 (2.32)\\\cmidrule{2-7}
             & \multicolumn{6}{c}{$\varepsilon_{\infty} = 2.33$ (2.63) \qquad $\varepsilon_{0} = 4.91$ (4.95)}\\
        \end{tabular}
    \end{ruledtabular}
\end{table}

\begin{figure}
\centering
\includegraphics[width=\columnwidth]{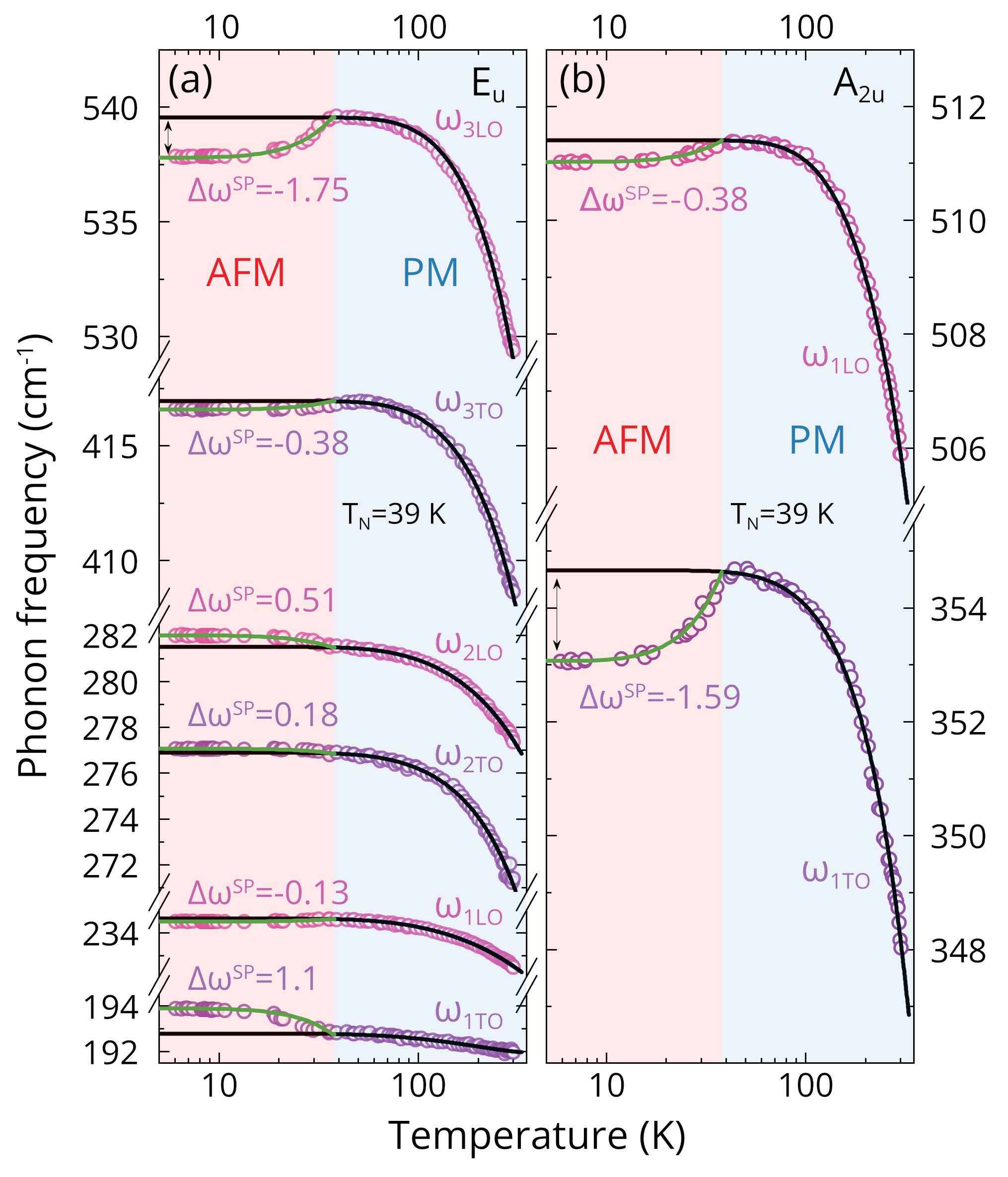}
\caption{\label{fig:phonon}
Temperature dependences of frequencies $\omega$ of the (a)~$E_{u}$ and (b)~$A_{2u}$ polar phonons in \CoF{} on a logarithmic scale.
The color circles are experimental data.
The black lines corresponds to the fit under the assumption of the anharmonic temperature behavior only according to Eq~\eqref{eq:omega_anharmonism}.
The green lines are fits of the frequency shift due to the spin-phonon coupling according to Eqs.~\eqref{eq:omega_SP}, \eqref{eq:brillouin}.
The antiferromagnetic and paramagnetic phases are shown in the red- and blue-filled backgrounds, respectively.
}
\end{figure}

To reveal the temperature evolution of the polar phonons in \CoF{}, we perform the measurements of the infrared reflectivity in the range from 5 to 300\,K for both polarizations which are shown by the color maps in Figs.~\ref{fig:reflectivity}(e) and~\ref{fig:reflectivity}(f).
The bright vertical brown bands correspond to polar phonons and it can be seen that their evolution with temperature is not significant.
\textcolor{newtext}{It should be noted that we did not detect in \CoF{} any evidences of the exchange driven phonon splitting below $T_{N}$ which was previously observed in the antiferromagnetic cubic transition-metal monoxides and oxide spinels~\cite{hemberger2006spin,rudolf2008magnetic,kant2009optical,kant2012universal}.
We attribute the absence of clear signs of the exchange driven phonon splitting in the antiferromagnetic rutiles \CoF{} and $\mathrm{MnF}_{2}$ from Ref.~\cite{schleck2010elastic} to the fact that they have tetragonal crystal structure and they have no magnetic frustration, which is apparently essential for this effect.}

Further, the obtained reflectivity spectra were analyzed using Eqs.~\eqref{eq:reflectivity} and~\eqref{eq:epsilon_TOLO}.
The derived temperature dependences of the TO and LO frequencies of $E_{u}$ and $A_{2u}$ polar phonons are shown on a logarithmic scale in Figs.~\ref{fig:phonon}(a) and~\ref{fig:phonon}(b).
In the paramagnetic state $T > T_{N}$, the TO and LO frequencies $\omega$ of all polar phonons increase ({\it{i.e.}} harden) at cooling. 
It should be noted that the hardening of phonon frequencies is typical for conventional insulator crystals due to anharmonic effects which manifest themselves in the compression of the crystal at cooling~\cite{wei2021phonon}.
In turn, a decrease in the distance between ions in a crystal leads to an increase of the interaction forces $k$ between them, which drives an increase of the phonon frequencies $\omega \sim \sqrt{{k}/{\mu}}$, where $\mu$ is a reduced mass~\cite{born1954dynamical}. 
According to the Ref.~\cite{chatterji2010magnetoelasticCoF2} the lattice parameters of \CoF{} decrease at cooling in the paramagnetic phase.
The observation in the isostructural antiferromagnet $\mathrm{MnF}_{2}$ presents a fascinating phenomenon where, contrary to typical behavior, the frequency $\omega_{1\mathrm{TO}}$ of the polar phonon exhibiting $E_{u}$ symmetry decreases on cooling.
This anomaly underscores a unique aspect of phonon behavior in antiferromagnetic materials, challenging conventional understandings and inviting a deeper investigation into the underlying mechanisms~\cite{schleck2010elastic}.

It is clearly seen in Fig.~\ref{fig:phonon} that the temperature dependencies of all polar phonons abruptly change the smooth behavior at the N{\'e}el temperature $T_{N} = 39$\,K evidencing the influence of the spin-phonon coupling.
For a deeper understanding of this effect, we fitted the temperature dependencies of phonon frequencies in the paramagnetic state using expression describing the three- and four-phonon anharmonic processes~\cite{balkanski1983anharmonic,lan2012phonon}
\begin{eqnarray}
\label{eq:omega_anharmonism}
\omega^{\mathrm{NM}}(T) = \omega_{0} + A \left( 1 + \frac{2}{e^{\hbar\omega_{0}/2k_{B}T} - 1} \right)
\nonumber\\
+ B \left( 1 + \frac{3}{e^{\hbar\omega_{0}/3k_{B}T} - 1} + \frac{3}{(e^{\hbar\omega_{0}/3k_{B}T} - 1)^2} \right),
\end{eqnarray}
where $\omega_{0}$ is harmonic phonon frequency, $A$ and $B$ are parameters.
It is worth noting that the phonon anharmonicity in the real crystals is very complicated, but Eq.~\eqref{eq:omega_anharmonism} gives reasonable fits above $T_{N}$ as shown by the black lines in the Fig.~\ref{fig:phonon}.
The frequency shifts of the phonon frequencies from extrapolated anharmonic behavior in the antiferromagnetic phase were fitted by the function~\cite{cottam2019spin}:
\begin{equation}
\label{eq:omega_SP}
\omega^{\mathrm{AFM}}(T) = \omega^{\mathrm{NM}}(T) + \Delta\omega^{\mathrm{SP}} M^{2}(T),
\end{equation}
where $\Delta\omega^{\mathrm{SP}}$ is the spin-phonon coupling constant which is equal to the phonon frequency at the low temperature and $M$ is the magnetic order parameter.
Neglecting the short-range magnetic ordering above $T_{N}$, the temperature dependence of $M$ can be estimated by using the Brillouin function~\cite{darby1967tables} such as:
\begin{equation}
\label{eq:brillouin}
M(x) = M_{0} \biggl[ \frac{2S + 1}{2S} \coth{\Bigl(\frac{2S + 1}{S}x\Bigr)} - \frac{1}{2S} \coth{\Bigl(\frac{x}{2S}\Bigr)} \biggr],
\end{equation}
where $x = \cfrac{3S}{S + 1} \cfrac{M}{M_{0}} \cfrac{T_{N}}{T}$, $S$ is spin value of the magnetic ion, and $M_{0}$ is the full magnetic order parameter.
There is a fair agreement between fits using Eqs.~\eqref{eq:omega_SP}, \eqref{eq:brillouin} and experimental data below $T_{N}$ as shown in Fig.~\ref{fig:phonon}.
The obtained values of the spin-phonon coupling constants $\Delta\omega^{\mathrm{SP}}$ given in Fig.~\ref{fig:phonon} are close to the published data for Raman-active phonons~\cite{cottam2019spin}.
It is worth nothing that spin-phonon coupling for IR-active TO phonons has been reported for the isostructural antiferromagnet \MnF{}~\cite{schleck2010elastic}.

Here, it can be observed that the temperature maps of the IR reflectivity spectra reveal features at about 37, 115, and 209\,cm$^{-1}$ appearing at antiferromagnetic ordering, as it can be seen in Figs.~\ref{fig:reflectivity}(e) and~\ref{fig:reflectivity}(h).
According to the numerous published data, the feature at 37\,cm$^{-1}$ corresponds to the magnon excitation with symmetry $\Gamma^{+}_{3} + \Gamma^{+}_{4}$, 115\,cm$^{-1}$ ($\Gamma^{+}_{4} + \Gamma^{+}_{5}$) is two-magnon mode, and 209\,cm$^{-1}$ ($\Gamma^{+}_{2}$) is magnetic exciton~\cite{moriya1966far,martel1968experimental,macfarlane1970light,ishikawa1971magnons,allen1971magnetic1,allen1971magnetic2,cipriani1971raman,natoli1973two,meloche2007two,meloche2014one}.
Magnetic excitons in \CoF{} are usually considered as a transition between the split by crystal and exchange fields, and the spin-orbit coupling of $\mathrm{Co}^{2+}$ ions levels~\cite{newman1959infrared,lines1965magnetic,martel1968experimental,ishikawa1971magnons}.
At that the magnetic exciton with the lowest frequency is a magnon.
It is interesting that the magnetic excitation $\mathrm{M}_{\mathrm{ex}}$ with frequency 209\,cm$^{-1}$ is located in the Reststrahlen band between 2TO and 2LO frequencies of the polar $E_{u}$ phonon and appears by a dip in the reflection spectra, as shown in Figs.~\ref{fig:reflectivity}(e) and~\ref{fig:reflectivity}(g), and it was previously discovered in Ref.~\cite{haussler1981infrared}.
Note that, this magnetic excitation $\mathrm{M}_{\mathrm{ex}}$ appears in the polarized IR reflectivity spectra on the polar $E_{u}$ phonon only at $\mathbf{E} \parallel a$, $\mathbf{H} \parallel c$ in the $ac$ plane and does not appear at $\mathbf{E} \parallel a$, $\mathbf{H} \parallel b$ in the $ab$ plane~\cite{haussler1981infrared,haussler1982far} which confirms its magnetodipole character.

To reveal the temperature dependence of the frequency of this magnetic excitation $\mathrm{M}_{\mathrm{ex}}$, the feature on the IR reflection spectra was fitted using Eqs.~\eqref{eq:reflectivity}, \eqref{eq:epsilon_TOLO} and magnetic permeability in the form which is close to the Eq.~\eqref{eq:epsilon_TOLO}~\cite{haussler1982far}
\begin{equation}
\label{eq:mu_TOLO}
\mu(\omega) = \mu_{1}(\omega) - i\mu_{2}(\omega) = \frac{{\omega^{2}_{\mathrm{LO}}} - {\omega}^2 + i\gamma_{\mathrm{LO}}\omega}{{\omega^{2}_{\mathrm{TO}}} - {\omega}^2 + i\gamma_{\mathrm{TO}}\omega},
\end{equation}
where $\omega_{\mathrm{LO}}$ and $\omega_{\mathrm{TO}}$ correspond to LO and TO frequencies of the magnetic excitation.
The recieved temperature dependences of $\omega_{\mathrm{LO}}$ and $\omega_{\mathrm{TO}}$ frequencies of the $\mathrm{M}_{\mathrm{ex}}$ and fit lines obtained using Eq.~\eqref{eq:brillouin} are shown in Fig.~\ref{fig:reflectivity}(h) according to Ref.~\cite{johnson1959antiferromagnetic}. 

\begin{figure}
\centering
\includegraphics[width=\columnwidth]{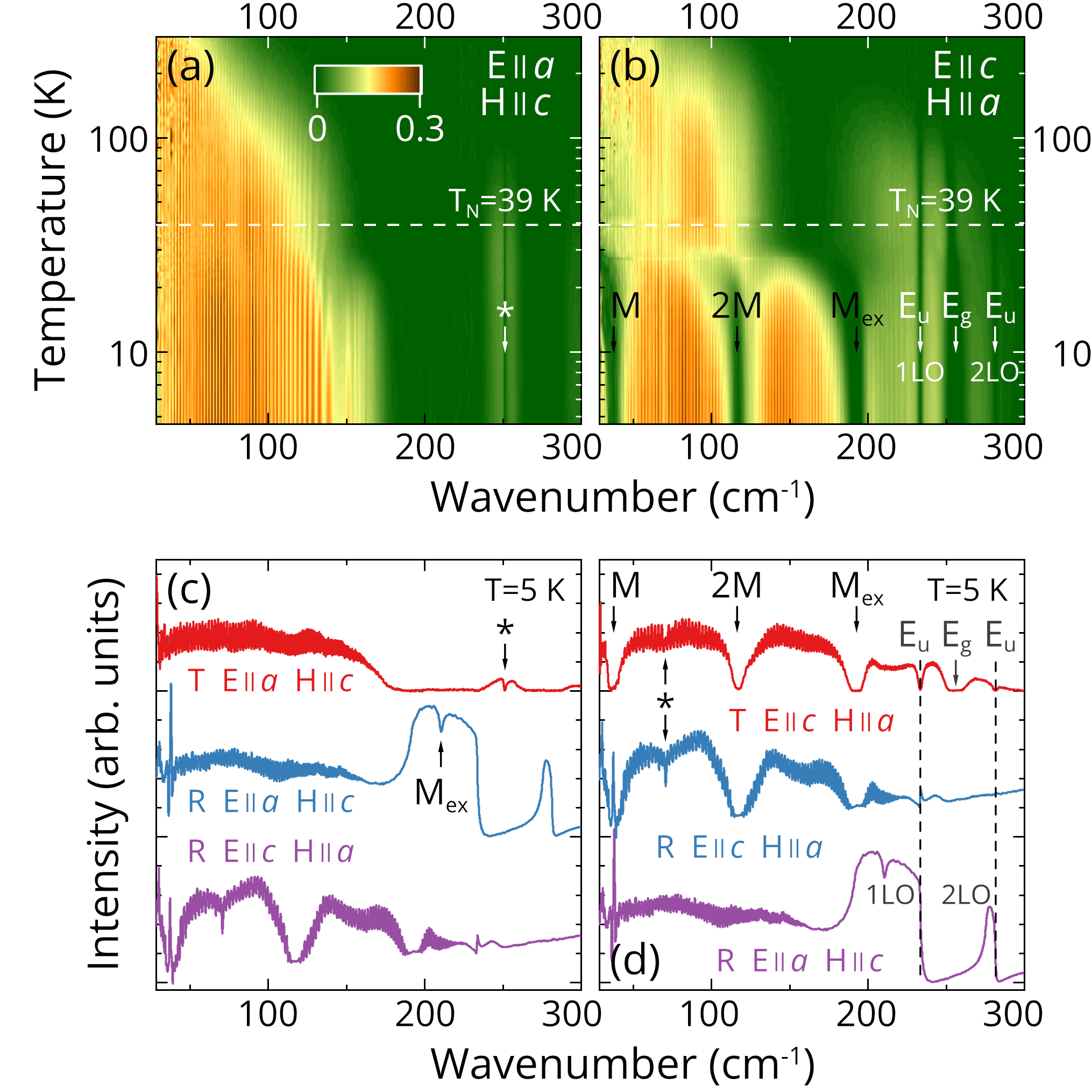}
\caption{\label{fig:transmissivity}
Temperature maps of the IR transmission spectra for \CoF{} for the light polarization (a)~$\mathbf{E} \parallel a$ and (b)~$\mathbf{E} \parallel c$ in the $ac$ plane.
Additionally, the transmission (T) and reflection (R) spectra at $T=5$\,K for the light polarization are presented for (c)~$\mathbf{E} \parallel a$ and (d)~$\mathbf{E} \parallel c$ along with the reflection spectra for another polarization in the $ac$ plane.
The detected excitations are described in the text. 
}
\end{figure}

At low temperatures, the \CoF{} crystal becomes more transparent in the low-frequency region below 200\,cm$^{-1}$ (see Fig.~\ref{fig:transmissivity}). The reflection from the back side of the sample with double absorption at its thickness and interference are added to the reflection from the front side as shown in Figs.~\ref{fig:reflectivity}(e) and~\ref{fig:reflectivity}(f).
Thus, the absorption related to magnon $\mathrm{M}$ at 37\,cm$^{-1}$ and two-magnon $\mathrm{2M}$ excitation at 115\,cm$^{-1}$ are observed in the spectra with  $\mathbf{E} \parallel c$, $\mathbf{H} \parallel a$ as can be seen in Figs.~\ref{fig:reflectivity}(f) and Figs.~\ref{fig:transmissivity}(b).
In addition to the magnon $\mathrm{M}$, and the two-magnon $\mathrm{2M}$ already described, four narrow absorption lines at temperatures both above and below $T_{N}$ were also observed in the transmission spectrum with $\mathbf{E} \parallel c$, $\mathbf{H} \parallel a$, as shown in Fig.~\ref{fig:transmissivity}(b).

Absorption at 193\,cm$^{-1}$ corresponds to another magnetic excitation $\mathrm{M}_{\mathrm{ex}}$ with symmetry $\Gamma^{+}_{3} + \Gamma^{+}_{4}$~\cite{martel1968experimental,ishikawa1971magnons,allen1971magnetic1,allen1971magnetic2,meloche2014one}.
Lines at 234 and 282\,cm$^{-1}$ are 1LO and 2LO frequencies of polar phonons with $E_{u}$ symmetry.
Note that the absorption bands corresponding to LO polar phonons begin to be observed in the spectra at temperatures significantly higher than $T_{N}$ [see Fig.~\ref{fig:transmissivity}(b)]. 
According to the crystal symmetry, the $E_{u}$ phonons are active at $\mathbf{E} \parallel a$.
In the reflection spectrum, these off-symmetry LO modes also appear in forbidden polarizations as shown in Fig.~\ref{fig:transmissivity}(d).
This effect occurs in uniaxial crystals in the infrared reflectivity spectra at the off-normal incidence of radiation with the $p$-polarization~\cite{duarte1987offnormal}.
We observe the appearance of this effect in the transmission spectra.

It is interesting to note that the absorption band at 256\,cm$^{-1}$ corresponds to the Raman-active $E_{g}$ phonon which is infrared inactive.
Magnetodipole absorption of this Raman-active phonon was previously experimentally observed in \CoF{} and it is due to the coupling of magnetic excitations with the $E_{g}$ mode~\cite{allen1968spin,allen1969magnon,mills1970exciton,allen1971spin,allen1971magnetic1,allen1971magnetic2}. 
Moreover, the magnetodipole phonon absorption begins at temperatures well above $T_{N}$, as in the previously reported papers [see Fig.~\ref{fig:transmissivity}(b)]. 
Also, a slight absorption is observed at low temperatures at 71\,cm$^{-1}$ as shown by an asterisk in Fig.~\ref{fig:transmissivity}(d).
The frequency of these features is close to the Raman-active $B_{1g}$ phonon at 65\,cm$^{-1}$, but it is most likely an artifact and is unlikely to be related. 
Weak but distinct absorption line at 251\,cm$^{-1}$ was detected on the transmission spectra with $\mathbf{E} \parallel a$, $\mathbf{H} \parallel c$ below $T=90$\,K [see Fig.~\ref{fig:transmissivity}(a)].
This line is close to the Raman-active $E_{g}$ phonon at 256\,cm$^{-1}$. Still, any absence of temperature dependence indicates that it is related to the small amount of water in the cryostat, and not with the sample under study.
Thus, the \CoF{} possesses several electrodipole and magnetodipole excitations with rich physics manifested in the infrared spectra. 

\subsection{Dielectric properties:}

\begin{figure}
\centering
\includegraphics[width=\columnwidth]{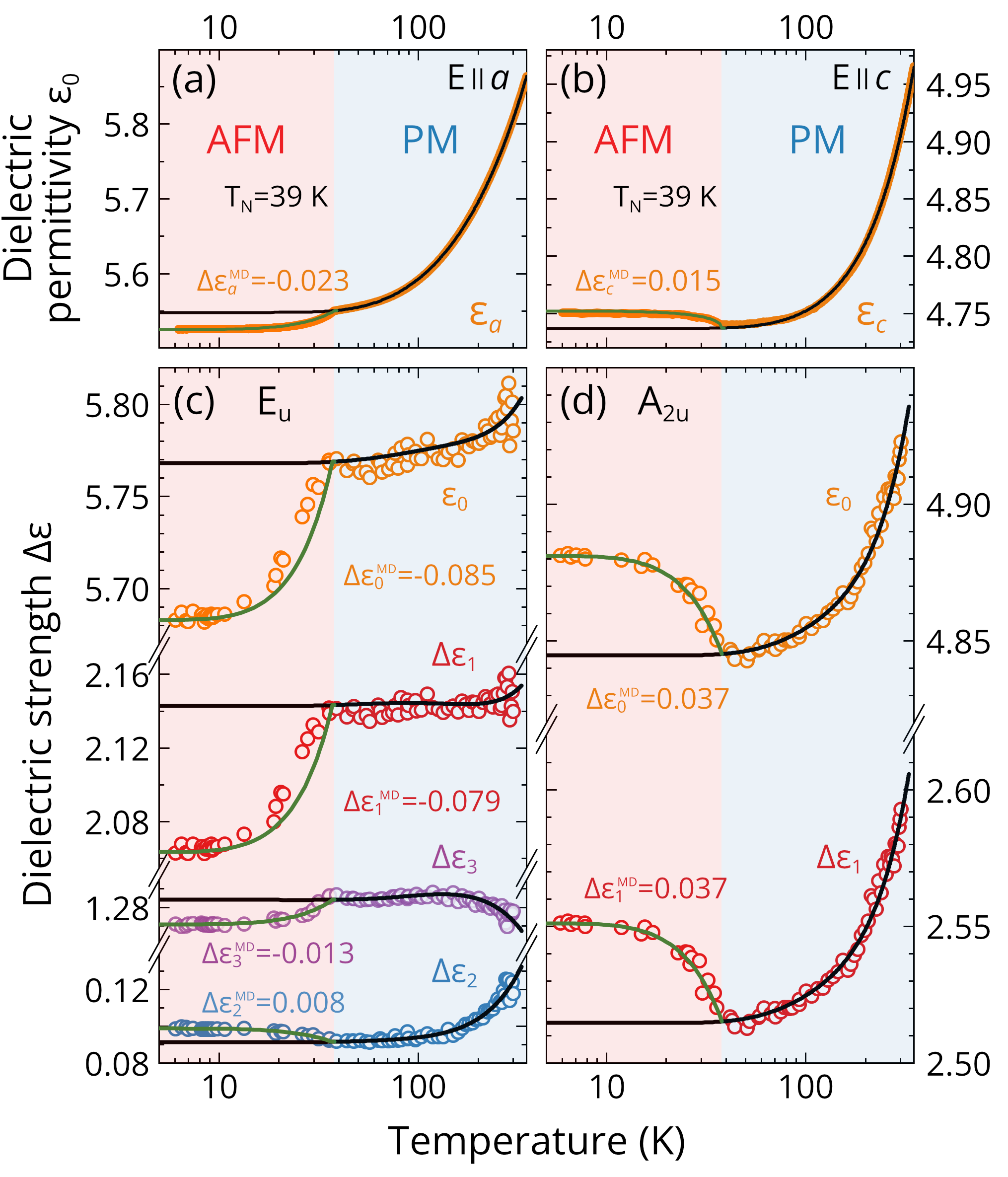}
\caption{\label{fig:dielectric}
Temperature dependences of the low-frequency dielectric permittivity $\varepsilon^{\mathrm{lf}}$ along the (a)~$a$ and (b)~$c$ axes, the dielectric strengths $\Delta\varepsilon_{j}$ and static dielectric permittivity $\varepsilon_{0}$ of the (c)~$E_{u}$ and (d)~$A_{2u}$ polar phonons.
The open color circles correspond to the experimental data.
The black and green lines are fitted, assuming the anharmonic and spontaneous magnetodielectric effects.
Values of the spontaneous magnetodielectric coupling parameter $\Delta\varepsilon^{\mathrm{MD}}$ are given.
The antiferromagnetic and paramagnetic phases are shown in the red- and blue-filled backgrounds, respectively. 
}
\end{figure}

It is well known that the static dielectric permittivity $\varepsilon_{0} = \varepsilon_{\infty} + \sum_{j}\Delta\varepsilon_{j}$ of crystals is related to the $j$th polar phonon thought its dielectric strength $\Delta\varepsilon_{j}$ according to the expression~\cite{gervais1983long}:
\begin{equation}
\label{eq:oscillator_strength_TOLO}
\Delta\varepsilon_{j}  =  \frac{\varepsilon_{\infty}}{{\omega^{2}_{j\mathrm{TO}}}}\frac{\prod\limits_{k}{\omega^{2}_{k\mathrm{LO}}}-{\omega^{2}_{j\mathrm{TO}}}}{\prod\limits_{k\neq{}j}{\omega^{2}_{k\mathrm{TO}}}-{\omega^{2}_{j\mathrm{TO}}}}.
\end{equation}
The static $\varepsilon_{0}$, high frequency $\varepsilon_{\infty}$ dielectric permittivities, and dielectric strengths $\Delta\varepsilon$ evaluated from reflectivity spectra at room temperature using Eq.~\eqref{eq:oscillator_strength_TOLO} are listed in Table~\ref{tab:phonon_parameters} and in Figs.~\ref{fig:reflectivity}(c) and~\ref{fig:reflectivity}(d).
We observed that our obtained values of the dielectric parameters are in fair agreement with previously published data~\cite{barker1965infrared,balkanski1966infrared}.

The temperature behavior of the dielectric permittivity $\varepsilon^{\mathrm{lf}}$ at frequencies (about 1\,MHz) much lower than lattice excitations represents the general temperature trend of the dielectric strengths of polar phonons.
Figures~\ref{fig:dielectric}(a) and~\ref{fig:dielectric}(b) show experimental temperature dependencies of the low-frequency dielectric permittivities $\varepsilon^{\mathrm{lf}}(T)$ measured in \CoF{} along the $a$ and $c$ axes, respectively. 
At all temperatures the relation $\varepsilon^{\mathrm{lf}}_{a}(T) > \varepsilon^{\mathrm{lf}}_{c}(T)$ holds. 
The values $\varepsilon^{\mathrm{lf}}_{a}$ and $\varepsilon^{\mathrm{lf}}_{c}$ at room temperature are close to those obtained from reflectivity spectra (see Table~\ref{tab:phonon_parameters}).
The low-frequency dielectric permittivities $\varepsilon^{\mathrm{lf}}_{a}$ and $\varepsilon^{\mathrm{lf}}_{c}$ decrease at cooling in the paramagnetic phase along both axes and the relative magnitude of these changes is \textcolor{newtext}{comparatively} tiny and amounts about 3\%.
Note that the similar temperature behavior of the low-frequency dielectric permittivity $\varepsilon^{\mathrm{lf}}$ was previously observed in the isostructural diamagnetic crystal $\mathrm{ZnF}_{2}$~\cite{vassiliou1986pressure}.
Moreover, this temperature behavior of the dielectric permittivity $\varepsilon^{\mathrm{lf}}(T)$ is typical for conventional dielectric crystals~\cite{lowndes1970dielectric,bartels1973pressure,wintersgill1979temperature,vassiliou1986pressure}.

At the N{\'e}el temperature $T_{N} = 39$\,K both temperature dependencies of the dielectric permittivity $\varepsilon^{\mathrm{lf}}(T)$ exhibit kinks due to the spontaneous magnetodielectric effect, as shown in Figs.~\ref{fig:dielectric}(a) and~\ref{fig:dielectric}(b).
Below $T_{N}$ the value of $\varepsilon^{\mathrm{lf}}_{a}$ decreases, while $\varepsilon^{\mathrm{lf}}_{c}$, on the contrary, increases at cooling [see Figs.~\ref{fig:dielectric}(a) and~\ref{fig:dielectric}(b)].  
It is worth mentioning that similar temperature behavior of the $\varepsilon^{\mathrm{lf}}_{a}$ and $\varepsilon^{\mathrm{lf}}_{c}$ below $T_{N}$ was previously observed in the isostructural antiferromagnet \MnF{}~\cite{seehra1984anomalous,seehra1986effect}.
To extract the spontaneous magnetodielectric effect, the temperature behavior above $T_{N}$ was fitted by an Einstein-type function~\cite{fox1980magnetoelectric,seehra1981dielectric,seehra1984anomalous,seehra1986effect}:
\begin{equation}
\label{eq:einstein}
\varepsilon^{\mathrm{lf}}_{\mathrm{NM}}(T) = \varepsilon^{\mathrm{lf}}_{0} + \frac{A}{e^{\hbar\omega^{*}/k_{B}T} - 1},
\end{equation}
where $\varepsilon^{\mathrm{lf}}_{0}$ is the temperature independent low-frequency dielectric permittivity, $A$ is constant, $\omega^{*}$ is the frequency of the ``effective'' polar phonon.
The fit results using Eq.~\eqref{eq:einstein} which are shown by black lines in Figs.~\ref{fig:dielectric}(a) and~\ref{fig:dielectric}(b) allow us to extrapolate the ``pure'' low-frequency dielectric permittivity to the antiferromagnetic phase.
The differences between experimental and fit lines give the magnetic contribution caused by the spontaneous magnetodielectric effect according to expression that is close to the Eq.~\eqref{eq:omega_SP}:
\begin{equation}
\label{eq:MD}
\varepsilon^{\mathrm{lf}}(T) = \varepsilon^{\mathrm{lf}}_{\mathrm{NM}}(T) + \Delta\varepsilon^{\mathrm{MD}} M^{2}(T),
\end{equation}
where $\Delta\varepsilon^{\mathrm{MD}}$ is a coefficient defining the spontaneous magnetodielecric effect, and $M$ is described by Eq.~\eqref{eq:brillouin}.
There is a good agreement between experimental results and fits using Eq.~\eqref{eq:MD} below $T_{N}$ as shown by green lines in Figs.~\ref{fig:dielectric}(a) and~\ref{fig:dielectric}(b).
The obtained values of the $\Delta\varepsilon^{\mathrm{MD}}_{a}$ and $\Delta\varepsilon^{\mathrm{MD}}_{c}$ listed in Figs.~\ref{fig:dielectric}(a) and~\ref{fig:dielectric}(b), respectively, are somewhat less than those for \MnF{} from literature~\cite{seehra1984anomalous} which we can attribute to the difference in the values of the ion spins $S=3/2$ for $\mathrm{Co}^{2+}$ and 5/2 for $\mathrm{Mn}^{2+}$.

To reveal the contribution of each polar phonon to the $\Delta\varepsilon^{\mathrm{MD}}$ we calculate the dielectric strengths $\Delta\varepsilon$ using experimental and fit TO and LO phonon frequencies at different temperatures by employing Eq.~\eqref{eq:oscillator_strength_TOLO}.
The obtained temperature dependencies of the dielectric strengths $\Delta\varepsilon(T)$ of the $E_{u}$ and $A_{2u}$ polar phonons and the static dielectric permittivities $\varepsilon_{0}$ along the $a$ and $c$ axes are shown in Figs.~\ref{fig:dielectric}(a) and~\ref{fig:dielectric}(b), respectively.
We assumed that the high-frequency dielectric permittivity temperature changes $\varepsilon_{\infty}$ are insignificant and can be neglected~\cite{jahn1973linear,markovin1979magnetic}. 
It is seen that at the antiferromagnetic ordering the clear shifts of the dielectric strengths $\Delta\varepsilon$ for all polar phonons are observed. 
These shifts $\Delta\varepsilon^{\mathrm{MD}}$ are a manifestation of the spontaneous magnetodielectric effect caused by frequency changes of the polar phonons $\Delta\omega^{\mathrm{SP}}$ due to the spin-phonon coupling.
Note that the $\Delta\varepsilon_{2}^{\mathrm{MD}}$ for the medium frequency $E_{u}$ and lowest frequency $A_{2u}$ polar phonons have a positive sign, while the $\Delta\varepsilon_{1}^{\mathrm{MD}}$ and $\Delta\varepsilon_{3}^{\mathrm{MD}}$, on the contrary, are positive as shown in Figs.~\ref{fig:dielectric}(a) and~\ref{fig:dielectric}(b). 
A fair qualitative agreement is observed between the temperature behavior of the low-frequency $\varepsilon^{\mathrm{lf}}(T)$ and the static $\varepsilon_{0}(T)$ dielectric permittivities and their shifts below $T_{N}$ due to the spontaneous magnetodielectric effect $\Delta\varepsilon^{\mathrm{MD}}$ as can be seen in Fig.~\ref{fig:dielectric}.
Thus, we have shown that the shifts of the macroscopic dielectric permittivity $\Delta\varepsilon^{\mathrm{MD}}$ at antiferromagnetic ordering are caused by changes of the frequencies $\Delta\omega^{\mathrm{SP}}$ of microscopic polar phonons. 

\subsection{Lattice dynamics:}

\begin{figure}[b]
\centering
\includegraphics[width=1\columnwidth]{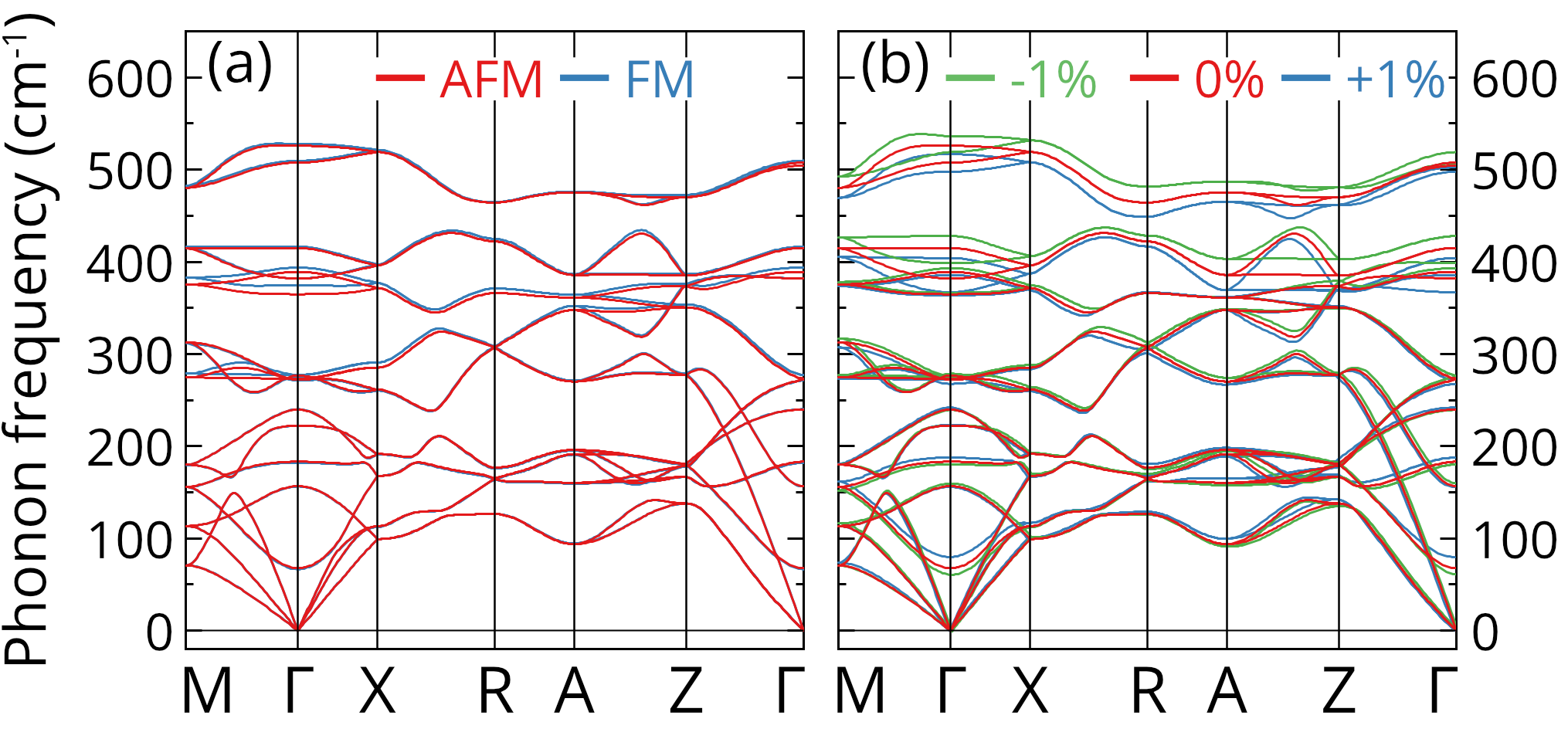}
\caption{\label{fig:phononDispersionCurve}
Computed phonon dispersion curves along the $\mathrm{M}$--$\Gamma$--$\mathrm{X}$--$\mathrm{R}$--$\mathrm{A}$--$\mathrm{Z}$--$\Gamma$ high-symmetry path of the Brillouin zone of the \CoF{} for (a)~antiferromagnetic (AFM) and ferromagnetic (FM) ordering, and (b)~$ab$ plane biaxial epitaxial strain -1\%, 0\%, +1\%.
Negative and positive signs of epitaxial strain correspond to the compression and expansion of the crystal, respectively.
}
\end{figure}

Aiming to explore the origin of the observed spin-phonon coupling, we performed several first-principles calculations of the lattice dynamics landscape in the \CoF{} in the antiferromagnetic (AFM) and ferromagnetic (FM) spin configurations considering the same lattice parameters as $a=b=4.678$\,\AA and $c=3.154$\,\AA. 
Because the nonmagnetic (NM) phase is not accessible for DFT calculations of magnetic crystals, we assumed as a reference that $\omega^{\mathrm{NM}} = (\omega^{\mathrm{AFM}} + \omega^{\mathrm{FM}}) / 2$~\cite{schleck2010elastic,dubrovin2021incipient}.
The calculated phonon dispersion curves in the whole Brillouin zone for AFM (in red curves) and FM (in blue curves) spin configurations are shown in Fig.~\ref{fig:phononDispersionCurve}(a).
The $\mathrm{M}$--$\Gamma$--$\mathrm{X}$--$\mathrm{R}$--$\mathrm{A}$--$\mathrm{Z}$--$\Gamma$ high-symmetry path shown in Fig.~\ref{fig:structure}(b) was used in the lattice dynamics calculations.
It is clearly seen that there are no significant differences between AFM and FM phonon dispersion curves which are consistent with the obtained values of the frequency shifts $\Delta\omega^{\mathrm{SP}}$ of about 1\,cm$^{-1}$ in the experiment.
\textcolor{newtext}{According to the Ref.~\cite{kumar2012spin}, the difference between FM and AFM phonon dispersion curves is related to the second derivative from exchange interactions with respect to the ionic displacements, which is presumably small in rutiles with respect to other magnetic materials.}
Note that the calculated phonon frequencies are real (positive) in the whole Brillouin zone and therefore \CoF{} in $P4_{2}/mnm$ space group is stable for both specified magnetic states.

\begin{figure*}
\centering
\includegraphics[width=2\columnwidth]{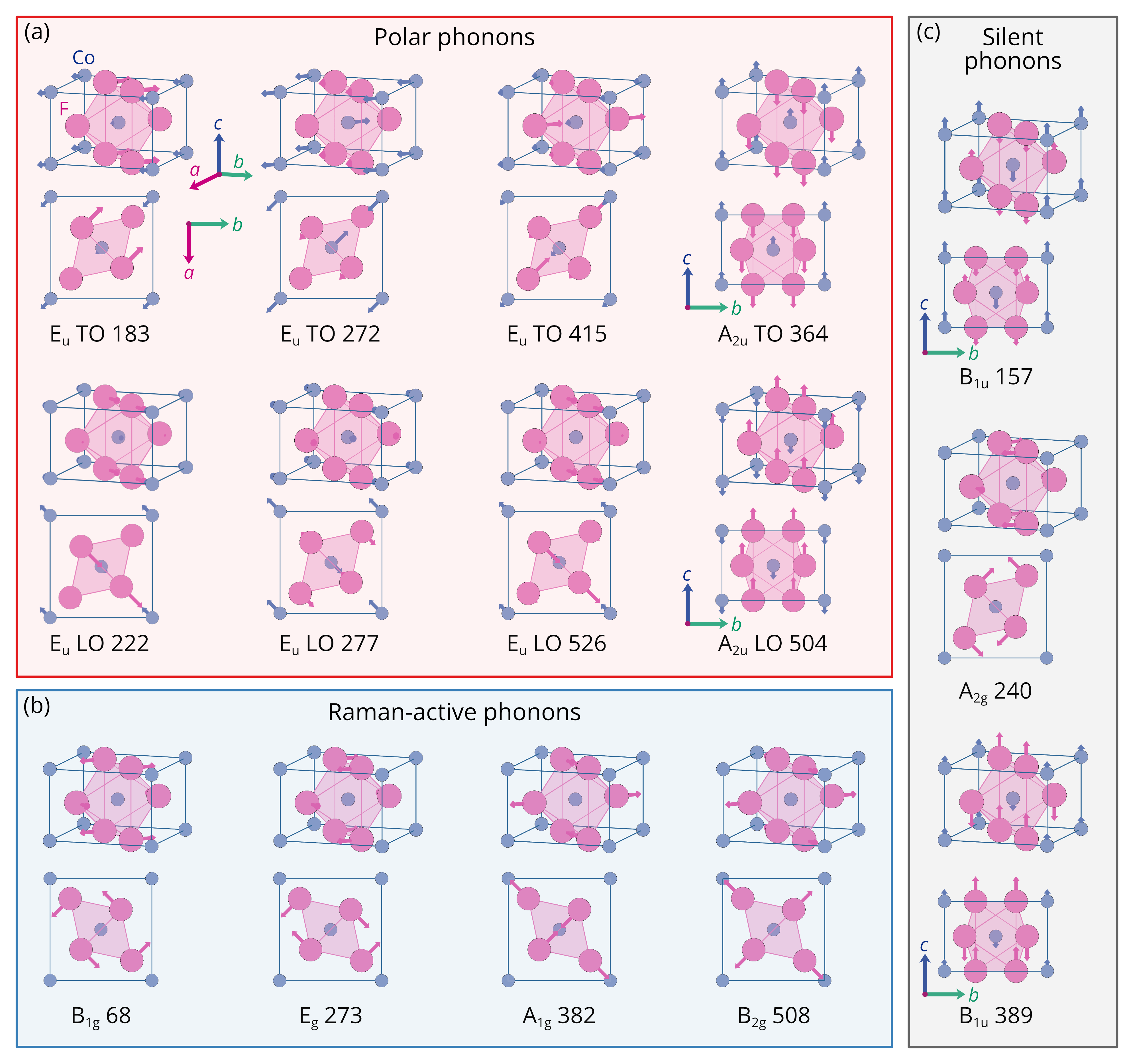}
\caption{\label{fig:DFT_displacement}
Sketch of the ion's displacements for (a)~polar, (b)~Raman-active, and (c)~silent optical phonons extracted for \CoF{} in the Brillouin zone center obtained from the DFT calculations. 
The numbers represent the phonon frequencies in cm$^{-1}$.
The picture was prepared using the \textsc{VESTA} software~\cite{momma2011vesta}.
}
\end{figure*}

\begin{table}[b]
    \caption{\label{tab:DFT_phonons} Calculated frequencies (cm$^{-1}$) of the infrared- (IR), Raman-active (R), and silent (S) optical phonons for antiferromagnetic (\textcolor{newtext2}{AFM}), ferromagnetic (FM), and nonmagnetic (NM) states together with spin-phonon coupling parameter (SP = AFM - NM) in \CoF.
    }
    \begin{ruledtabular}
            \begin{tabular}{ccccccc}
             Act. & Sym. & Phonon & AFM & FM & NM & SP\\
             \hline
             \multirow{8}{*}{IR} & \multirow{6}{*}{$E_{u}$} & $\omega_{\mathrm{1TO}}$ & 183.26 & 182.08 & 182.67 &  0.59\\
                                 &                          & $\omega_{\mathrm{1LO}}$ & 222.29 & 222.22 & 222.25 &  0.04\\
                                 &                          & $\omega_{\mathrm{2TO}}$ & 272.32 & 271.32 & 271.82 &  0.5\\
                                 &                          & $\omega_{\mathrm{2LO}}$ & 276.65 & 275.72 & 276.19 &  0.47\\
                                 &                          & $\omega_{\mathrm{3TO}}$ & 414.95 & 416.52 & 415.74 &  -0.78\\
                                 &                          & $\omega_{\mathrm{3LO}}$ & 525.81 & 527.93 & 526.87 &  -1.06\\
            \cmidrule{2-7}
                                 & \multirow{2}{*}{$A_{2u}$} & $\omega_{\mathrm{1TO}}$ & 364.82 & 374.72 & 369.77 & -4.95\\
                                 &                           & $\omega_{\mathrm{1LO}}$ & 504.47 & 509.74 & 507.11 & -2.63\\
            \hline
            \multirow{4}{*}{R}   &                  $B_{1g}$ &                         &  67.78 &  66.63 &   67.2 &  0.58\\
                                 &                   $E_{g}$ &                         & 273.16 & 277.35 & 275.25 & -2.1\\
                                 &                  $A_{1g}$ &                         & 381.8  & 381.88 & 381.84 & -0.04\\
                                 &                  $B_{2g}$ &                         & 507.65 & 509.51 & 508.58 & -0.93\\
            \hline
            \multirow{3}{*}{S}   &                  $B_{1u}$ &                         & 156.74 & 156.34 & 156.54 &  0.2\\
                                 &                  $A_{2g}$ &                         & 239.8  & 240.13 & 239.96 & -0.17\\
                                 &                  $B_{1u}$ &                         & 389    & 393.78 & 391.4  & -2.39\\
        \end{tabular}
    \end{ruledtabular}
\end{table}

The calculated frequencies $\omega$ of the optical phonons in AFM, FM, and NM states at the center of the Brillouin zone and estimated values of frequency shifts below $T_{N}$ ($\Delta\omega^{\mathrm{SP}} = \omega^{\mathrm{AFM}} - \omega^{\mathrm{NM}}$) are listed in Table~\ref{tab:DFT_phonons}.
The phonon frequencies are in fair agreement with previously reported calculations~\cite{barreda2013pressure}, experiments~\cite{barker1965infrared,balkanski1966infrared,macfarlane1970light,haussler1981infrared,cottam2019spin}, and our data from Table~\ref{tab:phonon_parameters}.
Signs and magnitudes of $\Delta\omega^{\mathrm{SP}}$ for polar phonons obtained in our calculation agree with our experimental findings shown in Fig.~\ref{fig:phonon}.
Thus, in these both cases $\Delta\omega^{\mathrm{SP}}$ have positive sign for $\omega_{\mathrm{1TO}}$, $\omega_{\mathrm{1LO}}$, $\omega_{\mathrm{2TO}}$, and $\omega_{\mathrm{2LO}}$ frequencies of the $E_{u}$ phonon and negative sign for other $E_{u}$ and $A_{2u}$ phonons.
For Raman-active phonons in \CoF{} there is fair agreement between the experiment from Ref.~\cite{cottam2019spin} and our simulation for $B_{1g}$, $A_{1g}$, and $B_{2g}$ phonons, whereas $\Delta\omega^{\mathrm{SP}}$ for $E_{g}$ phonon has different sign.
However, it is worth noting that the spin-phonon frequency shift $\Delta\omega^{\mathrm{SP}}$ for the $E_{g}$ phonon in \CoF{} is different from those in $\mathrm{MnF}_{2}$, $\mathrm{FeF}_{2}$, and $\mathrm{NiF}_{2}$~\cite{cottam2019spin}.
Besides, the $\Delta\omega^{\mathrm{SP}}$ for the polar 3TO phonon of $E_{u}$ symmetry has an opposite sign in \CoF{} and \MnF{}~\cite{schleck2010elastic}, which also indicates a difference in the spin-phonon coupling mechanism in these antiferromagnets.
Therefore, the lattice dynamic simulation shows that the magnetic ordering shifts the phonon frequencies without changing unit cell volume of \CoF{}.

\begin{figure}
\centering
\includegraphics[width=\columnwidth]{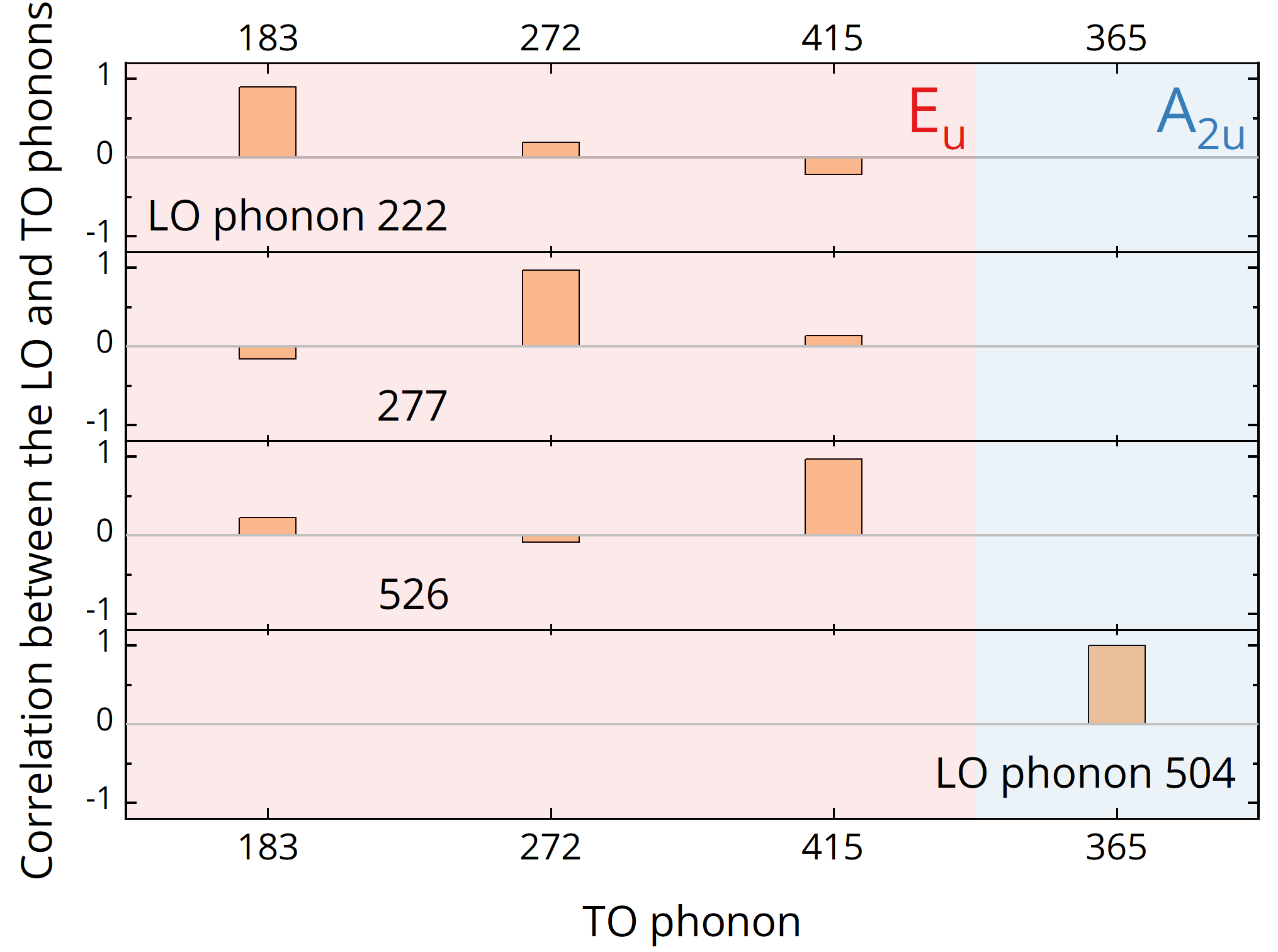}
\caption{\label{fig:LOTO_correlation}
Correlation between eigendisplacements of LO and TO polar phonons with $E_{u}$ (red background) and $A_{2u}$ (blue background) symmetry.
The latter is extracted according to the DFT calculations at the $\Gamma$ point of the Brillouin zone and it shows that the ``LO-TO'' rule is fully satisfied in \CoF.
The values of the LO phonon frequencies are given in cm$^{-1}$.
}
\end{figure}

Next, we calculate eigenvectors for all-optical phonons in the Brillouin zone's center, shown in Fig.~\ref{fig:DFT_displacement}.
The obtained ion displacements for phonons in \CoF{} are close to the same ones in isostructural rutile $\mathrm{TiO}_{2}$~\cite{lee1994lattice}.
\textcolor{newtext}{As can be seen in Fig.~\ref{fig:DFT_displacement}(b), the Co ions remain unmoved tor all Raman-active and $A_{2g}$ silent phonons.
This is due to the fact that from symmetry the Co (2a) ions are active only for \textit{ungerade} $A_{2u} \oplus E_{u} \oplus B_{1u}$ modes, while the F (4f) ions are involved for both \textit{gerade} and \textit{ungerade} phonons~\cite{kroumova2003bilbao}.
The ion displacement for most phonons are in the $ab$ plane and only for the $A_{2u}$ and $B_{1u}$ vibrations are they along the $c$ axis, as shown in Fig.~\ref{fig:DFT_displacement}. 
Besides, we see that the phonon displacements do not change the distances between the nearest-neighborhood Co ions and thus does not significantly modulate the exchange interaction. 
This in turn manifests itself in a small frequency shift due to the spin-phonon coupling and an insignificant difference between FM and AFM phonon dispersion curves in Fig.~\ref{fig:phononDispersionCurve}.
Thus, we can conclude that the rutile structure does not favor significant spin-phonon interaction, which is confirmed experimentally by us on \CoF{} and on $\mathrm{MnF}_{2}$ from Ref.~\cite{schleck2010elastic}.}

To assign LO to TO polar phonons we expand the eigenvector of the $m$th LO mode to a linear combination of the TO normal modes $| \xi^{\mathrm{LO}}_{m} \rangle = \sum_{n} C_{mn} | \xi^{\mathrm{TO}}_{n} \rangle$~\cite{raeliarijaona2015mode,fredrickson2016theoretical}. 
Figure~\ref{fig:LOTO_correlation} shows the overlap matrix $C_{mn} = \langle \xi^{\mathrm{LO}}_{m} | \xi^{\mathrm{TO}}_{n} \rangle$, which represents the degree of correlation between eigenvectors of the $m$th LO and $n$th TO polar phonons.
\textcolor{newtext}{There is the ``LO-TO rule'' according to which in most crystals the sequence of polar phonons is such that the TO frequency is followed by LO frequency, with $\omega_{\mathrm{LO}} > \omega_{\mathrm{TO}}$ for each main crystallographic axis~\cite{schubert2019phonon}.
In the isostructural rutile $\mathrm{TiO}_{2}$, which is an incipient ferroelectric, a more complex pattern of the correlation matrix $C_{mn}$ is observed, polar phonons are significantly mixed by the strong Coulomb interaction, and the ``LO-TO rule'' was not fully obeyed~\cite{lee1994lattice}.
Conversely, the ``LO-TO rule'' is satisfied for all polar phonons in \CoF{} and LO is more correlated with the nearest TO mode of \textcolor{newtext}{the same symmetry} with $\omega_{\mathrm{TO}} < \omega_{\mathrm{LO}}$.}
Moreover, since the $A_{2u}$ polar phonon is \textcolor{newtext}{the single} one, the eigenvectors $| \xi^{\mathrm{TO}} \rangle = | \xi^{\mathrm{LO}} \rangle$ ($C = 1$) are equal and ion displacements for this TO and LO modes are the same \textcolor{newtext}{as can be seen in Fig.~\ref{fig:DFT_displacement}(a)}.
Nonetheless, despite this fact, the frequency shifts $\Delta\omega^{\mathrm{SP}}$ are different in magnitude for TO and LO frequencies of the $A_{2u}$ polar phonons ($\Delta\omega^{\mathrm{SP}}_{\mathrm{TO}} > \Delta\omega^{\mathrm{SP}}_{\mathrm{LO}}$) which proves that the frequency shift is inversely proportional to the phonon frequency~\cite{granado1999magnetic}.
In contrast, for $E_{u}$ polar phonons, there are slight differences in ion displacements for TO and LO modes [see Fig.~\ref{fig:DFT_displacement}(a)] due to the slight mode mixing caused by the Coulomb interaction~\cite{zhong1994giant,lee1994lattice,khedidji2021microscopic}.

\begin{figure}
\centering
\includegraphics[width=\columnwidth]{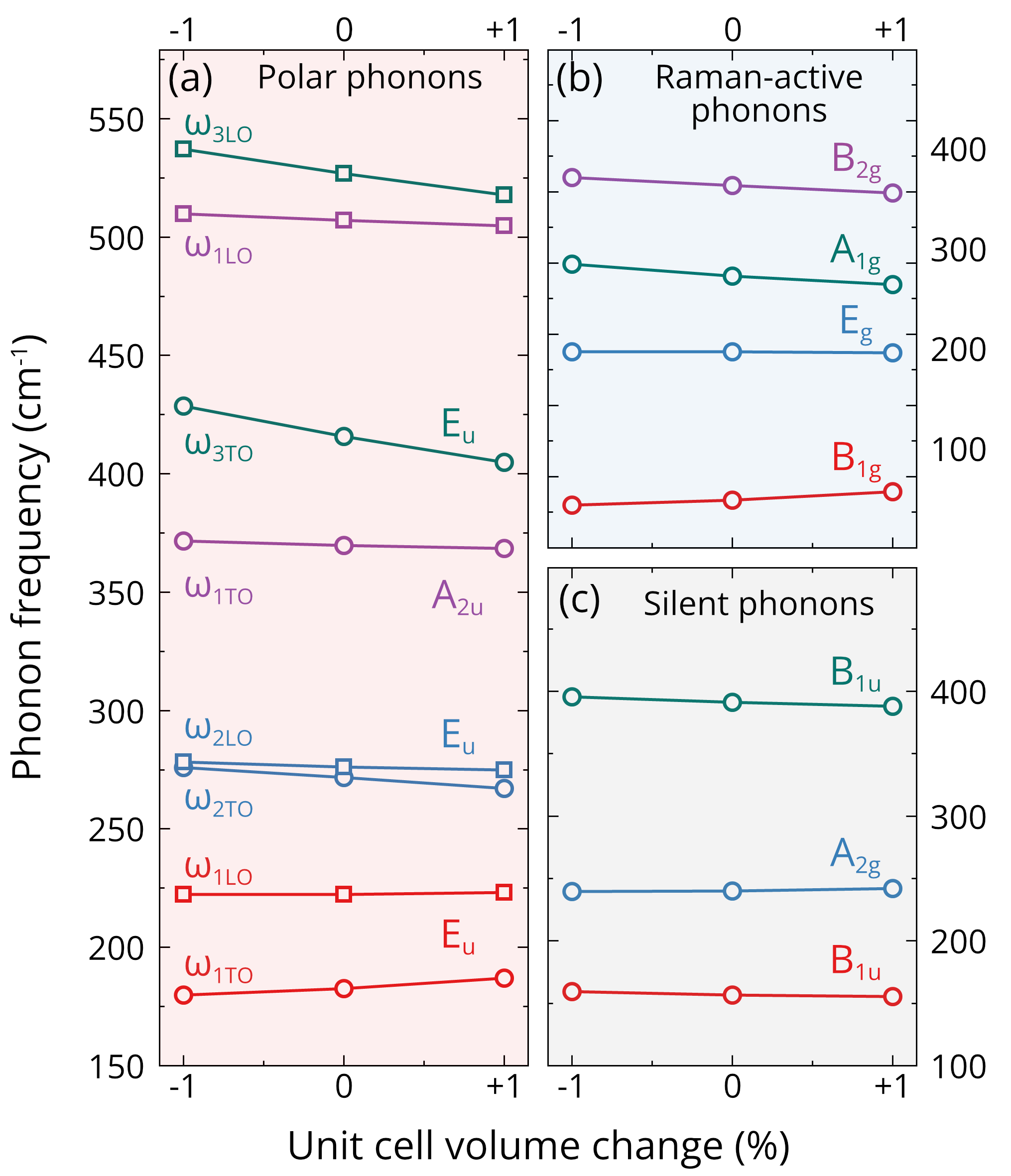}
\caption{\label{fig:phonon_DFT}
Calculated frequencies of the (a)~polar, (b)~Raman-active, and (c)~silent optical phonons as a function of the unit cell volume change in the nonmagnetic phase.
}
\end{figure}

Aiming to reveal the volume effect on the phonons, we performed lattice dynamics simulations with the unit cell volume change of $\pm1\%$ as shown in Fig.~\ref{fig:phononDispersionCurve}(b).
The dispersion curves of phonons qualitatively have the same behavior in the whole Brillouin zone.
Calculated frequencies of phonons at the center of the Brillouin zone as a function of the unit cell volume are shown in Fig.~\ref{fig:phonon_DFT}.
According to Ref.~\cite{chatterji2010magnetoelasticCoF2}, the lattice parameters decrease at cooling in the paramagnetic phase, thereby reducing the unit cell volume in \CoF{}.
Figure~\ref{fig:phonon_DFT}(a) shows that frequencies of not all polar phonons increase at the compression of the unit cell as in our experiment (see Fig.~\ref{fig:phonon}).
So, the $\omega_{1\mathrm{TO}}$ frequency of the $E_{u}$ phonon decreases with compression of the unit cell, whereas in our experiment, on the contrary, it increases.
Thus, the temperature dependences of not all polar phonons can be explained only by a harmonic change in the unit cell volume. 
The frequencies of Raman-active phonons in the calculation, see Fig.~\ref{fig:phonon_DFT}(b), and experiment~\cite{cottam2019spin} behave in the same way at the volume changes.
Note that the frequencies of the silent phonons in our calculations change rather weakly when the unit cell volume changes.

According to Ref.~\cite{chatterji2010magnetoelasticCoF2}, the lattice parameters $a$ and $b$ increase, whereas $c$ decreases at cooling below the N{\'e}el temperature in the antiferromagnetic phase due to the magnetostriction effect.
It can be seen that the change in the volume of the unit cell as a result of magnetostriction cannot explain the sign of frequency shift at the antiferromagnetic ordering since $\Delta\omega^{\mathrm{SP}}$ is negative for $A_{2u}$ phonons. At the same time, the compression of the $c$ axis leads to an increase in its TO and LO frequencies, as shown in Fig.~\ref{fig:phonon_DFT}(a).
The observed phonon frequency shifts at the antiferromagnetic ordering in \CoF{} cannot be explained only due to observed magnetostriction.

\begin{table}[b]
    \caption{\label{tab:DFT_dielectric} Calculated dielectric strengths $\Delta\varepsilon_{j}$ of the polar phonons, static $\varepsilon_{0}$ and high-frequency $\varepsilon_{\infty}$ dielectric permittivities for antiferromagnetic (AFM), ferromagnetic (FM), and nonmagnetic (NM) states \textcolor{newtext}{along} with spontaneous magnetodielectric coupling parameter (MD = AFM - NM) in \CoF.}
    \begin{ruledtabular}
            \begin{tabular}{cccccc}
             Sym. & Mode $j$ & AFM & FM & NM & MD\\
             \hline
              \multirow{5}{*}{$E_{u}$} & $\Delta\varepsilon_{1}$      & 2.23  & 2.30  & 2.26  & -0.031\\
                                       & $\Delta\varepsilon_{2}$      & 0.103 & 0.102 & 0.103 &  0.001\\
                                       & $\Delta\varepsilon_{3}$      & 1.34  & 1.32  & 1.33  &  0.006\\
                                       & $\varepsilon_{\infty}$       & 2.55  & 2.53  & 2.54  &  0.011\\
                                       & $\varepsilon_{0}$            & 6.23  & 6.25  & 6.24  & -0.013\\
             \hline
             \multirow{3}{*}{$A_{2u}$} & $\Delta\varepsilon_{1}$      & 2.41  & 2.23  & 2.32  & 0.09\\
                                       & $\varepsilon_{\infty}$       & 2.64  & 2.62  & 2.63  & 0.008\\
                                       & $\varepsilon_{0}$            & 5.05  & 4.85  & 4.95  & 0.099\\
        \end{tabular}
    \end{ruledtabular}
\end{table}

\begin{figure*}
\centering
\includegraphics[width=2\columnwidth]{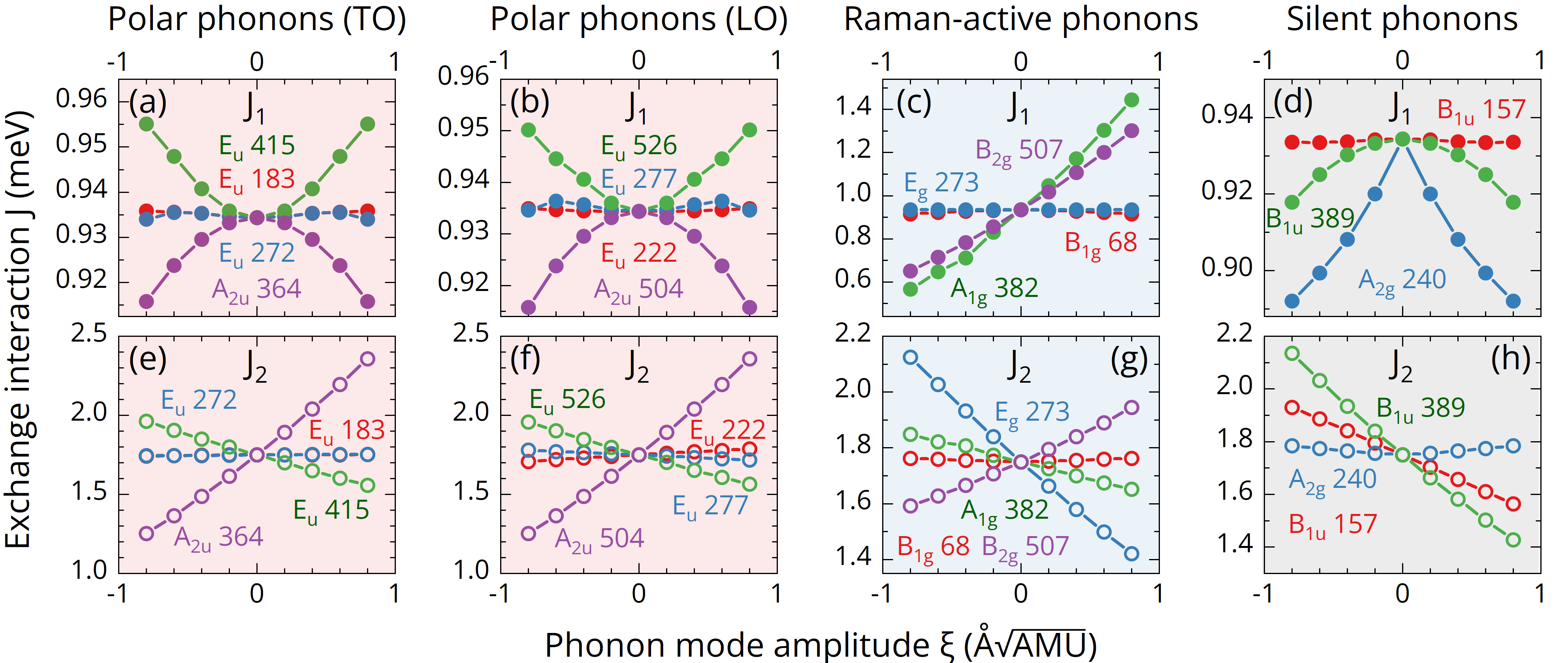}
\caption{\label{fig:J_phonon}
Computed exchange interactions $J_{1}$ and $J_{2, +}$ as a function of the phonon eigendisplacement $\xi$ ($\textrm{\AA}\sqrt{\mathrm{AMU}}$) for [(a) and (e)]~polar TO, [(b) and (f)] polar LO, [(c) and (g)] Raman-active, and [(d) and (h)]~silent phonons at the center of the Brillouin zone in \CoF, respectively.
\textcolor{newtext}{The numbers represent the phonon frequencies in cm$^{-1}$.}
}
\end{figure*}

We have also estimated the dielectric strengths $\Delta\varepsilon$ and static dielectric permittivity $\varepsilon_{0}$ using Eq.~\eqref{eq:oscillator_strength_TOLO} and phonons frequencies for AFM, FM and NM phases from DFT calculations, which are listed in Table~\ref{tab:DFT_dielectric}.
There is a good agreement with experimental data from Table~\ref{tab:phonon_parameters}.
Moreover, the calculated signs and relative values of the shifts of the dielectric strengths $\Delta\varepsilon^{\mathrm{MD}}$ correspond with the experimental ones for the most polar phonons, as can be seen in Figs.~\ref{fig:dielectric}(c) and~\ref{fig:dielectric}(d).
The values of the spontaneous magnetodielectric effect $\Delta\varepsilon^{\mathrm{MD}}_{0}$ obtained from DFT calculations also are in fair agreement with those for low-frequency dielectric permittivity and infrared spectroscopy. 
Thus, we conclude that the results of our DFT calculations adequately reflect the features of the lattice dynamics related to the antiferromagnetic ordering in \CoF.





\subsection{Magnon-phonon coupling}

After observing compelling evidence of spin-phonon coupling, our investigation delves into exploring the effects of phonon displacements on the magnetic properties of \CoF.
Our approach begins with formulating the effective Heisenberg model, as:
\begin{equation}
    H = \sum_{i \neq j} J_{ij} \mathbf{S}_i \mathbf{S}_j.
\end{equation}
Here, $\mathbf{S}_{i}$ ($\mathbf{S}_{j}$) represents the unit-length spin operator of the magnetic site $i$ ($j$), while $J_{ij}$ denotes the isotropic exchange interaction quantifying interactions between magnetic sites $i$ and $j$.
Once the exchange interactions are known, we can obtain the magnon frequencies by diagonalizing the Hesigneberg Hamiltonian.
Our primary objective centers on scrutinizing the impact of lattice vibrations on the magnon spectra.
To achieve this, we compute the perturbations in the exchange interactions elicited by the displacements induced by each distinct phonon mode.

In the case of \CoF, the magnetic properties are primarily influenced by the interactions among nearest and next-nearest-neighbors. 
These interactions are quantified by their respective exchange integrals, denoted as $J_{1}$ and $J_{2}$, \textcolor{newtext}{which are shown in Fig.~\ref{fig:structure}(a)}.
Table~\ref{tab:J} presents the values of $J_{1}$ and $J_{2}$ derived from first principles computations for the ground structure of \CoF{} \textcolor{newtext}{in comparison with experimental data from Refs.~\cite{cowley1975magnetic,meloche2007two}.}
The calculated value of $J_{2}$ aligns closely with experimental observations, playing a pivotal role in inducing antiferromagnetic behavior in \CoF.
In contrast, our computation of $J_{1}$ reveals a discrepancy in sign compared to experiments.
Notably, this sign mismatch persists across variations in DFT parameters, exchange-correlation functionals, and methodologies employed to determine the exchange constants.
Nevertheless, our findings are consistent with those reported for the isostructural antiferromagnet $\mathrm{MnF}_{2}$ in the existing literature~\cite{lopez2016first}.
Despite the ambiguity surrounding the sign of $J_{1}$, we proceed with our analysis based on the consensus regarding the accuracy of the $J_{2}$ values. 

\begin{table}
\caption{\label{tab:J} Isotropic exchange interactions of \CoF.}
\begin{ruledtabular}
\begin{tabular}{cccc} 
Method             & $J_{1}$ (meV) & $J_{2}$ (meV) & Reference \\ \hline
Green's function   & 0.9344      & 1.7500      & This work \\
Energy mapping     & 1.034       & 1.5998      & This work \\
Raman scattering   & -0.15       & 1.6         & Ref.~\cite{meloche2007two} \\ 
Neutron scattering & -0.207      & 1.53        & Ref.~\cite{cowley1975magnetic} \\ 
\end{tabular}
\end{ruledtabular}
\end{table}

The exchange interactions in the Heisenberg model are described by fixed values.
However, the interaction between two magnetic sites depends on the distance that separates them.
Consequently, the displacements induced by the phonon modes are anticipated to influence the values of $J_{1}$ and $J_{2}$. 
This motivates us to express each exchange constant as
\begin{equation}
    J_{i}^{\omega}(\xi) = J_{i}^{0} + \alpha_i^{\omega} \xi + \beta_i^{\omega} \xi^2 + o(\xi^3),
    \label{eq:J_mod}
\end{equation}
where $J_{i}^{0}$ is the exchange interaction computed by using the equilibrium distances, $\xi$ is the phonon amplitude, and $\omega$ labels a particular phonon mode.
\textcolor{newtext}{Furthermore, $\xi$ represents lattice distortions that modify the exchange interactions, which could be induced by strain or optical excitations~\cite{fechner2016effects,fechner2018magnetophotonics}.}

The coefficients $\alpha$ and $\beta$ on the right-hand side of Eq.~\eqref{eq:J_mod} quantitatively describe the effect of a particular phonon mode on the exchange interactions.
To obtain them, we calculate the exchange interactions of the distorted structures given by frozen phonon modulations for different amplitudes and fit the results to Eq.~\eqref{eq:J_mod}.
In Fig.~\ref{fig:J_phonon}, we plot the exchange constants as a function of the phonon amplitudes for each phonon mode.
\textcolor{newtext}{The exchange interactions $J_{1}$ and $J_{2}$ are affected significantly under the influence of the phonon modes.
A similar strong modulation of the exchange interaction by phonons has been previously reported in the literature~\cite{disa2023photo}.
On the other hand, the frequency shifts $\Delta\omega^{\mathrm{SP}}$ of phonons at the antiferromagnetic ordering in \CoF{} are rather small, which is indicated by the competing effects of different sign from the dynamical modulation of exchange interactions $J_{1}$ and $J_{2}$ by phonons due to the coupling with spins.}
Besides, we see for \CoF{} that the first- and second-order terms of Eq.~\eqref{eq:J_mod} suffice to account for the exchange interaction \textcolor{newtext}{variations}, and therefore we neglect higher-order terms.

\begin{table}[b]
    \caption{\label{tab:coefficients}
    \textcolor{newtext}{Calculated magnon-phonon coupling coefficients $\alpha_{i}$ (meV\,AMU$^{-\frac{1}{2}}\textrm{\AA}^{-1}$) and $\beta_{i}$ (meV\,AMU$^{-1}\textrm{\AA}^{-2}$) of the nearest $J_{1}$ and next-nearest-neighbor $J_{2, +}$ exchange interactions as a function of the phonon amplitude.
    AMU is the atomic mass unit.}
    }
    \begin{ruledtabular}
            \begin{tabular}{ccccccc}
             Act. & Sym. & Phonon & $\alpha_1$ & $\beta_1$ & $\alpha_2$ & $\beta_2$\\
             \hline
             \multirow{8}{*}{IR} & \multirow{6}{*}{$E_{u}$} & $\omega_{\mathrm{1TO}}$ &  0.000 & 0.002 & 0.005 &  0.004\\
                                 &                          & $\omega_{\mathrm{1LO}}$ &  0.000 & 0.002 &  -0.049 &  -0.004\\
                                 &                          & $\omega_{\mathrm{2TO}}$ &  0.000 &  -0.001 & 0.006 &  -0.001\\
                                 &                          & $\omega_{\mathrm{2LO}}$ &  0.000 &  -0.000 & 0.039 &  -0.002\\
                                 &                          & $\omega_{\mathrm{3TO}}$ &  0.000 & 0.032 &  -0.252 & 0.017\\
                                 &                          & $\omega_{\mathrm{3LO}}$ &  0.000 & 0.024 &  -0.245 & 0.020\\
            \cmidrule{2-7}
                                 & \multirow{2}{*}{$A_{2u}$} & $\omega_{\mathrm{1TO}}$ & 0.000 &  -0.029 &  0.691 & 0.085\\
                                 &                           & $\omega_{\mathrm{1LO}}$ & 0.000 &  -0.029 &  0.691 &  0.085\\
            \hline
            \multirow{4}{*}{R}   &                  $B_{1g}$ &                         & 0.000 &  -0.028 &  0.000 & 0.018\\
                                 &                   $E_{g}$ &                         & 0.000 & 0.001 &  -0.440 & 0.035\\
                                 &                  $A_{1g}$ &                         &0.551 & 0.118 &  -0.125 &  -0.002\\
                                 &                  $B_{2g}$ &                         &0.406 & 0.065 & 0.219 &  -0.030\\
            \hline
            \multirow{3}{*}{S}   &                  $B_{1u}$ &                         & 0.000 &  -0.001 & -0.229 &  -0.004\\
                                 &                  $A_{2g}$ &                         & 0.000 &  -0.053 &  0.000 & 0.050\\
                                 &                  $B_{1u}$ &                         & 0.000 &  -0.026 &  -0.442 &  -0.050\\
        \end{tabular}
    \end{ruledtabular}
\end{table}

\begin{figure}
\centering
\includegraphics[width=1\columnwidth]{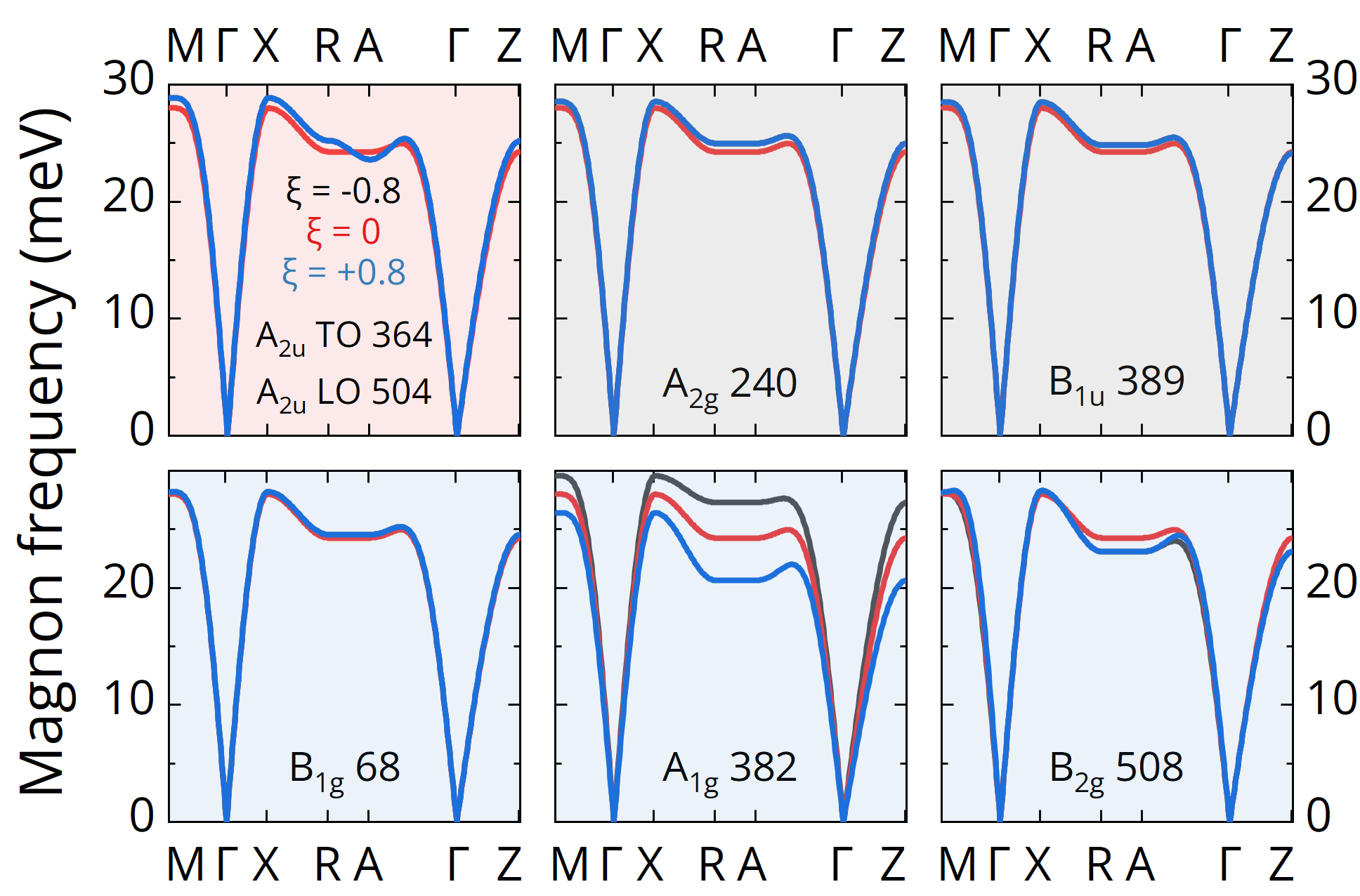}
\caption{\label{fig:Magnon_Phonons}
\textcolor{newtext}{Computed spin wave dispersion curves along the $\mathrm{M}$--$\Gamma$--$\mathrm{X}$--$\mathrm{R}$--$\mathrm{A}$--$\Gamma$--$\mathrm{Z}$ high-symmetry path of the Brillouin zone for distorted \CoF{} by a frozen polar $A_{2u}$ (red-filled background), silent $A_{2g}$, $B_{1u}$ (gray-filled background), and Raman-active $B_{1g}$, $A_{1g}$, and $B_{2g}$ (blue-filled background) phonons with eigendisplacements $\xi = - 0.8$, 0, and +0.8~$\textrm{\AA}\sqrt{\mathrm{AMU}}$.}
The numbers represent the phonon frequencies in cm$^{-1}$.
}
\end{figure}

In the unperturbed structure of \CoF, $J_{1}$ and $J_{2}$ have the same value for all possible nearest and next-nearest-neighbor pairs of the Co ions, respectively.
This is, generally, not true for the displaced structures since the distances and thus the interactions between neighbors can be different for equivalent pairs.
In our case, we see that $J_{1}^{\omega}(\xi)$ is the same for different nearest neighbor pairs (4 in total).
This is because the phonons at the Brillouin zone center only produce intra-cell displacements, whereas $J_{1}$ corresponds to an inter-cell interaction. 
In contrast, the next-nearest-neighbor pairs (16 in total) are split into two equal groups characterized by the expression:
\begin{equation}
    J_{2,\pm}^{\omega}(\xi) = J_{2}^{0} \pm \alpha_{2}^{\omega} \xi + \beta_{2}^{\omega} \xi^{2} + o(\xi^{3}),
\end{equation}
where $J_{2, +}$ ($J_{2, -}$) describes the interaction between two neighbors when their distance is increased (reduced) during a positive amplitude modulation.
These two groups of interactions appear for every phonon mode except for $B_{1g}$, $A_{1g}$, and $A_{2g}$.
In Table~\ref{tab:coefficients}, we show the coupling coefficients for the nearest $J_{1}$ and next-nearest interactions $J_{2, +}$. 

The preceding discussion underscores the anticipated dependence of magnon frequencies on phonon modulations.
To scrutinize this relationship, we calculate the magnon spectra corresponding to the distorted structures induced by each phonon mode.
The \textcolor{newtext}{most representative} outcomes of these analyses are depicted in Fig.~\ref{fig:Magnon_Phonons}.
\textcolor{newtext}{Note that, in general, the magnon (M) frequency at the Brillouin zone center $\omega_{M} = \gamma \sqrt{2 H_{A} H_{E}}$, where $\gamma$ is a gyromagnetic ratio, $H_{A}$ is a magnetic anisotropy field, and $H_{E}$ is an exchange field~\cite{gurevich1996magnetization}.
Since accounting for magnetic anisotropy is a challenging task, we did not take it into account in the calculations and assumed that {$H_{A} = 0$}.
Thus, the magnon frequency at the $\Gamma$ point in our calculations is 0\,meV whereas in our experiment it was 4.6\,meV (37\,cm$^{-1}$).
At the Brillouin zone boundaries, the magnon frequencies is determined only by the exchange interactions.
The two-magnon (2M) mode frequency 14.2\,meV (115\,cm$^{-1}$) allow us to estimate the magnon frequency at the Brillouin zone boundaries as 7.1\,meV.
This exceeds the calculated values of 24--28\,meV (see Fig.~\ref{fig:Magnon_Phonons}), which we attribute to a significant discrepancy in the value of the exchange interaction $J_{1}$ between experiment and theory.
Other studies analyze the magnon-phonon hybridization based on the overlap of the magnon and phonon spectra~\cite{chen2021one,wang2020first}.
In contrast, this study only considers the modulations of the magnon spectra coming from phonons at the center of the Brillouin zone and does not discuss of any avoided crossing between phonons and magnons~\cite{metzger2023impulsive}.}

\textcolor{newtext}{For most phonons slight variations in the magnon spectra as a result of modulation are observed as shown, for example, for the Raman-active $B_{1g}$ phonon in Fig.~\ref{fig:Magnon_Phonons}.}
Intriguingly, we observe that the most significant alterations in magnon frequencies stem from the $A_{1g}$ mode despite it not possessing the most significant coupling coefficients \textcolor{newtext}{as can be seen in Fig.~\ref{fig:Magnon_Phonons} and Table~\ref{tab:coefficients}, respectively.}
Conversely, modes characterized by more significant coupling coefficients, such as $A_{2u}$ or $E_{g}$, exhibit marginal deviations compared to the ground structure.
This phenomenon can be elucidated by examining how the Heisenberg Hamiltonian evolves under a given phonon distortion.
Specifically, for a designated phonon amplitude $\xi$, we construct a modified Hamiltonian considering the interactions among the 4 and 16 nearest and next-nearest-neighbors as follows:
\begin{align}
    \label{eq:H_distorted}
    \notag
    H^{\omega}(\xi) &= -2J_1^{\omega}(\xi)\left(\mathbf{S}_1^2 + \mathbf{S}_2^2\right) \\
    & - 8J_{2,-}^{\omega} (\xi)\mathbf{S}_1\mathbf{S}_2 - 8J_{2,+}^{\omega}(\xi)\mathbf{S}_1\mathbf{S}_2 \\ \notag
    &= -2J_1^{\omega}(\xi)\left(\mathbf{S}_1^2 + \mathbf{S}_2^2\right) -8 \left[ J_{2,-}^{\omega}(\xi) + J_{2,+}^{\omega}(\xi)\right]\mathbf{S}_1\mathbf{S}_2 \\ \notag
    &= -2J_1^{\omega}(\xi)\left(\mathbf{S}_1^2 + \mathbf{S}_2^2\right) -8 \left( J_2^0 + 2\beta_2^{\omega} \xi^2 \right)\mathbf{S}_1\mathbf{S}_2.
    \notag
\end{align}
The Hamiltonian's actual form~\eqref{eq:H_distorted} reveals that the linear coefficient, $\alpha_2^{\omega}$, does not affect the magnon dispersion energies, as two types of interactions, $J_{2,-}^{\omega}$ and $J_{2,+}^{\omega}$, emerge.
Consequently, in \CoF, next-nearest-neighbor interactions $J_{2}$ contribute to magnon-phonon coupling only if they respond quadratically to lattice distortions.
This clarifies why many phonon modes do not couple to magnetic response, as they primarily induce linear changes in $J_{2}$ (i.e., $\beta_2^{\omega} \approx 0$).
Only distortions from $A_{1g}$ and $B_{2g}$ modes notably modify $J_{1}$, thereby influencing the magnon spectra as can be seen in Table~\ref{tab:coefficients}.
Therefore, even if a phonon mode can individually alter the exchange interactions, it may not measurably affect the magnetic properties.
\textcolor{newtext}{It should be mentioned that it has been shown experimentally that magnon is strongly coupled to the Raman-active $B_{1g}$ phonon in \CoF~\cite{mashkovich2021terahertz}.
But according to our calculation results the effect of influence of the $B_{1g}$ mode to the magnon spectrum is rather small as can be seen in Fig.~\ref{fig:Magnon_Phonons}.
The discrepancy arises because the experiment focused on studying the magnon at the center of the Brillouin zone, while our calculations did not account for magnetic anisotropy, which is critical for capturing the coupling between magnons and phonons at the $\Gamma$ point.
}

\section{Conclusions and General Remarks}
In summary, we have studied in detail the polar phonons at the $\Gamma$ point of the Brillouin zone in the rutile antiferromagnet \CoF{} in a wide temperature range, including the antiferromagnetic phase transition by far-infrared and dielectric spectroscopic techniques.
We have experimentally observed that all polar phonons' TO and LO frequencies shift at the antiferromagnetic phase transition.
Moreover, our measurements have shown that the low-frequency dielectric permittivity also shifts at $T_{N}$ due to the spontaneous magnetodielectric effect.
Still, the signs of these changes are opposite for the $a$ and $c$ axes.
The combined analysis of these experimental results showed that the observed spontaneous magnetodielectric effect is mostly caused by changes in the frequencies of polar phonons at antiferromagnetic ordering.

Based on our first-principles calculations, we have found that observed shifts of frequencies of all polar phonons are fairly reproduced with antiferromagnetic ordering and unrelated to the unit cell changes due to the magnetostriction in \CoF.
Thus, according to the calculations, the observed frequency changes of polar phonons and the low-frequency dielectric permittivity results due to spin-phonon coupling and spontaneous magnetodielectric effect are related to the dynamic modulation of the exchange interaction by phonons.
In addition, we have observed magnetic excitations in the infrared spectra and identified their magnetodipole origin.
\textcolor{newtext}{First-principles calculations corroborate \textcolor{newtext2}{these} observations, emphasizing the coupling between polar phonons and magnons through the dynamic modulation of the exchange interactions.}

Notably, the most pronounced impact on the nearest-neighbor coupling $J_{1}$ is discernible in the Raman phonon $A_{1g}$ mode, evident across both linear and quadratic regimes, albeit to a lesser extent on $E_{g}$ modes.
While certain phonons may lack linear contributions to $J_{1}$, they exhibit discernible quadratic contributions, which define the strength of the phonon-magnon coupling~\cite{wang2023magnon}.
Conversely, concerning the next-nearest-neighbor coupling $J_{2}$, myriad infrared, silent, and Raman modes exhibit linear coupling, with the quadratic contributions typically on par with or subordinate to those observed in the $J_{1}$ scenario.
Regardless, significant phonon coupling in magnon spectra manifests predominantly when the quadratic term attains substantial magnitude, notably exemplified in the $A_{1g}$ mode and, to a lesser degree, in the $B_{2g}$ mode.
Intriguingly, specific infrared and silent modes exhibit more significant linear coupling than their Raman counterparts.

We believe our results will stimulate further experimental and theoretical studies of nonlinear phononics in the antiferromagnet \CoF, in which a nontrivial nonlinear coupling was observed between polar, Raman-active phonons and magnons. 
\textcolor{newtext}{
The other attractive materials based on the $3d^{7}$ $\mathrm{Co}^{2+}$ ions with the unquenched orbital angular momentum and strong spin-orbit effects are insulating honeycomb magnets, e.g., $\mathrm{Na}_{2}\mathrm{Co}_{2}\mathrm{TeO}_{6}$ and $\mathrm{Na}_{3}\mathrm{Co}_{2}\mathrm{SbO}_{6}$, which are of considerable scientific interest at the present time as a potential Kitaev system with nontrivial spin-phonon couplings which has yet to be revealed~\cite{liu2020kitaev,hong2021strongly,li2022giant,feng2022footprints,li2023magnon,zhang2023electronic}.
}

\section*{Acknowledgments}
We are grateful to V.A.\,Chernyshev for fruitful scientific discussion.
We thank D.A.\,Andronikova and M.P.\,Scheglov for the help with the x-ray orientation of single crystal.
The single crystal grown by S.V.\,Petrov was used in experiments.
This work was supported by the Russian Science Foundation under grant no.\,22-72-00025, https://rscf.ru/en/project/22-72-00025/.
R.M.D. acknowledges the support of the Ministry of Science and Higher Education of the Russian Federation (FSWR-2024-0003).
N.N.N. and K.N.B. acknowledge support by the research Project No.~FFUU-2022-0003 of the Institute of Spectroscopy of the Russian Academy of Sciences.
Some calculations included in this article were carried out using the GridUIS-2 experimental testbed.
The latter was developed under the Universidad Industrial de Santander (SC3-UIS) High Performance and Scientific Computing Centre with support from UIS Vicerrector\'ia de Investigaci\'on y Extensi\'on (VIE-UIS) and several UIS research groups.
A.C.G.C. acknowledge grant No.\,202303059C entitled “Optimización de las Propiedades Termoeléctricas Mediante Tensión Biaxial en la Familia de Materiales $\mathrm{Bi}_{4}\mathrm{O}_{4}\mathrm{Se}X_{2}$ ($X$ = $\mathrm{Cl}$, $\mathrm{Br}$, $\mathrm{I}$) Desde Primeros Principios” supported by the LNS - BUAP.
The group at West Virginia University thanks the Pittsburgh Supercomputer Center (Bridges2) and San Diego Supercomputer Center (Expanse) through allocation DMR140031 from the Advanced Cyberinfrastructure Coordination Ecosystem: Services \& Support (ACCESS) program, which is supported by National Science Foundation grants \#2138259, \#2138286, \#2138307, \#2137603, and \#2138296.
They also recognize the support of West Virginia Research under the call Research Challenge Grand Program 2022 and NASA EPSCoR Award 80NSSC22M0173.

\bibliography{bibliography}

\begin{thebibliography}{146}%
\makeatletter
\providecommand \@ifxundefined [1]{%
 \@ifx{#1\undefined}
}%
\providecommand \@ifnum [1]{%
 \ifnum #1\expandafter \@firstoftwo
 \else \expandafter \@secondoftwo
 \fi
}%
\providecommand \@ifx [1]{%
 \ifx #1\expandafter \@firstoftwo
 \else \expandafter \@secondoftwo
 \fi
}%
\providecommand \natexlab [1]{#1}%
\providecommand \enquote  [1]{``#1''}%
\providecommand \bibnamefont  [1]{#1}%
\providecommand \bibfnamefont [1]{#1}%
\providecommand \citenamefont [1]{#1}%
\providecommand \href@noop [0]{\@secondoftwo}%
\providecommand \href [0]{\begingroup \@sanitize@url \@href}%
\providecommand \@href[1]{\@@startlink{#1}\@@href}%
\providecommand \@@href[1]{\endgroup#1\@@endlink}%
\providecommand \@sanitize@url [0]{\catcode `\\12\catcode `\$12\catcode
  `\&12\catcode `\#12\catcode `\^12\catcode `\_12\catcode `\%12\relax}%
\providecommand \@@startlink[1]{}%
\providecommand \@@endlink[0]{}%
\providecommand \url  [0]{\begingroup\@sanitize@url \@url }%
\providecommand \@url [1]{\endgroup\@href {#1}{\urlprefix }}%
\providecommand \urlprefix  [0]{URL }%
\providecommand \Eprint [0]{\href }%
\providecommand \doibase [0]{https://doi.org/}%
\providecommand \selectlanguage [0]{\@gobble}%
\providecommand \bibinfo  [0]{\@secondoftwo}%
\providecommand \bibfield  [0]{\@secondoftwo}%
\providecommand \translation [1]{[#1]}%
\providecommand \BibitemOpen [0]{}%
\providecommand \bibitemStop [0]{}%
\providecommand \bibitemNoStop [0]{.\EOS\space}%
\providecommand \EOS [0]{\spacefactor3000\relax}%
\providecommand \BibitemShut  [1]{\csname bibitem#1\endcsname}%
\let\auto@bib@innerbib\@empty
\bibitem [{\citenamefont {Jungwirth}\ \emph {et~al.}(2016)\citenamefont
  {Jungwirth}, \citenamefont {Marti}, \citenamefont {Wadley},\ and\
  \citenamefont {Wunderlich}}]{jungwirth2016antiferromagnetic}%
  \BibitemOpen
  \bibfield  {author} {\bibinfo {author} {\bibfnamefont {T.}~\bibnamefont
  {Jungwirth}}, \bibinfo {author} {\bibfnamefont {X.}~\bibnamefont {Marti}},
  \bibinfo {author} {\bibfnamefont {P.}~\bibnamefont {Wadley}},\ and\ \bibinfo
  {author} {\bibfnamefont {J.}~\bibnamefont {Wunderlich}},\ }\bibfield  {title}
  {\bibinfo {title} {Antiferromagnetic spintronics},\ }\href
  {https://doi.org/10.1038/nnano.2016.18} {\bibfield  {journal} {\bibinfo
  {journal} {Nat. Nanotechnol.}\ }\textbf {\bibinfo {volume} {11}},\ \bibinfo
  {pages} {231} (\bibinfo {year} {2016})}\BibitemShut {NoStop}%
\bibitem [{\citenamefont {Baltz}\ \emph {et~al.}(2018)\citenamefont {Baltz},
  \citenamefont {Manchon}, \citenamefont {Tsoi}, \citenamefont {Moriyama},
  \citenamefont {Ono},\ and\ \citenamefont
  {Tserkovnyak}}]{baltz2018antiferromagnetic}%
  \BibitemOpen
  \bibfield  {author} {\bibinfo {author} {\bibfnamefont {V.}~\bibnamefont
  {Baltz}}, \bibinfo {author} {\bibfnamefont {A.}~\bibnamefont {Manchon}},
  \bibinfo {author} {\bibfnamefont {M.}~\bibnamefont {Tsoi}}, \bibinfo {author}
  {\bibfnamefont {T.}~\bibnamefont {Moriyama}}, \bibinfo {author}
  {\bibfnamefont {T.}~\bibnamefont {Ono}},\ and\ \bibinfo {author}
  {\bibfnamefont {Y.}~\bibnamefont {Tserkovnyak}},\ }\bibfield  {title}
  {\bibinfo {title} {Antiferromagnetic spintronics},\ }\href
  {https://doi.org/10.1103/RevModPhys.90.015005} {\bibfield  {journal}
  {\bibinfo  {journal} {Rev. Mod. Phys.}\ }\textbf {\bibinfo {volume} {90}},\
  \bibinfo {pages} {015005} (\bibinfo {year} {2018})}\BibitemShut {NoStop}%
\bibitem [{\citenamefont {Jungwirth}\ \emph {et~al.}(2018)\citenamefont
  {Jungwirth}, \citenamefont {Sinova}, \citenamefont {Manchon}, \citenamefont
  {Marti}, \citenamefont {Wunderlich},\ and\ \citenamefont
  {Felser}}]{jungwirth2018multiple}%
  \BibitemOpen
  \bibfield  {author} {\bibinfo {author} {\bibfnamefont {T.}~\bibnamefont
  {Jungwirth}}, \bibinfo {author} {\bibfnamefont {J.}~\bibnamefont {Sinova}},
  \bibinfo {author} {\bibfnamefont {A.}~\bibnamefont {Manchon}}, \bibinfo
  {author} {\bibfnamefont {X.}~\bibnamefont {Marti}}, \bibinfo {author}
  {\bibfnamefont {J.}~\bibnamefont {Wunderlich}},\ and\ \bibinfo {author}
  {\bibfnamefont {C.}~\bibnamefont {Felser}},\ }\bibfield  {title} {\bibinfo
  {title} {The multiple directions of antiferromagnetic spintronics},\ }\href
  {https://doi.org/10.1038/s41567-018-0063-6} {\bibfield  {journal} {\bibinfo
  {journal} {Nat. Phys.}\ }\textbf {\bibinfo {volume} {14}},\ \bibinfo {pages}
  {200} (\bibinfo {year} {2018})}\BibitemShut {NoStop}%
\bibitem [{\citenamefont {N{\v{e}}mec}\ \emph {et~al.}(2018)\citenamefont
  {N{\v{e}}mec}, \citenamefont {Fiebig}, \citenamefont {Kampfrath},\ and\
  \citenamefont {Kimel}}]{nemec2018antiferromagnetic}%
  \BibitemOpen
  \bibfield  {author} {\bibinfo {author} {\bibfnamefont {P.}~\bibnamefont
  {N{\v{e}}mec}}, \bibinfo {author} {\bibfnamefont {M.}~\bibnamefont {Fiebig}},
  \bibinfo {author} {\bibfnamefont {T.}~\bibnamefont {Kampfrath}},\ and\
  \bibinfo {author} {\bibfnamefont {A.~V.}\ \bibnamefont {Kimel}},\ }\bibfield
  {title} {\bibinfo {title} {Antiferromagnetic opto-spintronics},\ }\href
  {https://doi.org/10.1038/s41567-018-0051-x} {\bibfield  {journal} {\bibinfo
  {journal} {Nat. Phys.}\ }\textbf {\bibinfo {volume} {14}},\ \bibinfo {pages}
  {229} (\bibinfo {year} {2018})}\BibitemShut {NoStop}%
\bibitem [{\citenamefont {Liu}\ \emph {et~al.}(2019)\citenamefont {Liu},
  \citenamefont {Feng}, \citenamefont {Yan}, \citenamefont {Wang},
  \citenamefont {Zhou}, \citenamefont {Qin}, \citenamefont {Guo}, \citenamefont
  {Yu},\ and\ \citenamefont {Jiang}}]{liu2019antiferromagnetic}%
  \BibitemOpen
  \bibfield  {author} {\bibinfo {author} {\bibfnamefont {Z.}~\bibnamefont
  {Liu}}, \bibinfo {author} {\bibfnamefont {Z.}~\bibnamefont {Feng}}, \bibinfo
  {author} {\bibfnamefont {H.}~\bibnamefont {Yan}}, \bibinfo {author}
  {\bibfnamefont {X.}~\bibnamefont {Wang}}, \bibinfo {author} {\bibfnamefont
  {X.}~\bibnamefont {Zhou}}, \bibinfo {author} {\bibfnamefont {P.}~\bibnamefont
  {Qin}}, \bibinfo {author} {\bibfnamefont {H.}~\bibnamefont {Guo}}, \bibinfo
  {author} {\bibfnamefont {R.}~\bibnamefont {Yu}},\ and\ \bibinfo {author}
  {\bibfnamefont {C.}~\bibnamefont {Jiang}},\ }\bibfield  {title} {\bibinfo
  {title} {Antiferromagnetic piezospintronics},\ }\href
  {https://doi.org/10.1002/aelm.201900176} {\bibfield  {journal} {\bibinfo
  {journal} {Adv. Electron. Mater.}\ }\textbf {\bibinfo {volume} {5}},\
  \bibinfo {pages} {1900176} (\bibinfo {year} {2019})}\BibitemShut {NoStop}%
\bibitem [{\citenamefont {Fukami}\ \emph {et~al.}(2020)\citenamefont {Fukami},
  \citenamefont {Lorenz},\ and\ \citenamefont
  {Gomonay}}]{fukami2020antiferromagnetic}%
  \BibitemOpen
  \bibfield  {author} {\bibinfo {author} {\bibfnamefont {S.}~\bibnamefont
  {Fukami}}, \bibinfo {author} {\bibfnamefont {V.~O.}\ \bibnamefont {Lorenz}},\
  and\ \bibinfo {author} {\bibfnamefont {O.}~\bibnamefont {Gomonay}},\
  }\bibfield  {title} {\bibinfo {title} {Antiferromagnetic spintronics},\
  }\href {https://doi.org/10.1063/5.0023614} {\bibfield  {journal} {\bibinfo
  {journal} {J. Appl. Phys.}\ }\textbf {\bibinfo {volume} {128}},\ \bibinfo
  {pages} {070401} (\bibinfo {year} {2020})}\BibitemShut {NoStop}%
\bibitem [{\citenamefont {Brataas}\ \emph {et~al.}(2020)\citenamefont
  {Brataas}, \citenamefont {van Wees}, \citenamefont {Klein}, \citenamefont
  {de~Loubens},\ and\ \citenamefont {Viret}}]{brataas2020spin}%
  \BibitemOpen
  \bibfield  {author} {\bibinfo {author} {\bibfnamefont {A.}~\bibnamefont
  {Brataas}}, \bibinfo {author} {\bibfnamefont {B.}~\bibnamefont {van Wees}},
  \bibinfo {author} {\bibfnamefont {O.}~\bibnamefont {Klein}}, \bibinfo
  {author} {\bibfnamefont {G.}~\bibnamefont {de~Loubens}},\ and\ \bibinfo
  {author} {\bibfnamefont {M.}~\bibnamefont {Viret}},\ }\bibfield  {title}
  {\bibinfo {title} {{S}pin {I}nsulatronics},\ }\href
  {https://doi.org/10.1016/j.physrep.2020.08.006} {\bibfield  {journal}
  {\bibinfo  {journal} {Phys. Rep.}\ }\textbf {\bibinfo {volume} {885}},\
  \bibinfo {pages} {1} (\bibinfo {year} {2020})}\BibitemShut {NoStop}%
\bibitem [{\citenamefont {Fl{\'o}rez-G{\'o}mez}\ \emph
  {et~al.}(2022)\citenamefont {Fl{\'o}rez-G{\'o}mez}, \citenamefont
  {Ibarra-Hern{\'a}ndez},\ and\ \citenamefont
  {Garcia-Castro}}]{florez2022lattice}%
  \BibitemOpen
  \bibfield  {author} {\bibinfo {author} {\bibfnamefont {L.}~\bibnamefont
  {Fl{\'o}rez-G{\'o}mez}}, \bibinfo {author} {\bibfnamefont {W.}~\bibnamefont
  {Ibarra-Hern{\'a}ndez}},\ and\ \bibinfo {author} {\bibfnamefont {A.~C.}\
  \bibnamefont {Garcia-Castro}},\ }\bibfield  {title} {\bibinfo {title}
  {Lattice dynamics and spin--phonon coupling in the noncollinear
  antiferromagnetic antiperovskite $\mathrm{Mn}_{3}\mathrm{NiN}$},\ }\href
  {https://doi.org/10.1016/j.jmmm.2022.169813} {\bibfield  {journal} {\bibinfo
  {journal} {J. Magn. Magn. Mater.}\ }\textbf {\bibinfo {volume} {562}},\
  \bibinfo {pages} {169813} (\bibinfo {year} {2022})}\BibitemShut {NoStop}%
\bibitem [{\citenamefont {Xiong}\ \emph {et~al.}(2022)\citenamefont {Xiong},
  \citenamefont {Jiang}, \citenamefont {Shi}, \citenamefont {Du}, \citenamefont
  {Yao}, \citenamefont {Guo}, \citenamefont {Zhu}, \citenamefont {Cao},
  \citenamefont {Peng}, \citenamefont {Cai} \emph
  {et~al.}}]{xiong2022antiferromagnetic}%
  \BibitemOpen
  \bibfield  {author} {\bibinfo {author} {\bibfnamefont {D.}~\bibnamefont
  {Xiong}}, \bibinfo {author} {\bibfnamefont {Y.}~\bibnamefont {Jiang}},
  \bibinfo {author} {\bibfnamefont {K.}~\bibnamefont {Shi}}, \bibinfo {author}
  {\bibfnamefont {A.}~\bibnamefont {Du}}, \bibinfo {author} {\bibfnamefont
  {Y.}~\bibnamefont {Yao}}, \bibinfo {author} {\bibfnamefont {Z.}~\bibnamefont
  {Guo}}, \bibinfo {author} {\bibfnamefont {D.}~\bibnamefont {Zhu}}, \bibinfo
  {author} {\bibfnamefont {K.}~\bibnamefont {Cao}}, \bibinfo {author}
  {\bibfnamefont {S.}~\bibnamefont {Peng}}, \bibinfo {author} {\bibfnamefont
  {W.}~\bibnamefont {Cai}}, \emph {et~al.},\ }\bibfield  {title} {\bibinfo
  {title} {{A}ntiferromagnetic spintronics: {A}n overview and outlook},\ }\href
  {https://doi.org/10.1016/j.fmre.2022.03.016} {\bibfield  {journal} {\bibinfo
  {journal} {Fundam. Res.}\ }\textbf {\bibinfo {volume} {2}},\ \bibinfo {pages}
  {522} (\bibinfo {year} {2022})}\BibitemShut {NoStop}%
\bibitem [{\citenamefont {Han}\ \emph {et~al.}(2023)\citenamefont {Han},
  \citenamefont {Cheng}, \citenamefont {Liu}, \citenamefont {Ohno},\ and\
  \citenamefont {Fukami}}]{han2023coherent}%
  \BibitemOpen
  \bibfield  {author} {\bibinfo {author} {\bibfnamefont {J.}~\bibnamefont
  {Han}}, \bibinfo {author} {\bibfnamefont {R.}~\bibnamefont {Cheng}}, \bibinfo
  {author} {\bibfnamefont {L.}~\bibnamefont {Liu}}, \bibinfo {author}
  {\bibfnamefont {H.}~\bibnamefont {Ohno}},\ and\ \bibinfo {author}
  {\bibfnamefont {S.}~\bibnamefont {Fukami}},\ }\bibfield  {title} {\bibinfo
  {title} {Coherent antiferromagnetic spintronics},\ }\href
  {https://doi.org/10.1038/s41563-023-01492-6} {\bibfield  {journal} {\bibinfo
  {journal} {Nat. Mater.}\ }\textbf {\bibinfo {volume} {22}},\ \bibinfo {pages}
  {684} (\bibinfo {year} {2023})}\BibitemShut {NoStop}%
\bibitem [{\citenamefont {Meer}\ \emph {et~al.}(2023)\citenamefont {Meer},
  \citenamefont {Gomonay}, \citenamefont {Wittmann},\ and\ \citenamefont
  {Kl{\"a}ui}}]{meer2023antiferromagnetic}%
  \BibitemOpen
  \bibfield  {author} {\bibinfo {author} {\bibfnamefont {H.}~\bibnamefont
  {Meer}}, \bibinfo {author} {\bibfnamefont {O.}~\bibnamefont {Gomonay}},
  \bibinfo {author} {\bibfnamefont {A.}~\bibnamefont {Wittmann}},\ and\
  \bibinfo {author} {\bibfnamefont {M.}~\bibnamefont {Kl{\"a}ui}},\ }\bibfield
  {title} {\bibinfo {title} {{A}ntiferromagnetic insulatronics: {S}pintronics
  in insulating 3d metal oxides with antiferromagnetic coupling},\ }\bibfield
  {journal} {\bibinfo  {journal} {Appl. Phys. Lett.}\ }\textbf {\bibinfo
  {volume} {122}},\ \href {https://doi.org/10.1063/5.0135079}
  {10.1063/5.0135079} (\bibinfo {year} {2023})\BibitemShut {NoStop}%
\bibitem [{\citenamefont {Nova}\ \emph {et~al.}(2017)\citenamefont {Nova},
  \citenamefont {Cartella}, \citenamefont {Cantaluppi}, \citenamefont
  {F{\"o}rst}, \citenamefont {Bossini}, \citenamefont {Mikhaylovskiy},
  \citenamefont {Kimel}, \citenamefont {Merlin},\ and\ \citenamefont
  {Cavalleri}}]{nova2017effective}%
  \BibitemOpen
  \bibfield  {author} {\bibinfo {author} {\bibfnamefont {T.~F.}\ \bibnamefont
  {Nova}}, \bibinfo {author} {\bibfnamefont {A.}~\bibnamefont {Cartella}},
  \bibinfo {author} {\bibfnamefont {A.}~\bibnamefont {Cantaluppi}}, \bibinfo
  {author} {\bibfnamefont {M.}~\bibnamefont {F{\"o}rst}}, \bibinfo {author}
  {\bibfnamefont {D.}~\bibnamefont {Bossini}}, \bibinfo {author} {\bibfnamefont
  {R.~V.}\ \bibnamefont {Mikhaylovskiy}}, \bibinfo {author} {\bibfnamefont
  {A.~V.}\ \bibnamefont {Kimel}}, \bibinfo {author} {\bibfnamefont
  {R.}~\bibnamefont {Merlin}},\ and\ \bibinfo {author} {\bibfnamefont
  {A.}~\bibnamefont {Cavalleri}},\ }\bibfield  {title} {\bibinfo {title} {An
  effective magnetic field from optically driven phonons},\ }\href
  {https://doi.org/10.1038/nphys3925} {\bibfield  {journal} {\bibinfo
  {journal} {Nat. Phys.}\ }\textbf {\bibinfo {volume} {13}},\ \bibinfo {pages}
  {132} (\bibinfo {year} {2017})}\BibitemShut {NoStop}%
\bibitem [{\citenamefont {Streib}\ \emph {et~al.}(2019)\citenamefont {Streib},
  \citenamefont {Vidal-Silva}, \citenamefont {Shen},\ and\ \citenamefont
  {Bauer}}]{streib2019magnon}%
  \BibitemOpen
  \bibfield  {author} {\bibinfo {author} {\bibfnamefont {S.}~\bibnamefont
  {Streib}}, \bibinfo {author} {\bibfnamefont {N.}~\bibnamefont {Vidal-Silva}},
  \bibinfo {author} {\bibfnamefont {K.}~\bibnamefont {Shen}},\ and\ \bibinfo
  {author} {\bibfnamefont {G.~E.~W.}\ \bibnamefont {Bauer}},\ }\bibfield
  {title} {\bibinfo {title} {Magnon-phonon interactions in magnetic
  insulators},\ }\href {https://doi.org/10.1103/PhysRevB.99.184442} {\bibfield
  {journal} {\bibinfo  {journal} {Phys. Rev. B}\ }\textbf {\bibinfo {volume}
  {99}},\ \bibinfo {pages} {184442} (\bibinfo {year} {2019})}\BibitemShut
  {NoStop}%
\bibitem [{\citenamefont {Juraschek}\ \emph {et~al.}(2020)\citenamefont
  {Juraschek}, \citenamefont {Narang},\ and\ \citenamefont
  {Spaldin}}]{juraschek2020phono}%
  \BibitemOpen
  \bibfield  {author} {\bibinfo {author} {\bibfnamefont {D.~M.}\ \bibnamefont
  {Juraschek}}, \bibinfo {author} {\bibfnamefont {P.}~\bibnamefont {Narang}},\
  and\ \bibinfo {author} {\bibfnamefont {N.~A.}\ \bibnamefont {Spaldin}},\
  }\bibfield  {title} {\bibinfo {title} {Phono-magnetic analogs to
  opto-magnetic effects},\ }\href
  {https://doi.org/10.1103/PhysRevResearch.2.043035} {\bibfield  {journal}
  {\bibinfo  {journal} {Phys. Rev. Research}\ }\textbf {\bibinfo {volume}
  {2}},\ \bibinfo {pages} {043035} (\bibinfo {year} {2020})}\BibitemShut
  {NoStop}%
\bibitem [{\citenamefont {Juraschek}\ \emph {et~al.}(2021)\citenamefont
  {Juraschek}, \citenamefont {Wang},\ and\ \citenamefont
  {Narang}}]{juraschek2021sum}%
  \BibitemOpen
  \bibfield  {author} {\bibinfo {author} {\bibfnamefont {D.~M.}\ \bibnamefont
  {Juraschek}}, \bibinfo {author} {\bibfnamefont {D.~S.}\ \bibnamefont
  {Wang}},\ and\ \bibinfo {author} {\bibfnamefont {P.}~\bibnamefont {Narang}},\
  }\bibfield  {title} {\bibinfo {title} {Sum-frequency excitation of coherent
  magnons},\ }\href {https://doi.org/10.1103/PhysRevB.103.094407} {\bibfield
  {journal} {\bibinfo  {journal} {Phys. Rev. B}\ }\textbf {\bibinfo {volume}
  {103}},\ \bibinfo {pages} {094407} (\bibinfo {year} {2021})}\BibitemShut
  {NoStop}%
\bibitem [{\citenamefont {Stupakiewicz}\ \emph {et~al.}(2021)\citenamefont
  {Stupakiewicz}, \citenamefont {Davies}, \citenamefont {Szerenos},
  \citenamefont {Afanasiev}, \citenamefont {Rabinovich}, \citenamefont {Boris},
  \citenamefont {Caviglia}, \citenamefont {Kimel},\ and\ \citenamefont
  {Kirilyuk}}]{stupakiewicz2021ultrafast}%
  \BibitemOpen
  \bibfield  {author} {\bibinfo {author} {\bibfnamefont {A.}~\bibnamefont
  {Stupakiewicz}}, \bibinfo {author} {\bibfnamefont {C.}~\bibnamefont
  {Davies}}, \bibinfo {author} {\bibfnamefont {K.}~\bibnamefont {Szerenos}},
  \bibinfo {author} {\bibfnamefont {D.}~\bibnamefont {Afanasiev}}, \bibinfo
  {author} {\bibfnamefont {K.}~\bibnamefont {Rabinovich}}, \bibinfo {author}
  {\bibfnamefont {A.}~\bibnamefont {Boris}}, \bibinfo {author} {\bibfnamefont
  {A.}~\bibnamefont {Caviglia}}, \bibinfo {author} {\bibfnamefont
  {A.}~\bibnamefont {Kimel}},\ and\ \bibinfo {author} {\bibfnamefont
  {A.}~\bibnamefont {Kirilyuk}},\ }\bibfield  {title} {\bibinfo {title}
  {Ultrafast phononic switching of magnetization},\ }\href
  {https://doi.org/10.1038/s41567-020-01124-9} {\bibfield  {journal} {\bibinfo
  {journal} {Nat. Phys.}\ }\textbf {\bibinfo {volume} {17}},\ \bibinfo {pages}
  {489} (\bibinfo {year} {2021})}\BibitemShut {NoStop}%
\bibitem [{\citenamefont {Afanasiev}\ \emph {et~al.}(2021)\citenamefont
  {Afanasiev}, \citenamefont {Hortensius}, \citenamefont {Ivanov},
  \citenamefont {Sasani}, \citenamefont {Bousquet}, \citenamefont {Blanter},
  \citenamefont {Mikhaylovskiy}, \citenamefont {Kimel},\ and\ \citenamefont
  {Caviglia}}]{afanasiev2021ultrafast}%
  \BibitemOpen
  \bibfield  {author} {\bibinfo {author} {\bibfnamefont {D.}~\bibnamefont
  {Afanasiev}}, \bibinfo {author} {\bibfnamefont {J.}~\bibnamefont
  {Hortensius}}, \bibinfo {author} {\bibfnamefont {B.}~\bibnamefont {Ivanov}},
  \bibinfo {author} {\bibfnamefont {A.}~\bibnamefont {Sasani}}, \bibinfo
  {author} {\bibfnamefont {E.}~\bibnamefont {Bousquet}}, \bibinfo {author}
  {\bibfnamefont {Y.}~\bibnamefont {Blanter}}, \bibinfo {author} {\bibfnamefont
  {R.}~\bibnamefont {Mikhaylovskiy}}, \bibinfo {author} {\bibfnamefont
  {A.}~\bibnamefont {Kimel}},\ and\ \bibinfo {author} {\bibfnamefont
  {A.}~\bibnamefont {Caviglia}},\ }\bibfield  {title} {\bibinfo {title}
  {Ultrafast control of magnetic interactions via light-driven phonons},\
  }\href {https://doi.org/10.1038/s41563-021-00922-7} {\bibfield  {journal}
  {\bibinfo  {journal} {Nat. Mater.}\ }\textbf {\bibinfo {volume} {20}},\
  \bibinfo {pages} {607} (\bibinfo {year} {2021})}\BibitemShut {NoStop}%
\bibitem [{\citenamefont {Disa}\ \emph {et~al.}(2021)\citenamefont {Disa},
  \citenamefont {Nova},\ and\ \citenamefont {Cavalleri}}]{disa2021engineering}%
  \BibitemOpen
  \bibfield  {author} {\bibinfo {author} {\bibfnamefont {A.~S.}\ \bibnamefont
  {Disa}}, \bibinfo {author} {\bibfnamefont {T.~F.}\ \bibnamefont {Nova}},\
  and\ \bibinfo {author} {\bibfnamefont {A.}~\bibnamefont {Cavalleri}},\
  }\bibfield  {title} {\bibinfo {title} {Engineering crystal structures with
  light},\ }\href {https://doi.org/10.1038/s41567-021-01366-1} {\bibfield
  {journal} {\bibinfo  {journal} {Nat. Phys.}\ }\textbf {\bibinfo {volume}
  {17}},\ \bibinfo {pages} {1087} (\bibinfo {year} {2021})}\BibitemShut
  {NoStop}%
\bibitem [{\citenamefont {Ueda}\ \emph {et~al.}(2023)\citenamefont {Ueda},
  \citenamefont {Mankowsky}, \citenamefont {Paris}, \citenamefont {Sander},
  \citenamefont {Deng}, \citenamefont {Liu}, \citenamefont {Leroy},
  \citenamefont {Nag}, \citenamefont {Skoropata}, \citenamefont {Wang},
  \citenamefont {Ukleev}, \citenamefont {Perren}, \citenamefont
  {D\"{o}ssegger}, \citenamefont {Gurung}, \citenamefont {Svetina},
  \citenamefont {Abreu}, \citenamefont {Savoini}, \citenamefont {Kimura},
  \citenamefont {Patthey}, \citenamefont {Razzoli}, \citenamefont {Lemke},
  \citenamefont {Johnson},\ and\ \citenamefont {Staub}}]{ueda2023non}%
  \BibitemOpen
  \bibfield  {author} {\bibinfo {author} {\bibfnamefont {H.}~\bibnamefont
  {Ueda}}, \bibinfo {author} {\bibfnamefont {R.}~\bibnamefont {Mankowsky}},
  \bibinfo {author} {\bibfnamefont {E.}~\bibnamefont {Paris}}, \bibinfo
  {author} {\bibfnamefont {M.}~\bibnamefont {Sander}}, \bibinfo {author}
  {\bibfnamefont {Y.}~\bibnamefont {Deng}}, \bibinfo {author} {\bibfnamefont
  {B.}~\bibnamefont {Liu}}, \bibinfo {author} {\bibfnamefont {L.}~\bibnamefont
  {Leroy}}, \bibinfo {author} {\bibfnamefont {A.}~\bibnamefont {Nag}}, \bibinfo
  {author} {\bibfnamefont {E.}~\bibnamefont {Skoropata}}, \bibinfo {author}
  {\bibfnamefont {C.}~\bibnamefont {Wang}}, \bibinfo {author} {\bibfnamefont
  {V.}~\bibnamefont {Ukleev}}, \bibinfo {author} {\bibfnamefont {G.~S.}\
  \bibnamefont {Perren}}, \bibinfo {author} {\bibfnamefont {J.}~\bibnamefont
  {D\"{o}ssegger}}, \bibinfo {author} {\bibfnamefont {S.}~\bibnamefont
  {Gurung}}, \bibinfo {author} {\bibfnamefont {C.}~\bibnamefont {Svetina}},
  \bibinfo {author} {\bibfnamefont {E.}~\bibnamefont {Abreu}}, \bibinfo
  {author} {\bibfnamefont {M.}~\bibnamefont {Savoini}}, \bibinfo {author}
  {\bibfnamefont {T.}~\bibnamefont {Kimura}}, \bibinfo {author} {\bibfnamefont
  {L.}~\bibnamefont {Patthey}}, \bibinfo {author} {\bibfnamefont
  {E.}~\bibnamefont {Razzoli}}, \bibinfo {author} {\bibfnamefont {H.~T.}\
  \bibnamefont {Lemke}}, \bibinfo {author} {\bibfnamefont {S.~L.}\ \bibnamefont
  {Johnson}},\ and\ \bibinfo {author} {\bibfnamefont {U.}~\bibnamefont
  {Staub}},\ }\bibfield  {title} {\bibinfo {title} {Non-equilibrium dynamics of
  spin-lattice coupling},\ }\href {https://doi.org/10.1038/s41467-023-43581-9}
  {\bibfield  {journal} {\bibinfo  {journal} {Nat. Commun.}\ }\textbf {\bibinfo
  {volume} {14}},\ \bibinfo {pages} {7778} (\bibinfo {year}
  {2023})}\BibitemShut {NoStop}%
\bibitem [{\citenamefont {Davies}\ \emph {et~al.}(2024)\citenamefont {Davies},
  \citenamefont {Fennema}, \citenamefont {Tsukamoto}, \citenamefont
  {Razdolski}, \citenamefont {Kimel},\ and\ \citenamefont
  {Kirilyuk}}]{davies2024phononis}%
  \BibitemOpen
  \bibfield  {author} {\bibinfo {author} {\bibfnamefont {C.}~\bibnamefont
  {Davies}}, \bibinfo {author} {\bibfnamefont {F.~G.~N.}\ \bibnamefont
  {Fennema}}, \bibinfo {author} {\bibfnamefont {A.}~\bibnamefont {Tsukamoto}},
  \bibinfo {author} {\bibfnamefont {I.}~\bibnamefont {Razdolski}}, \bibinfo
  {author} {\bibfnamefont {A.~V.}\ \bibnamefont {Kimel}},\ and\ \bibinfo
  {author} {\bibfnamefont {A.}~\bibnamefont {Kirilyuk}},\ }\bibfield  {title}
  {\bibinfo {title} {{P}hononic switching of magnetization by the ultrafast
  {B}arnett effect},\ }\href {https://doi.org/10.1038/s41586-024-07200-x}
  {\bibfield  {journal} {\bibinfo  {journal} {Nature}\ }\textbf {\bibinfo
  {volume} {628}},\ \bibinfo {pages} {540} (\bibinfo {year}
  {2024})}\BibitemShut {NoStop}%
\bibitem [{\citenamefont {Basini}\ \emph {et~al.}(2024)\citenamefont {Basini},
  \citenamefont {Pancaldi}, \citenamefont {Wehinger}, \citenamefont {Udina},
  \citenamefont {Unikandanunni}, \citenamefont {Tadano}, \citenamefont
  {Hoffmann}, \citenamefont {Balatsky},\ and\ \citenamefont
  {Bonetti}}]{basini2024terahertz}%
  \BibitemOpen
  \bibfield  {author} {\bibinfo {author} {\bibfnamefont {M.}~\bibnamefont
  {Basini}}, \bibinfo {author} {\bibfnamefont {M.}~\bibnamefont {Pancaldi}},
  \bibinfo {author} {\bibfnamefont {B.}~\bibnamefont {Wehinger}}, \bibinfo
  {author} {\bibfnamefont {M.}~\bibnamefont {Udina}}, \bibinfo {author}
  {\bibfnamefont {V.}~\bibnamefont {Unikandanunni}}, \bibinfo {author}
  {\bibfnamefont {T.}~\bibnamefont {Tadano}}, \bibinfo {author} {\bibfnamefont
  {M.~C.}\ \bibnamefont {Hoffmann}}, \bibinfo {author} {\bibfnamefont {A.~V.}\
  \bibnamefont {Balatsky}},\ and\ \bibinfo {author} {\bibfnamefont
  {S.}~\bibnamefont {Bonetti}},\ }\bibfield  {title} {\bibinfo {title}
  {{T}erahertz electric-field-driven dynamical multiferroicity in
  $\mathrm{SrTiO}_{3}$},\ }\href {https://doi.org/10.1038/s41586-024-07175-9}
  {\bibfield  {journal} {\bibinfo  {journal} {Nature}\ }\textbf {\bibinfo
  {volume} {628}},\ \bibinfo {pages} {534} (\bibinfo {year}
  {2024})}\BibitemShut {NoStop}%
\bibitem [{\citenamefont {Gu}\ \emph {et~al.}(2022)\citenamefont {Gu},
  \citenamefont {Bai}, \citenamefont {Zhang},\ and\ \citenamefont
  {George}}]{gu2022spin}%
  \BibitemOpen
  \bibfield  {author} {\bibinfo {author} {\bibfnamefont {M.}~\bibnamefont
  {Gu}}, \bibinfo {author} {\bibfnamefont {Y.~H.}\ \bibnamefont {Bai}},
  \bibinfo {author} {\bibfnamefont {G.~P.}\ \bibnamefont {Zhang}},\ and\
  \bibinfo {author} {\bibfnamefont {T.~F.}\ \bibnamefont {George}},\ }\bibfield
   {title} {\bibinfo {title} {Spin--phonon dispersion in magnetic materials},\
  }\href {https://doi.org/10.1088/1361-648X/ac7f17} {\bibfield  {journal}
  {\bibinfo  {journal} {J. Phys. Condens. Matter}\ }\textbf {\bibinfo {volume}
  {34}},\ \bibinfo {pages} {375802} (\bibinfo {year} {2022})}\BibitemShut
  {NoStop}%
\bibitem [{\citenamefont {Guo}\ \emph {et~al.}(2023)\citenamefont {Guo},
  \citenamefont {Liu}, \citenamefont {Janson}, \citenamefont {Fulga},
  \citenamefont {van~den Brink},\ and\ \citenamefont {Facio}}]{guo2023spin}%
  \BibitemOpen
  \bibfield  {author} {\bibinfo {author} {\bibfnamefont {Y.}~\bibnamefont
  {Guo}}, \bibinfo {author} {\bibfnamefont {H.}~\bibnamefont {Liu}}, \bibinfo
  {author} {\bibfnamefont {O.}~\bibnamefont {Janson}}, \bibinfo {author}
  {\bibfnamefont {I.~C.}\ \bibnamefont {Fulga}}, \bibinfo {author}
  {\bibfnamefont {J.}~\bibnamefont {van~den Brink}},\ and\ \bibinfo {author}
  {\bibfnamefont {J.~I.}\ \bibnamefont {Facio}},\ }\bibfield  {title} {\bibinfo
  {title} {Spin-split collinear antiferromagnets: a large-scale ab-initio
  study},\ }\href {https://doi.org/10.1016/j.mtphys.2023.100991} {\bibfield
  {journal} {\bibinfo  {journal} {Mater. Today Phys.}\ }\textbf {\bibinfo
  {volume} {32}},\ \bibinfo {pages} {100991} (\bibinfo {year}
  {2023})}\BibitemShut {NoStop}%
\bibitem [{\citenamefont {Wu}\ \emph {et~al.}(2016)\citenamefont {Wu},
  \citenamefont {Zhang}, \citenamefont {KC}, \citenamefont {Borisov},
  \citenamefont {Pearson}, \citenamefont {Jiang}, \citenamefont {Lederman},
  \citenamefont {Hoffmann},\ and\ \citenamefont
  {Bhattacharya}}]{wu2016antiferromagnetic}%
  \BibitemOpen
  \bibfield  {author} {\bibinfo {author} {\bibfnamefont {S.~M.}\ \bibnamefont
  {Wu}}, \bibinfo {author} {\bibfnamefont {W.}~\bibnamefont {Zhang}}, \bibinfo
  {author} {\bibfnamefont {A.}~\bibnamefont {KC}}, \bibinfo {author}
  {\bibfnamefont {P.}~\bibnamefont {Borisov}}, \bibinfo {author} {\bibfnamefont
  {J.~E.}\ \bibnamefont {Pearson}}, \bibinfo {author} {\bibfnamefont {J.~S.}\
  \bibnamefont {Jiang}}, \bibinfo {author} {\bibfnamefont {D.}~\bibnamefont
  {Lederman}}, \bibinfo {author} {\bibfnamefont {A.}~\bibnamefont {Hoffmann}},\
  and\ \bibinfo {author} {\bibfnamefont {A.}~\bibnamefont {Bhattacharya}},\
  }\bibfield  {title} {\bibinfo {title} {Antiferromagnetic spin seebeck
  effect},\ }\href {https://doi.org/10.1103/PhysRevLett.116.097204} {\bibfield
  {journal} {\bibinfo  {journal} {Phys. Rev. Lett.}\ }\textbf {\bibinfo
  {volume} {116}},\ \bibinfo {pages} {097204} (\bibinfo {year}
  {2016})}\BibitemShut {NoStop}%
\bibitem [{\citenamefont {Li}\ \emph {et~al.}(2019)\citenamefont {Li},
  \citenamefont {Shi}, \citenamefont {Ortiz}, \citenamefont {Aldosary},
  \citenamefont {Chen}, \citenamefont {Aji}, \citenamefont {Wei},\ and\
  \citenamefont {Shi}}]{li2019spin}%
  \BibitemOpen
  \bibfield  {author} {\bibinfo {author} {\bibfnamefont {J.}~\bibnamefont
  {Li}}, \bibinfo {author} {\bibfnamefont {Z.}~\bibnamefont {Shi}}, \bibinfo
  {author} {\bibfnamefont {V.~H.}\ \bibnamefont {Ortiz}}, \bibinfo {author}
  {\bibfnamefont {M.}~\bibnamefont {Aldosary}}, \bibinfo {author}
  {\bibfnamefont {C.}~\bibnamefont {Chen}}, \bibinfo {author} {\bibfnamefont
  {V.}~\bibnamefont {Aji}}, \bibinfo {author} {\bibfnamefont {P.}~\bibnamefont
  {Wei}},\ and\ \bibinfo {author} {\bibfnamefont {J.}~\bibnamefont {Shi}},\
  }\bibfield  {title} {\bibinfo {title} {{S}pin {S}eebeck {E}ffect from
  {A}ntiferromagnetic {M}agnons and {C}ritical {S}pin {F}luctuations in
  {E}pitaxial $\mathrm{FeF}_{2}$ films},\ }\href
  {https://doi.org/10.1103/PhysRevLett.122.217204} {\bibfield  {journal}
  {\bibinfo  {journal} {Phys. Rev. Lett.}\ }\textbf {\bibinfo {volume} {122}},\
  \bibinfo {pages} {217204} (\bibinfo {year} {2019})}\BibitemShut {NoStop}%
\bibitem [{\citenamefont {Vaidya}\ \emph {et~al.}(2020)\citenamefont {Vaidya},
  \citenamefont {Morley}, \citenamefont {van Tol}, \citenamefont {Liu},
  \citenamefont {Cheng}, \citenamefont {Brataas}, \citenamefont {Lederman},\
  and\ \citenamefont {Del~Barco}}]{vaidya2020subterahertz}%
  \BibitemOpen
  \bibfield  {author} {\bibinfo {author} {\bibfnamefont {P.}~\bibnamefont
  {Vaidya}}, \bibinfo {author} {\bibfnamefont {S.~A.}\ \bibnamefont {Morley}},
  \bibinfo {author} {\bibfnamefont {J.}~\bibnamefont {van Tol}}, \bibinfo
  {author} {\bibfnamefont {Y.}~\bibnamefont {Liu}}, \bibinfo {author}
  {\bibfnamefont {R.}~\bibnamefont {Cheng}}, \bibinfo {author} {\bibfnamefont
  {A.}~\bibnamefont {Brataas}}, \bibinfo {author} {\bibfnamefont
  {D.}~\bibnamefont {Lederman}},\ and\ \bibinfo {author} {\bibfnamefont
  {E.}~\bibnamefont {Del~Barco}},\ }\bibfield  {title} {\bibinfo {title}
  {Subterahertz spin pumping from an insulating antiferromagnet},\ }\href
  {https://doi.org/10.1126/science.aaz4247} {\bibfield  {journal} {\bibinfo
  {journal} {Science}\ }\textbf {\bibinfo {volume} {368}},\ \bibinfo {pages}
  {160} (\bibinfo {year} {2020})}\BibitemShut {NoStop}%
\bibitem [{\citenamefont {Lockwood}\ and\ \citenamefont
  {Cottam}(1988)}]{lockwood1988spin}%
  \BibitemOpen
  \bibfield  {author} {\bibinfo {author} {\bibfnamefont {D.~J.}\ \bibnamefont
  {Lockwood}}\ and\ \bibinfo {author} {\bibfnamefont {M.~G.}\ \bibnamefont
  {Cottam}},\ }\bibfield  {title} {\bibinfo {title} {The spin-phonon
  interaction in $\mathrm{FeF}_{2}$ and $\mathrm{MnF}_{2}$ studied by {R}aman
  spectroscopy},\ }\href {https://doi.org/10.1063/1.342186} {\bibfield
  {journal} {\bibinfo  {journal} {J. Appl. Phys.}\ }\textbf {\bibinfo {volume}
  {64}},\ \bibinfo {pages} {5876} (\bibinfo {year} {1988})}\BibitemShut
  {NoStop}%
\bibitem [{\citenamefont {Schleck}\ \emph {et~al.}(2010)\citenamefont
  {Schleck}, \citenamefont {Nahas}, \citenamefont {Lobo}, \citenamefont
  {Varignon}, \citenamefont {Lepetit}, \citenamefont {Nelson},\ and\
  \citenamefont {Moreira}}]{schleck2010elastic}%
  \BibitemOpen
  \bibfield  {author} {\bibinfo {author} {\bibfnamefont {R.}~\bibnamefont
  {Schleck}}, \bibinfo {author} {\bibfnamefont {Y.}~\bibnamefont {Nahas}},
  \bibinfo {author} {\bibfnamefont {R.~P. S.~M.}\ \bibnamefont {Lobo}},
  \bibinfo {author} {\bibfnamefont {J.}~\bibnamefont {Varignon}}, \bibinfo
  {author} {\bibfnamefont {M.~B.}\ \bibnamefont {Lepetit}}, \bibinfo {author}
  {\bibfnamefont {C.~S.}\ \bibnamefont {Nelson}},\ and\ \bibinfo {author}
  {\bibfnamefont {R.~L.}\ \bibnamefont {Moreira}},\ }\bibfield  {title}
  {\bibinfo {title} {Elastic and magnetic effects on the infrared phonon
  spectra of $\mathrm{MnF}_{2}$},\ }\href
  {https://doi.org/10.1103/PhysRevB.82.054412} {\bibfield  {journal} {\bibinfo
  {journal} {Phys. Rev. B}\ }\textbf {\bibinfo {volume} {82}},\ \bibinfo
  {pages} {054412} (\bibinfo {year} {2010})}\BibitemShut {NoStop}%
\bibitem [{\citenamefont {Seehra}\ and\ \citenamefont
  {Helmick}(1984)}]{seehra1984anomalous}%
  \BibitemOpen
  \bibfield  {author} {\bibinfo {author} {\bibfnamefont {M.~S.}\ \bibnamefont
  {Seehra}}\ and\ \bibinfo {author} {\bibfnamefont {R.~E.}\ \bibnamefont
  {Helmick}},\ }\bibfield  {title} {\bibinfo {title} {Anomalous changes in the
  dielectric constants of $\mathrm{MnF}_{2}$ near its {N}{\'e}el temperature},\
  }\href {https://doi.org/10.1063/1.333652} {\bibfield  {journal} {\bibinfo
  {journal} {J. Appl. Phys.}\ }\textbf {\bibinfo {volume} {55}},\ \bibinfo
  {pages} {2330} (\bibinfo {year} {1984})}\BibitemShut {NoStop}%
\bibitem [{\citenamefont {Seehra}\ \emph {et~al.}(1986)\citenamefont {Seehra},
  \citenamefont {Helmick},\ and\ \citenamefont
  {Srinivasan}}]{seehra1986effect}%
  \BibitemOpen
  \bibfield  {author} {\bibinfo {author} {\bibfnamefont {M.~S.}\ \bibnamefont
  {Seehra}}, \bibinfo {author} {\bibfnamefont {R.~E.}\ \bibnamefont
  {Helmick}},\ and\ \bibinfo {author} {\bibfnamefont {G.}~\bibnamefont
  {Srinivasan}},\ }\bibfield  {title} {\bibinfo {title} {Effect of temperature
  and antiferromagnetic ordering on the dielectric constants of $\mathrm{MnO}$
  and $\mathrm{MnF}_{2}$},\ }\href
  {https://doi.org/10.1088/0022-3719/19/10/016} {\bibfield  {journal} {\bibinfo
   {journal} {J. Phys. C: Solid State Phys.}\ }\textbf {\bibinfo {volume}
  {19}},\ \bibinfo {pages} {1627} (\bibinfo {year} {1986})}\BibitemShut
  {NoStop}%
\bibitem [{\citenamefont {Moriya}\ \emph {et~al.}(1956)\citenamefont {Moriya},
  \citenamefont {Motizuki}, \citenamefont {Kanamori},\ and\ \citenamefont
  {Nagamiya}}]{moriya1956magnetic}%
  \BibitemOpen
  \bibfield  {author} {\bibinfo {author} {\bibfnamefont {T.}~\bibnamefont
  {Moriya}}, \bibinfo {author} {\bibfnamefont {K.}~\bibnamefont {Motizuki}},
  \bibinfo {author} {\bibfnamefont {J.}~\bibnamefont {Kanamori}},\ and\
  \bibinfo {author} {\bibfnamefont {T.}~\bibnamefont {Nagamiya}},\ }\bibfield
  {title} {\bibinfo {title} {On the {M}agnetic {A}nisotropy of
  $\mathrm{FeF}_{2}$ and $\mathrm{CoF}_{2}$},\ }\href
  {https://doi.org/10.1143/JPSJ.11.211} {\bibfield  {journal} {\bibinfo
  {journal} {J. Phys. Soc. Jpn.}\ }\textbf {\bibinfo {volume} {11}},\ \bibinfo
  {pages} {211} (\bibinfo {year} {1956})}\BibitemShut {NoStop}%
\bibitem [{\citenamefont {Corr\^ea}\ and\ \citenamefont
  {V\'yborn\'y}(2018)}]{correa2018electronic}%
  \BibitemOpen
  \bibfield  {author} {\bibinfo {author} {\bibfnamefont {C.~A.}\ \bibnamefont
  {Corr\^ea}}\ and\ \bibinfo {author} {\bibfnamefont {K.}~\bibnamefont
  {V\'yborn\'y}},\ }\bibfield  {title} {\bibinfo {title} {Electronic structure
  and magnetic anisotropies of antiferromagnetic transition-metal
  difluorides},\ }\href {https://doi.org/10.1103/PhysRevB.97.235111} {\bibfield
   {journal} {\bibinfo  {journal} {Phys. Rev. B}\ }\textbf {\bibinfo {volume}
  {97}},\ \bibinfo {pages} {235111} (\bibinfo {year} {2018})}\BibitemShut
  {NoStop}%
\bibitem [{\citenamefont {Jauch}\ \emph {et~al.}(2004)\citenamefont {Jauch},
  \citenamefont {Reehuis},\ and\ \citenamefont {Schultz}}]{jauch2004gamma}%
  \BibitemOpen
  \bibfield  {author} {\bibinfo {author} {\bibfnamefont {W.}~\bibnamefont
  {Jauch}}, \bibinfo {author} {\bibfnamefont {M.}~\bibnamefont {Reehuis}},\
  and\ \bibinfo {author} {\bibfnamefont {A.~J.}\ \bibnamefont {Schultz}},\
  }\bibfield  {title} {\bibinfo {title} {$\gamma$-ray and neutron diffraction
  studies of $\mathrm{CoF}_{2}$: magnetostriction, electron density and
  magnetic moments},\ }\href {https://doi.org/10.1107/S0108767303022803}
  {\bibfield  {journal} {\bibinfo  {journal} {Acta Cryst.}\ }\textbf {\bibinfo
  {volume} {A60}},\ \bibinfo {pages} {51} (\bibinfo {year} {2004})}\BibitemShut
  {NoStop}%
\bibitem [{\citenamefont {Chatterji}\ \emph {et~al.}(2010)\citenamefont
  {Chatterji}, \citenamefont {Ouladdiaf},\ and\ \citenamefont
  {Hansen}}]{chatterji2010magnetoelasticCoF2}%
  \BibitemOpen
  \bibfield  {author} {\bibinfo {author} {\bibfnamefont {T.}~\bibnamefont
  {Chatterji}}, \bibinfo {author} {\bibfnamefont {B.}~\bibnamefont
  {Ouladdiaf}},\ and\ \bibinfo {author} {\bibfnamefont {T.~C.}\ \bibnamefont
  {Hansen}},\ }\bibfield  {title} {\bibinfo {title} {The magnetoelastic effect
  in $\mathrm{CoF}_{2}$ investigated by means of neutron powder diffraction},\
  }\href {https://doi.org/10.1088/0953-8984/22/9/096001} {\bibfield  {journal}
  {\bibinfo  {journal} {J. Phys. Condens. Matter}\ }\textbf {\bibinfo {volume}
  {22}},\ \bibinfo {pages} {096001} (\bibinfo {year} {2010})}\BibitemShut
  {NoStop}%
\bibitem [{\citenamefont {Borovik-Romanov}(1960)}]{borovik1960piezomagnetism}%
  \BibitemOpen
  \bibfield  {author} {\bibinfo {author} {\bibfnamefont {A.}~\bibnamefont
  {Borovik-Romanov}},\ }\bibfield  {title} {\bibinfo {title} {Piezomagnetism in
  the antiferromagnetic fluorides of cobalt and manganese},\ }\href
  {http://www.jetp.ras.ru/cgi-bin/dn/e_011_04_0786.pdf} {\bibfield  {journal}
  {\bibinfo  {journal} {Sov. Phys. JETP}\ }\textbf {\bibinfo {volume} {11}},\
  \bibinfo {pages} {786} (\bibinfo {year} {1960})}\BibitemShut {NoStop}%
\bibitem [{\citenamefont {Moriya}(1959)}]{moriya1959piezomagnetism}%
  \BibitemOpen
  \bibfield  {author} {\bibinfo {author} {\bibfnamefont {T.}~\bibnamefont
  {Moriya}},\ }\bibfield  {title} {\bibinfo {title} {Piezomagnetism in
  $\mathrm{CoF}_{2}$},\ }\href {https://doi.org/10.1016/0022-3697(59)90043-5}
  {\bibfield  {journal} {\bibinfo  {journal} {J. Phys. Chem. Solids}\ }\textbf
  {\bibinfo {volume} {11}},\ \bibinfo {pages} {73} (\bibinfo {year}
  {1959})}\BibitemShut {NoStop}%
\bibitem [{\citenamefont {Phillips}\ \emph {et~al.}(1967)\citenamefont
  {Phillips}, \citenamefont {Townsend},\ and\ \citenamefont
  {White}}]{phillips1967piezomagnetism}%
  \BibitemOpen
  \bibfield  {author} {\bibinfo {author} {\bibfnamefont {T.~G.}\ \bibnamefont
  {Phillips}}, \bibinfo {author} {\bibfnamefont {R.~L.}\ \bibnamefont
  {Townsend}},\ and\ \bibinfo {author} {\bibfnamefont {R.~L.}\ \bibnamefont
  {White}},\ }\bibfield  {title} {\bibinfo {title} {Piezomagnetism of
  $\mathrm{CoF}_{2}$ and $\alpha$-$\mathrm{Fe}_{2}\mathrm{O}_{3}$ from
  electron-paramagnetic-resonance pressure experiments},\ }\href
  {https://doi.org/10.1103/PhysRevLett.18.646} {\bibfield  {journal} {\bibinfo
  {journal} {Phys. Rev. Lett.}\ }\textbf {\bibinfo {volume} {18}},\ \bibinfo
  {pages} {646} (\bibinfo {year} {1967})}\BibitemShut {NoStop}%
\bibitem [{\citenamefont {Disa}\ \emph {et~al.}(2020)\citenamefont {Disa},
  \citenamefont {Fechner}, \citenamefont {Nova}, \citenamefont {Liu},
  \citenamefont {F\"{o}rst}, \citenamefont {Prabhakaran}, \citenamefont
  {Radaelli},\ and\ \citenamefont {Cavalleri}}]{disa2020polarizing}%
  \BibitemOpen
  \bibfield  {author} {\bibinfo {author} {\bibfnamefont {A.~S.}\ \bibnamefont
  {Disa}}, \bibinfo {author} {\bibfnamefont {M.}~\bibnamefont {Fechner}},
  \bibinfo {author} {\bibfnamefont {T.~F.}\ \bibnamefont {Nova}}, \bibinfo
  {author} {\bibfnamefont {B.}~\bibnamefont {Liu}}, \bibinfo {author}
  {\bibfnamefont {M.}~\bibnamefont {F\"{o}rst}}, \bibinfo {author}
  {\bibfnamefont {D.}~\bibnamefont {Prabhakaran}}, \bibinfo {author}
  {\bibfnamefont {P.~G.}\ \bibnamefont {Radaelli}},\ and\ \bibinfo {author}
  {\bibfnamefont {A.}~\bibnamefont {Cavalleri}},\ }\bibfield  {title} {\bibinfo
  {title} {Polarizing an antiferromagnet by optical engineering of the crystal
  field},\ }\href {https://doi.org/10.1038/s41567-020-0936-3} {\bibfield
  {journal} {\bibinfo  {journal} {Nat. Phys.}\ } (\bibinfo {year}
  {2020})}\BibitemShut {NoStop}%
\bibitem [{\citenamefont {Formisano}\ \emph
  {et~al.}(2022{\natexlab{a}})\citenamefont {Formisano}, \citenamefont
  {Dubrovin}, \citenamefont {Pisarev}, \citenamefont {Kalashnikova},\ and\
  \citenamefont {Kimel}}]{formisano2022laser_JPCM}%
  \BibitemOpen
  \bibfield  {author} {\bibinfo {author} {\bibfnamefont {F.}~\bibnamefont
  {Formisano}}, \bibinfo {author} {\bibfnamefont {R.~M.}\ \bibnamefont
  {Dubrovin}}, \bibinfo {author} {\bibfnamefont {R.~V.}\ \bibnamefont
  {Pisarev}}, \bibinfo {author} {\bibfnamefont {A.~M.}\ \bibnamefont
  {Kalashnikova}},\ and\ \bibinfo {author} {\bibfnamefont {A.~V.}\ \bibnamefont
  {Kimel}},\ }\bibfield  {title} {\bibinfo {title} {{L}aser-induced {THz}
  magnetism of antiferromagnetic $\mathrm{CoF}_{2}$},\ }\href
  {https://doi.org/10.1088/1361-648X/ac5c20} {\bibfield  {journal} {\bibinfo
  {journal} {J. Phys. Condens. Matter}\ }\textbf {\bibinfo {volume} {34}},\
  \bibinfo {pages} {225801} (\bibinfo {year} {2022}{\natexlab{a}})}\BibitemShut
  {NoStop}%
\bibitem [{\citenamefont {Formisano}\ \emph
  {et~al.}(2022{\natexlab{b}})\citenamefont {Formisano}, \citenamefont
  {Dubrovin}, \citenamefont {Pisarev}, \citenamefont {Zvezdin}, \citenamefont
  {Kalashnikova},\ and\ \citenamefont {Kimel}}]{formisano2022laser_AP}%
  \BibitemOpen
  \bibfield  {author} {\bibinfo {author} {\bibfnamefont {F.}~\bibnamefont
  {Formisano}}, \bibinfo {author} {\bibfnamefont {R.~M.}\ \bibnamefont
  {Dubrovin}}, \bibinfo {author} {\bibfnamefont {R.~V.}\ \bibnamefont
  {Pisarev}}, \bibinfo {author} {\bibfnamefont {A.~K.}\ \bibnamefont
  {Zvezdin}}, \bibinfo {author} {\bibfnamefont {A.~M.}\ \bibnamefont
  {Kalashnikova}},\ and\ \bibinfo {author} {\bibfnamefont {A.~V.}\ \bibnamefont
  {Kimel}},\ }\bibfield  {title} {\bibinfo {title} {{L}aser-induced {THz}
  piezomagnetism and lattice dynamics of antiferromagnets $\mathrm{MnF}_{2}$
  and $\mathrm{CoF}_{2}$},\ }\href {https://doi.org/10.1016/j.aop.2022.169041}
  {\bibfield  {journal} {\bibinfo  {journal} {Ann. Phys.}\ ,\ \bibinfo {pages}
  {169041}} (\bibinfo {year} {2022}{\natexlab{b}})}\BibitemShut {NoStop}%
\bibitem [{\citenamefont {Mashkovich}\ \emph {et~al.}(2021)\citenamefont
  {Mashkovich}, \citenamefont {Grishunin}, \citenamefont {Dubrovin},
  \citenamefont {Zvezdin}, \citenamefont {Pisarev},\ and\ \citenamefont
  {Kimel}}]{mashkovich2021terahertz}%
  \BibitemOpen
  \bibfield  {author} {\bibinfo {author} {\bibfnamefont {E.~A.}\ \bibnamefont
  {Mashkovich}}, \bibinfo {author} {\bibfnamefont {K.~A.}\ \bibnamefont
  {Grishunin}}, \bibinfo {author} {\bibfnamefont {R.~M.}\ \bibnamefont
  {Dubrovin}}, \bibinfo {author} {\bibfnamefont {A.~K.}\ \bibnamefont
  {Zvezdin}}, \bibinfo {author} {\bibfnamefont {R.~V.}\ \bibnamefont
  {Pisarev}},\ and\ \bibinfo {author} {\bibfnamefont {A.~V.}\ \bibnamefont
  {Kimel}},\ }\bibfield  {title} {\bibinfo {title} {Terahertz light-driven
  coupling of antiferromagnetic spins to lattice},\ }\href
  {https://doi.org/10.1126/science.abk1121} {\bibfield  {journal} {\bibinfo
  {journal} {Science}\ }\textbf {\bibinfo {volume} {374}},\ \bibinfo {pages}
  {1608} (\bibinfo {year} {2021})}\BibitemShut {NoStop}%
\bibitem [{\citenamefont {Metzger}\ \emph {et~al.}(2023)\citenamefont
  {Metzger}, \citenamefont {Grishunin}, \citenamefont {Reinhoffer},
  \citenamefont {Dubrovin}, \citenamefont {Arshad}, \citenamefont {Ilyakov},
  \citenamefont {de~Oliveira}, \citenamefont {Ponomaryov}, \citenamefont
  {Deinert}, \citenamefont {Kovalev}, \citenamefont {Pisarev}, \citenamefont
  {Katsnelson}, \citenamefont {Ivanov}, \citenamefont {van Loosdrecht},
  \citenamefont {Kimel},\ and\ \citenamefont
  {Mashkovich}}]{metzger2023impulsive}%
  \BibitemOpen
  \bibfield  {author} {\bibinfo {author} {\bibfnamefont {T.~W.}\ \bibnamefont
  {Metzger}}, \bibinfo {author} {\bibfnamefont {K.~A.}\ \bibnamefont
  {Grishunin}}, \bibinfo {author} {\bibfnamefont {C.}~\bibnamefont
  {Reinhoffer}}, \bibinfo {author} {\bibfnamefont {R.~M.}\ \bibnamefont
  {Dubrovin}}, \bibinfo {author} {\bibfnamefont {A.}~\bibnamefont {Arshad}},
  \bibinfo {author} {\bibfnamefont {I.}~\bibnamefont {Ilyakov}}, \bibinfo
  {author} {\bibfnamefont {T.~V.}\ \bibnamefont {de~Oliveira}}, \bibinfo
  {author} {\bibfnamefont {A.}~\bibnamefont {Ponomaryov}}, \bibinfo {author}
  {\bibfnamefont {J.-C.}\ \bibnamefont {Deinert}}, \bibinfo {author}
  {\bibfnamefont {S.}~\bibnamefont {Kovalev}}, \bibinfo {author} {\bibfnamefont
  {R.~V.}\ \bibnamefont {Pisarev}}, \bibinfo {author} {\bibfnamefont {M.~I.}\
  \bibnamefont {Katsnelson}}, \bibinfo {author} {\bibfnamefont {B.~A.}\
  \bibnamefont {Ivanov}}, \bibinfo {author} {\bibfnamefont {P.~H.}\
  \bibnamefont {van Loosdrecht}}, \bibinfo {author} {\bibfnamefont {A.~V.}\
  \bibnamefont {Kimel}},\ and\ \bibinfo {author} {\bibfnamefont {E.~A.}\
  \bibnamefont {Mashkovich}},\ }\bibfield  {title} {\bibinfo {title} {Impulsive
  fermi magnon-phonon resonance in antiferromagnetic $\mathrm{CoF}_{2}$},\
  }\bibfield  {journal} {\bibinfo  {journal} {arXiv preprint arXiv:2308.01052}\
  }\href {https://doi.org/10.48550/arXiv.2308.01052}
  {10.48550/arXiv.2308.01052} (\bibinfo {year} {2023})\BibitemShut {NoStop}%
\bibitem [{\citenamefont {Ozhogin}(1964)}]{ozhogin1964antiferromagnets}%
  \BibitemOpen
  \bibfield  {author} {\bibinfo {author} {\bibfnamefont {V.}~\bibnamefont
  {Ozhogin}},\ }\bibfield  {title} {\bibinfo {title} {The antiferromagnets
  $\mathrm{CoCO}_{3}$, $\mathrm{CoF}_{2}$, and $\mathrm{FeCO}_{3}$ in strong
  fields},\ }\href {http://www.jetp.ras.ru/cgi-bin/dn/e_018_04_1156.pdf}
  {\bibfield  {journal} {\bibinfo  {journal} {Soviet Physics JETP-USSR}\
  }\textbf {\bibinfo {volume} {18}},\ \bibinfo {pages} {1156} (\bibinfo {year}
  {1964})}\BibitemShut {NoStop}%
\bibitem [{\citenamefont {Ozhogin}(1968)}]{ozhogin1968behavior}%
  \BibitemOpen
  \bibfield  {author} {\bibinfo {author} {\bibfnamefont {V.~I.}\ \bibnamefont
  {Ozhogin}},\ }\bibfield  {title} {\bibinfo {title} {On the {B}ehavior of
  $\mathrm{CoF}_{2}$ in a {H}igh {M}agnetic {F}ield {N}ormal to the $c$
  {A}xis},\ }\href {https://doi.org/10.1063/1.1656157} {\bibfield  {journal}
  {\bibinfo  {journal} {J. Appl. Phys.}\ }\textbf {\bibinfo {volume} {39}},\
  \bibinfo {pages} {1029} (\bibinfo {year} {1968})}\BibitemShut {NoStop}%
\bibitem [{\citenamefont {Gufan}\ \emph {et~al.}(1974)\citenamefont {Gufan},
  \citenamefont {Kocharyan}, \citenamefont {Prokhorov},\ and\ \citenamefont
  {Rudashevskii}}]{gufan1974dependence}%
  \BibitemOpen
  \bibfield  {author} {\bibinfo {author} {\bibfnamefont {Y.~M.}\ \bibnamefont
  {Gufan}}, \bibinfo {author} {\bibfnamefont {K.}~\bibnamefont {Kocharyan}},
  \bibinfo {author} {\bibfnamefont {A.}~\bibnamefont {Prokhorov}},\ and\
  \bibinfo {author} {\bibfnamefont {E.}~\bibnamefont {Rudashevskii}},\
  }\bibfield  {title} {\bibinfo {title} {Dependence of the resonant frequencies
  of antiferromagnets on the magnetic field, and antiferromagnetic resonance in
  $\mathrm{CoF}_{2}$},\ }\href
  {http://jetp.ras.ru/cgi-bin/dn/e_039_03_0565.pdf} {\bibfield  {journal}
  {\bibinfo  {journal} {JETP}\ }\textbf {\bibinfo {volume} {66}},\ \bibinfo
  {pages} {1155} (\bibinfo {year} {1974})}\BibitemShut {NoStop}%
\bibitem [{\citenamefont {Kharchenko}\ \emph {et~al.}(1982)\citenamefont
  {Kharchenko}, \citenamefont {Eremenko},\ and\ \citenamefont
  {Belyi}}]{kharchenko1982magnetooptical}%
  \BibitemOpen
  \bibfield  {author} {\bibinfo {author} {\bibfnamefont {N.~F.}\ \bibnamefont
  {Kharchenko}}, \bibinfo {author} {\bibfnamefont {V.~V.}\ \bibnamefont
  {Eremenko}},\ and\ \bibinfo {author} {\bibfnamefont {L.}~\bibnamefont
  {Belyi}},\ }\bibfield  {title} {\bibinfo {title} {Magnetooptical
  investigation of the noncollinear state induced in antiferromagnetic cobalt
  fluoride by a longitudinal magnetic field},\ }\href
  {http://www.jetp.ras.ru/cgi-bin/dn/e_055_03_0490.pdf} {\bibfield  {journal}
  {\bibinfo  {journal} {Sov. Phys. JETP}\ }\textbf {\bibinfo {volume} {55}},\
  \bibinfo {pages} {490} (\bibinfo {year} {1982})}\BibitemShut {NoStop}%
\bibitem [{\citenamefont {Gurtovoi}\ \emph {et~al.}(1982)\citenamefont
  {Gurtovoi}, \citenamefont {Lagutin},\ and\ \citenamefont
  {Ozhogin}}]{gurtovoi1982noncolinear}%
  \BibitemOpen
  \bibfield  {author} {\bibinfo {author} {\bibfnamefont {K.~G.}\ \bibnamefont
  {Gurtovoi}}, \bibinfo {author} {\bibfnamefont {A.~S.}\ \bibnamefont
  {Lagutin}},\ and\ \bibinfo {author} {\bibfnamefont {V.~I.}\ \bibnamefont
  {Ozhogin}},\ }\bibfield  {title} {\bibinfo {title} {Noncollinear magnetic
  phases in a strongly anisotropic antiferromagnet $\mathrm{CoF}_{2}$ with
  large dzyaloshinskii interaction},\ }\href
  {http://www.jetp.ras.ru/cgi-bin/dn/e_056_05_1122.pdf} {\bibfield  {journal}
  {\bibinfo  {journal} {Sov. Phys. JETP}\ }\textbf {\bibinfo {volume} {26}},\
  \bibinfo {pages} {1122} (\bibinfo {year} {1982})}\BibitemShut {NoStop}%
\bibitem [{\citenamefont {Macfarlane}(1970)}]{macflane1970raman}%
  \BibitemOpen
  \bibfield  {author} {\bibinfo {author} {\bibfnamefont {R.~M.}\ \bibnamefont
  {Macfarlane}},\ }\bibfield  {title} {\bibinfo {title} {{R}aman {L}ight
  {S}cattering from {E}xcitons and {M}agnons in {C}obalt {F}luoride},\ }\href
  {https://doi.org/10.1103/PhysRevLett.25.1454} {\bibfield  {journal} {\bibinfo
   {journal} {Phys. Rev. Lett.}\ }\textbf {\bibinfo {volume} {25}},\ \bibinfo
  {pages} {1454} (\bibinfo {year} {1970})}\BibitemShut {NoStop}%
\bibitem [{\citenamefont {Ishikawa}\ and\ \citenamefont
  {Moriya}(1971)}]{ishikawa1971magnons}%
  \BibitemOpen
  \bibfield  {author} {\bibinfo {author} {\bibfnamefont {A.}~\bibnamefont
  {Ishikawa}}\ and\ \bibinfo {author} {\bibfnamefont {T.}~\bibnamefont
  {Moriya}},\ }\bibfield  {title} {\bibinfo {title} {{M}agnons, {E}xcitons and
  {L}ight {S}cattering in {A}ntiferromagnetic $\mathrm{CoF}_{2}$},\ }\href
  {https://doi.org/10.1143/JPSJ.30.117} {\bibfield  {journal} {\bibinfo
  {journal} {J. Phys. Soc. Jpn.}\ }\textbf {\bibinfo {volume} {30}},\ \bibinfo
  {pages} {117} (\bibinfo {year} {1971})}\BibitemShut {NoStop}%
\bibitem [{\citenamefont {Meloche}\ \emph {et~al.}(2014)\citenamefont
  {Meloche}, \citenamefont {Cottam},\ and\ \citenamefont
  {Lockwood}}]{meloche2014one}%
  \BibitemOpen
  \bibfield  {author} {\bibinfo {author} {\bibfnamefont {E.}~\bibnamefont
  {Meloche}}, \bibinfo {author} {\bibfnamefont {M.~G.}\ \bibnamefont
  {Cottam}},\ and\ \bibinfo {author} {\bibfnamefont {D.~J.}\ \bibnamefont
  {Lockwood}},\ }\bibfield  {title} {\bibinfo {title} {One-magnon and exciton
  inelastic light scattering in the antiferromagnet $\mathrm{CoF}_{2}$},\
  }\href {https://doi.org/10.1063/1.4865560} {\bibfield  {journal} {\bibinfo
  {journal} {Low Temp. Phys.}\ }\textbf {\bibinfo {volume} {40}},\ \bibinfo
  {pages} {134} (\bibinfo {year} {2014})}\BibitemShut {NoStop}%
\bibitem [{\citenamefont {Cipriani}\ \emph {et~al.}(1971)\citenamefont
  {Cipriani}, \citenamefont {Racine},\ and\ \citenamefont
  {Dupeyrat}}]{cipriani1971raman}%
  \BibitemOpen
  \bibfield  {author} {\bibinfo {author} {\bibfnamefont {J.}~\bibnamefont
  {Cipriani}}, \bibinfo {author} {\bibfnamefont {S.}~\bibnamefont {Racine}},\
  and\ \bibinfo {author} {\bibfnamefont {R.}~\bibnamefont {Dupeyrat}},\
  }\bibfield  {title} {\bibinfo {title} {Raman scattering by two-magnon
  excitations in $\mathrm{CoF}_{2}$},\ }\href
  {https://doi.org/10.1016/0375-9601(71)90819-X} {\bibfield  {journal}
  {\bibinfo  {journal} {Phys. Lett. A}\ }\textbf {\bibinfo {volume} {34}},\
  \bibinfo {pages} {187} (\bibinfo {year} {1971})}\BibitemShut {NoStop}%
\bibitem [{\citenamefont {Natoli}\ and\ \citenamefont
  {Ranninger}(1973)}]{natoli1973two}%
  \BibitemOpen
  \bibfield  {author} {\bibinfo {author} {\bibfnamefont {C.~R.}\ \bibnamefont
  {Natoli}}\ and\ \bibinfo {author} {\bibfnamefont {J.}~\bibnamefont
  {Ranninger}},\ }\bibfield  {title} {\bibinfo {title} {Two-magnon neutron
  scattering in $\mathrm{CoF}_{2}$},\ }\href
  {https://doi.org/10.1088/0022-3719/6/2/015} {\bibfield  {journal} {\bibinfo
  {journal} {J. Phys. C: Solid State Phys.}\ }\textbf {\bibinfo {volume} {6}},\
  \bibinfo {pages} {370} (\bibinfo {year} {1973})}\BibitemShut {NoStop}%
\bibitem [{\citenamefont {Meloche}\ \emph {et~al.}(2007)\citenamefont
  {Meloche}, \citenamefont {Cottam},\ and\ \citenamefont
  {Lockwood}}]{meloche2007two}%
  \BibitemOpen
  \bibfield  {author} {\bibinfo {author} {\bibfnamefont {E.}~\bibnamefont
  {Meloche}}, \bibinfo {author} {\bibfnamefont {M.~G.}\ \bibnamefont
  {Cottam}},\ and\ \bibinfo {author} {\bibfnamefont {D.~J.}\ \bibnamefont
  {Lockwood}},\ }\bibfield  {title} {\bibinfo {title} {Two-magnon inelastic
  light scattering in the antiferromagnets $\mathrm{CoF}_{2}$ and
  $\mathrm{NiF}_{2}$: Experiment and theory},\ }\href
  {https://doi.org/10.1103/PhysRevB.76.104406} {\bibfield  {journal} {\bibinfo
  {journal} {Phys. Rev. B}\ }\textbf {\bibinfo {volume} {76}},\ \bibinfo
  {pages} {104406} (\bibinfo {year} {2007})}\BibitemShut {NoStop}%
\bibitem [{\citenamefont {Cowley}\ \emph {et~al.}(1973)\citenamefont {Cowley},
  \citenamefont {Buyers}, \citenamefont {Martel},\ and\ \citenamefont
  {Stevenson}}]{cowley1973magnetic}%
  \BibitemOpen
  \bibfield  {author} {\bibinfo {author} {\bibfnamefont {R.}~\bibnamefont
  {Cowley}}, \bibinfo {author} {\bibfnamefont {W.}~\bibnamefont {Buyers}},
  \bibinfo {author} {\bibfnamefont {P.}~\bibnamefont {Martel}},\ and\ \bibinfo
  {author} {\bibfnamefont {R.}~\bibnamefont {Stevenson}},\ }\bibfield  {title}
  {\bibinfo {title} {Magnetic excitations and magnetic critical scattering in
  cobalt fluoride},\ }\href {https://doi.org/10.1088/0022-3719/6/20/014}
  {\bibfield  {journal} {\bibinfo  {journal} {J. Phys. C: Solid State Phys.}\
  }\textbf {\bibinfo {volume} {6}},\ \bibinfo {pages} {2997} (\bibinfo {year}
  {1973})}\BibitemShut {NoStop}%
\bibitem [{\citenamefont {Macfarlane}\ and\ \citenamefont
  {Ushioda}(1970)}]{macfarlane1970light}%
  \BibitemOpen
  \bibfield  {author} {\bibinfo {author} {\bibfnamefont {R.~M.}\ \bibnamefont
  {Macfarlane}}\ and\ \bibinfo {author} {\bibfnamefont {S.}~\bibnamefont
  {Ushioda}},\ }\bibfield  {title} {\bibinfo {title} {Light scattering from
  phonons in $\mathrm{CoF}_{2}$},\ }\href
  {https://doi.org/10.1016/0038-1098(70)90264-4} {\bibfield  {journal}
  {\bibinfo  {journal} {Solid State Commun.}\ }\textbf {\bibinfo {volume}
  {8}},\ \bibinfo {pages} {1081} (\bibinfo {year} {1970})}\BibitemShut
  {NoStop}%
\bibitem [{\citenamefont {Barker~Jr}\ and\ \citenamefont
  {Ditzenberger}(1965)}]{barker1965infrared}%
  \BibitemOpen
  \bibfield  {author} {\bibinfo {author} {\bibfnamefont {A.~S.}\ \bibnamefont
  {Barker~Jr}}\ and\ \bibinfo {author} {\bibfnamefont {J.~A.}\ \bibnamefont
  {Ditzenberger}},\ }\bibfield  {title} {\bibinfo {title} {Infrared lattice
  vibrations in $\mathrm{CoF}_{2}$},\ }\href
  {https://doi.org/10.1016/0038-1098(65)90388-1} {\bibfield  {journal}
  {\bibinfo  {journal} {Solid State Commun.}\ }\textbf {\bibinfo {volume}
  {3}},\ \bibinfo {pages} {131} (\bibinfo {year} {1965})}\BibitemShut {NoStop}%
\bibitem [{\citenamefont {Balkanski}\ \emph {et~al.}(1966)\citenamefont
  {Balkanski}, \citenamefont {Moch},\ and\ \citenamefont
  {Parisot}}]{balkanski1966infrared}%
  \BibitemOpen
  \bibfield  {author} {\bibinfo {author} {\bibfnamefont {M.}~\bibnamefont
  {Balkanski}}, \bibinfo {author} {\bibfnamefont {P.}~\bibnamefont {Moch}},\
  and\ \bibinfo {author} {\bibfnamefont {G.}~\bibnamefont {Parisot}},\
  }\bibfield  {title} {\bibinfo {title} {Infrared lattice-vibration spectra in
  $\mathrm{NiF}_{2}$, $\mathrm{CoF}_{2}$, and $\mathrm{FeF}_{2}$},\ }\href
  {https://doi.org/10.1063/1.1726845} {\bibfield  {journal} {\bibinfo
  {journal} {J. Chem. Phys.}\ }\textbf {\bibinfo {volume} {44}},\ \bibinfo
  {pages} {940} (\bibinfo {year} {1966})}\BibitemShut {NoStop}%
\bibitem [{\citenamefont {H{\"a}ussler}(1981)}]{haussler1981infrared}%
  \BibitemOpen
  \bibfield  {author} {\bibinfo {author} {\bibfnamefont {K.~M.}\ \bibnamefont
  {H{\"a}ussler}},\ }\bibfield  {title} {\bibinfo {title} {Infrared spectra of
  magnetic excitations in $\mathrm{CoF}_{2}$},\ }\href
  {https://doi.org/10.1002/pssb.2221050245} {\bibfield  {journal} {\bibinfo
  {journal} {Phys. Status Solidi B}\ }\textbf {\bibinfo {volume} {105}},\
  \bibinfo {pages} {K81} (\bibinfo {year} {1981})}\BibitemShut {NoStop}%
\bibitem [{\citenamefont {Stout}\ and\ \citenamefont
  {Reed}(1954)}]{stout1954crystal}%
  \BibitemOpen
  \bibfield  {author} {\bibinfo {author} {\bibfnamefont {J.~W.}\ \bibnamefont
  {Stout}}\ and\ \bibinfo {author} {\bibfnamefont {S.~A.}\ \bibnamefont
  {Reed}},\ }\bibfield  {title} {\bibinfo {title} {The crystal structure of
  $\mathrm{MnF}_{2}$, $\mathrm{FeF}_{2}$, $\mathrm{CoF}_{2}$,
  $\mathrm{NiF}_{2}$ and $\mathrm{ZnF}_{2}$},\ }\href
  {https://doi.org/10.1021/ja01650a005} {\bibfield  {journal} {\bibinfo
  {journal} {J. Am. Chem. Soc.}\ }\textbf {\bibinfo {volume} {76}},\ \bibinfo
  {pages} {5279} (\bibinfo {year} {1954})}\BibitemShut {NoStop}%
\bibitem [{\citenamefont {Costa}\ \emph {et~al.}(1993)\citenamefont {Costa},
  \citenamefont {Paixao}, \citenamefont {De~Almeida},\ and\ \citenamefont
  {Andrade}}]{costa1993charge}%
  \BibitemOpen
  \bibfield  {author} {\bibinfo {author} {\bibfnamefont {M.~M.~R.}\
  \bibnamefont {Costa}}, \bibinfo {author} {\bibfnamefont {J.~A.}\ \bibnamefont
  {Paixao}}, \bibinfo {author} {\bibfnamefont {M.~J.~M.}\ \bibnamefont
  {De~Almeida}},\ and\ \bibinfo {author} {\bibfnamefont {L.~C.~R.}\
  \bibnamefont {Andrade}},\ }\bibfield  {title} {\bibinfo {title} {Charge
  densities of two rutile structures: $\mathrm{NiF}_{2}$ and
  $\mathrm{CoF}_{2}$},\ }\href {https://doi.org/10.1107/S0108768193001624}
  {\bibfield  {journal} {\bibinfo  {journal} {Acta Cryst.}\ }\textbf {\bibinfo
  {volume} {B49}},\ \bibinfo {pages} {591} (\bibinfo {year}
  {1993})}\BibitemShut {NoStop}%
\bibitem [{\citenamefont {Thomson}\ \emph {et~al.}(2014)\citenamefont
  {Thomson}, \citenamefont {Chatterji},\ and\ \citenamefont
  {Carpenter}}]{thomson2014cof2}%
  \BibitemOpen
  \bibfield  {author} {\bibinfo {author} {\bibfnamefont {R.~I.}\ \bibnamefont
  {Thomson}}, \bibinfo {author} {\bibfnamefont {T.}~\bibnamefont {Chatterji}},\
  and\ \bibinfo {author} {\bibfnamefont {M.~A.}\ \bibnamefont {Carpenter}},\
  }\bibfield  {title} {\bibinfo {title} {$\mathrm{CoF}_{2}$: a model system for
  magnetoelastic coupling and elastic softening mechanisms associated with
  paramagnetic $\leftrightarrow$ antiferromagnetic phase transitions},\ }\href
  {https://doi.org/10.1088/0953-8984/26/14/146001} {\bibfield  {journal}
  {\bibinfo  {journal} {J. Phys. Condens. Matter}\ }\textbf {\bibinfo {volume}
  {26}},\ \bibinfo {pages} {146001} (\bibinfo {year} {2014})}\BibitemShut
  {NoStop}%
\bibitem [{\citenamefont {Stout}\ and\ \citenamefont
  {Matarrese}(1953)}]{stout1953magnetic}%
  \BibitemOpen
  \bibfield  {author} {\bibinfo {author} {\bibfnamefont {J.~W.}\ \bibnamefont
  {Stout}}\ and\ \bibinfo {author} {\bibfnamefont {L.~M.}\ \bibnamefont
  {Matarrese}},\ }\bibfield  {title} {\bibinfo {title} {{M}agnetic {A}nisotropy
  of the {I}ron-{G}roup {F}luorides},\ }\href
  {https://doi.org/10.1103/RevModPhys.25.338} {\bibfield  {journal} {\bibinfo
  {journal} {Rev. Mod. Phys.}\ }\textbf {\bibinfo {volume} {25}},\ \bibinfo
  {pages} {338} (\bibinfo {year} {1953})}\BibitemShut {NoStop}%
\bibitem [{\citenamefont {Erickson}(1953)}]{erickson1953neutron}%
  \BibitemOpen
  \bibfield  {author} {\bibinfo {author} {\bibfnamefont {R.~A.}\ \bibnamefont
  {Erickson}},\ }\bibfield  {title} {\bibinfo {title} {{N}eutron {D}iffraction
  {S}tudies of {A}ntiferromagnetism in {M}anganous {F}luoride and {S}ome
  {I}somorphous {C}ompounds},\ }\href {https://doi.org/10.1103/PhysRev.90.779}
  {\bibfield  {journal} {\bibinfo  {journal} {Phys. Rev.}\ }\textbf {\bibinfo
  {volume} {90}},\ \bibinfo {pages} {779} (\bibinfo {year} {1953})}\BibitemShut
  {NoStop}%
\bibitem [{\citenamefont {Strempfer}\ \emph {et~al.}(2004)\citenamefont
  {Strempfer}, \citenamefont {R\"utt}, \citenamefont {Bayrakci}, \citenamefont
  {Br\"uckel},\ and\ \citenamefont {Jauch}}]{strempfer2004magnetic}%
  \BibitemOpen
  \bibfield  {author} {\bibinfo {author} {\bibfnamefont {J.}~\bibnamefont
  {Strempfer}}, \bibinfo {author} {\bibfnamefont {U.}~\bibnamefont {R\"utt}},
  \bibinfo {author} {\bibfnamefont {S.~P.}\ \bibnamefont {Bayrakci}}, \bibinfo
  {author} {\bibfnamefont {T.}~\bibnamefont {Br\"uckel}},\ and\ \bibinfo
  {author} {\bibfnamefont {W.}~\bibnamefont {Jauch}},\ }\bibfield  {title}
  {\bibinfo {title} {Magnetic properties of transition metal fluorides
  $\mathrm{MF}_{2}$ ($\mathrm{M}$ = $\mathrm{Mn}$, $\mathrm{Fe}$,
  $\mathrm{Co}$, $\mathrm{Ni}$) via high-energy photon diffraction},\ }\href
  {https://doi.org/10.1103/PhysRevB.69.014417} {\bibfield  {journal} {\bibinfo
  {journal} {Phys. Rev. B}\ }\textbf {\bibinfo {volume} {69}},\ \bibinfo
  {pages} {014417} (\bibinfo {year} {2004})}\BibitemShut {NoStop}%
\bibitem [{\citenamefont {Eremenko}\ \emph {et~al.}(1982)\citenamefont
  {Eremenko}, \citenamefont {Naumenko}, \citenamefont {Petrov},\ and\
  \citenamefont {Pishko}}]{eremenko1982rearrangement}%
  \BibitemOpen
  \bibfield  {author} {\bibinfo {author} {\bibfnamefont {V.~V.}\ \bibnamefont
  {Eremenko}}, \bibinfo {author} {\bibfnamefont {V.~M.}\ \bibnamefont
  {Naumenko}}, \bibinfo {author} {\bibfnamefont {S.~V.}\ \bibnamefont
  {Petrov}},\ and\ \bibinfo {author} {\bibfnamefont {V.~V.}\ \bibnamefont
  {Pishko}},\ }\bibfield  {title} {\bibinfo {title} {Rearrangement of the
  spectrum of magnetic excitations of antiferromagnetic $\mathrm{CoF}_{2}$ with
  low-concentration $\mathrm{MnF}_{2}$ impurity},\ }\href
  {http://jetp.ras.ru/cgi-bin/dn/e_055_03_0481.pdf} {\bibfield  {journal}
  {\bibinfo  {journal} {JETP}\ }\textbf {\bibinfo {volume} {55}},\ \bibinfo
  {pages} {481} (\bibinfo {year} {1982})}\BibitemShut {NoStop}%
\bibitem [{\citenamefont {Hohenberg}\ and\ \citenamefont
  {Kohn}(1964)}]{hohenberg1964inhomogeneous}%
  \BibitemOpen
  \bibfield  {author} {\bibinfo {author} {\bibfnamefont {P.}~\bibnamefont
  {Hohenberg}}\ and\ \bibinfo {author} {\bibfnamefont {W.}~\bibnamefont
  {Kohn}},\ }\bibfield  {title} {\bibinfo {title} {{I}nhomogeneous {E}lectron
  {G}as},\ }\href {https://doi.org/10.1103/PhysRev.136.B864} {\bibfield
  {journal} {\bibinfo  {journal} {Phys. Rev.}\ }\textbf {\bibinfo {volume}
  {136}},\ \bibinfo {pages} {B864} (\bibinfo {year} {1964})}\BibitemShut
  {NoStop}%
\bibitem [{\citenamefont {Kohn}\ and\ \citenamefont
  {Sham}(1965)}]{kohn1965self}%
  \BibitemOpen
  \bibfield  {author} {\bibinfo {author} {\bibfnamefont {W.}~\bibnamefont
  {Kohn}}\ and\ \bibinfo {author} {\bibfnamefont {L.~J.}\ \bibnamefont
  {Sham}},\ }\bibfield  {title} {\bibinfo {title} {{S}elf-{C}onsistent
  {E}quations {I}ncluding {E}xchange and {C}orrelation {E}ffects},\ }\href
  {https://doi.org/10.1103/PhysRev.140.A1133} {\bibfield  {journal} {\bibinfo
  {journal} {Phys. Rev.}\ }\textbf {\bibinfo {volume} {140}},\ \bibinfo {pages}
  {A1133} (\bibinfo {year} {1965})}\BibitemShut {NoStop}%
\bibitem [{\citenamefont {Bl\"ochl}(1994)}]{blochl1994projector}%
  \BibitemOpen
  \bibfield  {author} {\bibinfo {author} {\bibfnamefont {P.~E.}\ \bibnamefont
  {Bl\"ochl}},\ }\bibfield  {title} {\bibinfo {title} {Projector augmented-wave
  method},\ }\href {https://doi.org/10.1103/PhysRevB.50.17953} {\bibfield
  {journal} {\bibinfo  {journal} {Phys. Rev. B}\ }\textbf {\bibinfo {volume}
  {50}},\ \bibinfo {pages} {17953} (\bibinfo {year} {1994})}\BibitemShut
  {NoStop}%
\bibitem [{\citenamefont {Kresse}\ and\ \citenamefont
  {Furthm\"uller}(1996)}]{kresse1996efficient}%
  \BibitemOpen
  \bibfield  {author} {\bibinfo {author} {\bibfnamefont {G.}~\bibnamefont
  {Kresse}}\ and\ \bibinfo {author} {\bibfnamefont {J.}~\bibnamefont
  {Furthm\"uller}},\ }\bibfield  {title} {\bibinfo {title} {Efficient iterative
  schemes for \textit{ab initio} total-energy calculations using a plane-wave
  basis set},\ }\href {https://doi.org/10.1103/PhysRevB.54.11169} {\bibfield
  {journal} {\bibinfo  {journal} {Phys. Rev. B}\ }\textbf {\bibinfo {volume}
  {54}},\ \bibinfo {pages} {11169} (\bibinfo {year} {1996})}\BibitemShut
  {NoStop}%
\bibitem [{\citenamefont {Kresse}\ and\ \citenamefont
  {Joubert}(1999)}]{kresse1999from}%
  \BibitemOpen
  \bibfield  {author} {\bibinfo {author} {\bibfnamefont {G.}~\bibnamefont
  {Kresse}}\ and\ \bibinfo {author} {\bibfnamefont {D.}~\bibnamefont
  {Joubert}},\ }\bibfield  {title} {\bibinfo {title} {From ultrasoft
  pseudopotentials to the projector augmented-wave method},\ }\href
  {https://doi.org/10.1103/PhysRevB.59.1758} {\bibfield  {journal} {\bibinfo
  {journal} {Phys. Rev. B}\ }\textbf {\bibinfo {volume} {59}},\ \bibinfo
  {pages} {1758} (\bibinfo {year} {1999})}\BibitemShut {NoStop}%
\bibitem [{\citenamefont {Monkhorst}\ and\ \citenamefont
  {Pack}(1976)}]{monkhorst1976special}%
  \BibitemOpen
  \bibfield  {author} {\bibinfo {author} {\bibfnamefont {H.~J.}\ \bibnamefont
  {Monkhorst}}\ and\ \bibinfo {author} {\bibfnamefont {J.~D.}\ \bibnamefont
  {Pack}},\ }\bibfield  {title} {\bibinfo {title} {{S}pecial points for
  {B}rillouin-zone integrations},\ }\href
  {https://doi.org/10.1103/PhysRevB.13.5188} {\bibfield  {journal} {\bibinfo
  {journal} {Phys. Rev. B}\ }\textbf {\bibinfo {volume} {13}},\ \bibinfo
  {pages} {5188} (\bibinfo {year} {1976})}\BibitemShut {NoStop}%
\bibitem [{\citenamefont {Perdew}\ \emph {et~al.}(2008)\citenamefont {Perdew},
  \citenamefont {Ruzsinszky}, \citenamefont {Csonka}, \citenamefont {Vydrov},
  \citenamefont {Scuseria}, \citenamefont {Constantin}, \citenamefont {Zhou},\
  and\ \citenamefont {Burke}}]{perdew2008restoring}%
  \BibitemOpen
  \bibfield  {author} {\bibinfo {author} {\bibfnamefont {J.~P.}\ \bibnamefont
  {Perdew}}, \bibinfo {author} {\bibfnamefont {A.}~\bibnamefont {Ruzsinszky}},
  \bibinfo {author} {\bibfnamefont {G.~I.}\ \bibnamefont {Csonka}}, \bibinfo
  {author} {\bibfnamefont {O.~A.}\ \bibnamefont {Vydrov}}, \bibinfo {author}
  {\bibfnamefont {G.~E.}\ \bibnamefont {Scuseria}}, \bibinfo {author}
  {\bibfnamefont {L.~A.}\ \bibnamefont {Constantin}}, \bibinfo {author}
  {\bibfnamefont {X.}~\bibnamefont {Zhou}},\ and\ \bibinfo {author}
  {\bibfnamefont {K.}~\bibnamefont {Burke}},\ }\bibfield  {title} {\bibinfo
  {title} {{R}estoring the {D}ensity-{G}radient {E}xpansion for {E}xchange in
  {S}olids and {S}urfaces},\ }\href
  {https://doi.org/10.1103/PhysRevLett.100.136406} {\bibfield  {journal}
  {\bibinfo  {journal} {Phys. Rev. Lett.}\ }\textbf {\bibinfo {volume} {100}},\
  \bibinfo {pages} {136406} (\bibinfo {year} {2008})}\BibitemShut {NoStop}%
\bibitem [{\citenamefont {Dudarev}\ \emph {et~al.}(1998)\citenamefont
  {Dudarev}, \citenamefont {Botton}, \citenamefont {Savrasov}, \citenamefont
  {Humphreys},\ and\ \citenamefont {Sutton}}]{dudarev1998electron}%
  \BibitemOpen
  \bibfield  {author} {\bibinfo {author} {\bibfnamefont {S.~L.}\ \bibnamefont
  {Dudarev}}, \bibinfo {author} {\bibfnamefont {G.~A.}\ \bibnamefont {Botton}},
  \bibinfo {author} {\bibfnamefont {S.~Y.}\ \bibnamefont {Savrasov}}, \bibinfo
  {author} {\bibfnamefont {C.~J.}\ \bibnamefont {Humphreys}},\ and\ \bibinfo
  {author} {\bibfnamefont {A.~P.}\ \bibnamefont {Sutton}},\ }\bibfield  {title}
  {\bibinfo {title} {Electron-energy-loss spectra and the structural stability
  of nickel oxide: {A}n {LSDA+U} study},\ }\href
  {https://doi.org/10.1103/PhysRevB.57.1505} {\bibfield  {journal} {\bibinfo
  {journal} {Phys. Rev. B}\ }\textbf {\bibinfo {volume} {57}},\ \bibinfo
  {pages} {1505} (\bibinfo {year} {1998})}\BibitemShut {NoStop}%
\bibitem [{\citenamefont {Bousquet}\ and\ \citenamefont
  {Spaldin}(2010)}]{bousquet2010dependence}%
  \BibitemOpen
  \bibfield  {author} {\bibinfo {author} {\bibfnamefont {E.}~\bibnamefont
  {Bousquet}}\ and\ \bibinfo {author} {\bibfnamefont {N.}~\bibnamefont
  {Spaldin}},\ }\bibfield  {title} {\bibinfo {title} {${J}$ dependence in the
  $\text{LSDA}+{U}$ treatment of noncollinear magnets},\ }\href
  {https://doi.org/10.1103/PhysRevB.82.220402} {\bibfield  {journal} {\bibinfo
  {journal} {Phys. Rev. B}\ }\textbf {\bibinfo {volume} {82}},\ \bibinfo
  {pages} {220402} (\bibinfo {year} {2010})}\BibitemShut {NoStop}%
\bibitem [{\citenamefont {Gonze}\ and\ \citenamefont
  {Lee}(1997)}]{gonze1997dynamical}%
  \BibitemOpen
  \bibfield  {author} {\bibinfo {author} {\bibfnamefont {X.}~\bibnamefont
  {Gonze}}\ and\ \bibinfo {author} {\bibfnamefont {C.}~\bibnamefont {Lee}},\
  }\bibfield  {title} {\bibinfo {title} {{D}ynamical matrices, {B}orn effective
  charges, dielectric permittivity tensors, and interatomic force constants
  from density-functional perturbation theory},\ }\href
  {https://doi.org/10.1103/PhysRevB.55.10355} {\bibfield  {journal} {\bibinfo
  {journal} {Phys. Rev. B}\ }\textbf {\bibinfo {volume} {55}},\ \bibinfo
  {pages} {10355} (\bibinfo {year} {1997})}\BibitemShut {NoStop}%
\bibitem [{\citenamefont {Togo}\ and\ \citenamefont
  {Tanaka}(2015)}]{togo2015first}%
  \BibitemOpen
  \bibfield  {author} {\bibinfo {author} {\bibfnamefont {A.}~\bibnamefont
  {Togo}}\ and\ \bibinfo {author} {\bibfnamefont {I.}~\bibnamefont {Tanaka}},\
  }\bibfield  {title} {\bibinfo {title} {First principles phonon calculations
  in materials science},\ }\href
  {https://doi.org/10.1016/j.scriptamat.2015.07.021} {\bibfield  {journal}
  {\bibinfo  {journal} {Scr. Mater.}\ }\textbf {\bibinfo {volume} {108}},\
  \bibinfo {pages} {1} (\bibinfo {year} {2015})}\BibitemShut {NoStop}%
\bibitem [{\citenamefont {Wang}\ \emph {et~al.}(2010)\citenamefont {Wang},
  \citenamefont {Wang}, \citenamefont {Wang}, \citenamefont {Mei},
  \citenamefont {Shang}, \citenamefont {Chen},\ and\ \citenamefont
  {Liu}}]{wang2010mixed}%
  \BibitemOpen
  \bibfield  {author} {\bibinfo {author} {\bibfnamefont {Y.}~\bibnamefont
  {Wang}}, \bibinfo {author} {\bibfnamefont {J.~J.}\ \bibnamefont {Wang}},
  \bibinfo {author} {\bibfnamefont {W.~Y.}\ \bibnamefont {Wang}}, \bibinfo
  {author} {\bibfnamefont {Z.~G.}\ \bibnamefont {Mei}}, \bibinfo {author}
  {\bibfnamefont {S.~L.}\ \bibnamefont {Shang}}, \bibinfo {author}
  {\bibfnamefont {L.~Q.}\ \bibnamefont {Chen}},\ and\ \bibinfo {author}
  {\bibfnamefont {Z.~K.}\ \bibnamefont {Liu}},\ }\bibfield  {title} {\bibinfo
  {title} {A mixed-space approach to first-principles calculations of phonon
  frequencies for polar materials},\ }\href
  {https://doi.org/10.1088/0953-8984/22/20/202201} {\bibfield  {journal}
  {\bibinfo  {journal} {J. Phys. Condens. Matter}\ }\textbf {\bibinfo {volume}
  {22}},\ \bibinfo {pages} {202201} (\bibinfo {year} {2010})}\BibitemShut
  {NoStop}%
\bibitem [{\citenamefont {ladyteam}()}]{ladyteam}%
  \BibitemOpen
  \bibfield  {author} {\bibinfo {author} {\bibnamefont {ladyteam}},\ }\href
  {https://github.com/ladyteam/LADYtools} {\bibinfo {title} {{LA}ttice
  {DY}namic tools}}\BibitemShut {NoStop}%
\bibitem [{\citenamefont {Liechtenstein}\ \emph {et~al.}(1987)\citenamefont
  {Liechtenstein}, \citenamefont {Katsnelson}, \citenamefont {Antropov},\ and\
  \citenamefont {Gubanov}}]{liechtenstein1987local}%
  \BibitemOpen
  \bibfield  {author} {\bibinfo {author} {\bibfnamefont {A.~I.}\ \bibnamefont
  {Liechtenstein}}, \bibinfo {author} {\bibfnamefont {M.~I.}\ \bibnamefont
  {Katsnelson}}, \bibinfo {author} {\bibfnamefont {V.~P.}\ \bibnamefont
  {Antropov}},\ and\ \bibinfo {author} {\bibfnamefont {V.~A.}\ \bibnamefont
  {Gubanov}},\ }\bibfield  {title} {\bibinfo {title} {Local spin density
  functional approach to the theory of exchange interactions in ferromagnetic
  metals and alloys},\ }\href {https://doi.org/10.1016/0304-8853(87)90721-9}
  {\bibfield  {journal} {\bibinfo  {journal} {J. Magn. Magn. Mater.}\ }\textbf
  {\bibinfo {volume} {67}},\ \bibinfo {pages} {65} (\bibinfo {year}
  {1987})}\BibitemShut {NoStop}%
\bibitem [{\citenamefont {He}\ \emph {et~al.}(2021)\citenamefont {He},
  \citenamefont {Helbig}, \citenamefont {Verstraete},\ and\ \citenamefont
  {Bousquet}}]{he2021tb2j}%
  \BibitemOpen
  \bibfield  {author} {\bibinfo {author} {\bibfnamefont {X.}~\bibnamefont
  {He}}, \bibinfo {author} {\bibfnamefont {N.}~\bibnamefont {Helbig}}, \bibinfo
  {author} {\bibfnamefont {M.~J.}\ \bibnamefont {Verstraete}},\ and\ \bibinfo
  {author} {\bibfnamefont {E.}~\bibnamefont {Bousquet}},\ }\bibfield  {title}
  {\bibinfo {title} {{TB2J}: {A} python package for computing magnetic
  interaction parameters},\ }\href {https://doi.org/10.1016/j.cpc.2021.107938}
  {\bibfield  {journal} {\bibinfo  {journal} {Comput. Phys. Commun.}\ }\textbf
  {\bibinfo {volume} {264}},\ \bibinfo {pages} {107938} (\bibinfo {year}
  {2021})}\BibitemShut {NoStop}%
\bibitem [{\citenamefont {Tellez-Mora}\ \emph {et~al.}(2024)\citenamefont
  {Tellez-Mora}, \citenamefont {He}, \citenamefont {Bousquet}, \citenamefont
  {Wirtz},\ and\ \citenamefont {Romero}}]{tellez2024systematic}%
  \BibitemOpen
  \bibfield  {author} {\bibinfo {author} {\bibfnamefont {A.}~\bibnamefont
  {Tellez-Mora}}, \bibinfo {author} {\bibfnamefont {X.}~\bibnamefont {He}},
  \bibinfo {author} {\bibfnamefont {E.}~\bibnamefont {Bousquet}}, \bibinfo
  {author} {\bibfnamefont {L.}~\bibnamefont {Wirtz}},\ and\ \bibinfo {author}
  {\bibfnamefont {A.~H.}\ \bibnamefont {Romero}},\ }\bibfield  {title}
  {\bibinfo {title} {Systematic determination of a material’s magnetic ground
  state from first principles},\ }\href
  {https://doi.org/10.1038/s41524-024-01202-z} {\bibfield  {journal} {\bibinfo
  {journal} {npj Comput. Mater.}\ }\textbf {\bibinfo {volume} {10}},\ \bibinfo
  {pages} {20} (\bibinfo {year} {2024})}\BibitemShut {NoStop}%
\bibitem [{\citenamefont {Soler}\ \emph {et~al.}(2002)\citenamefont {Soler},
  \citenamefont {Artacho}, \citenamefont {Gale}, \citenamefont {Garc{\'\i}a},
  \citenamefont {Junquera}, \citenamefont {Ordej{\'o}n},\ and\ \citenamefont
  {S{\'a}nchez-Portal}}]{soler2002siesta}%
  \BibitemOpen
  \bibfield  {author} {\bibinfo {author} {\bibfnamefont {J.~M.}\ \bibnamefont
  {Soler}}, \bibinfo {author} {\bibfnamefont {E.}~\bibnamefont {Artacho}},
  \bibinfo {author} {\bibfnamefont {J.~D.}\ \bibnamefont {Gale}}, \bibinfo
  {author} {\bibfnamefont {A.}~\bibnamefont {Garc{\'\i}a}}, \bibinfo {author}
  {\bibfnamefont {J.}~\bibnamefont {Junquera}}, \bibinfo {author}
  {\bibfnamefont {P.}~\bibnamefont {Ordej{\'o}n}},\ and\ \bibinfo {author}
  {\bibfnamefont {D.}~\bibnamefont {S{\'a}nchez-Portal}},\ }\bibfield  {title}
  {\bibinfo {title} {The {SIESTA} method for \textit{ab initio} order-{N}
  materials simulation},\ }\href {https://doi.org/10.1088/0953-8984/14/11/302}
  {\bibfield  {journal} {\bibinfo  {journal} {J. Phys. Condens. Matter}\
  }\textbf {\bibinfo {volume} {14}},\ \bibinfo {pages} {2745} (\bibinfo {year}
  {2002})}\BibitemShut {NoStop}%
\bibitem [{\citenamefont {Toth}\ and\ \citenamefont
  {Lake}(2015)}]{toth2015linear}%
  \BibitemOpen
  \bibfield  {author} {\bibinfo {author} {\bibfnamefont {S.}~\bibnamefont
  {Toth}}\ and\ \bibinfo {author} {\bibfnamefont {B.}~\bibnamefont {Lake}},\
  }\bibfield  {title} {\bibinfo {title} {Linear spin wave theory for single-q
  incommensurate magnetic structures},\ }\href
  {https://doi.org/10.1088/0953-8984/27/16/166002} {\bibfield  {journal}
  {\bibinfo  {journal} {J. Phys.: Condens. Matter}\ }\textbf {\bibinfo {volume}
  {27}},\ \bibinfo {pages} {166002} (\bibinfo {year} {2015})}\BibitemShut
  {NoStop}%
\bibitem [{\citenamefont {Kroumova}\ \emph {et~al.}(2003)\citenamefont
  {Kroumova}, \citenamefont {Aroyo}, \citenamefont {Perez-Mato}, \citenamefont
  {Kirov}, \citenamefont {Capillas}, \citenamefont {Ivantchev},\ and\
  \citenamefont {Wondratschek}}]{kroumova2003bilbao}%
  \BibitemOpen
  \bibfield  {author} {\bibinfo {author} {\bibfnamefont {E.}~\bibnamefont
  {Kroumova}}, \bibinfo {author} {\bibfnamefont {M.~I.}\ \bibnamefont {Aroyo}},
  \bibinfo {author} {\bibfnamefont {J.~M.}\ \bibnamefont {Perez-Mato}},
  \bibinfo {author} {\bibfnamefont {A.}~\bibnamefont {Kirov}}, \bibinfo
  {author} {\bibfnamefont {C.}~\bibnamefont {Capillas}}, \bibinfo {author}
  {\bibfnamefont {S.}~\bibnamefont {Ivantchev}},\ and\ \bibinfo {author}
  {\bibfnamefont {H.}~\bibnamefont {Wondratschek}},\ }\bibfield  {title}
  {\bibinfo {title} {{B}ilbao {C}rystallographic {S}erver: {U}seful {D}atabases
  and {T}ools for {P}hase-{T}ransition {S}tudies},\ }\href
  {https://doi.org/10.1080/0141159031000076110} {\bibfield  {journal} {\bibinfo
   {journal} {Phase Transit.}\ }\textbf {\bibinfo {volume} {76}},\ \bibinfo
  {pages} {155} (\bibinfo {year} {2003})}\BibitemShut {NoStop}%
\bibitem [{\citenamefont {Born}\ and\ \citenamefont
  {Wolf}(2013)}]{born2013principles}%
  \BibitemOpen
  \bibfield  {author} {\bibinfo {author} {\bibfnamefont {M.}~\bibnamefont
  {Born}}\ and\ \bibinfo {author} {\bibfnamefont {E.}~\bibnamefont {Wolf}},\
  }\href@noop {} {\emph {\bibinfo {title} {Principles of Optics:
  Electromagnetic Theory of Propagation, Interference and Diffraction of
  Light}}}\ (\bibinfo  {publisher} {Elsevier},\ \bibinfo {year}
  {2013})\BibitemShut {NoStop}%
\bibitem [{\citenamefont {Gervais}\ and\ \citenamefont
  {Piriou}(1974)}]{gervais1974anharmonicity}%
  \BibitemOpen
  \bibfield  {author} {\bibinfo {author} {\bibfnamefont {F.}~\bibnamefont
  {Gervais}}\ and\ \bibinfo {author} {\bibfnamefont {B.}~\bibnamefont
  {Piriou}},\ }\bibfield  {title} {\bibinfo {title} {Anharmonicity in
  several-polar-mode crystals: adjusting phonon self-energy of {LO} and {TO}
  modes in $\mathrm{Al}_{2}\mathrm{O}_{3}$ and $\mathrm{TiO}_{2}$ to fit
  infrared reflectivity},\ }\href {https://doi.org/10.1088/0022-3719/7/13/017}
  {\bibfield  {journal} {\bibinfo  {journal} {J. Phys. C: Solid State Phys.}\
  }\textbf {\bibinfo {volume} {7}},\ \bibinfo {pages} {2374} (\bibinfo {year}
  {1974})}\BibitemShut {NoStop}%
\bibitem [{\citenamefont {Benoit}\ and\ \citenamefont
  {Giordano}(1988)}]{benoit1988dynamical}%
  \BibitemOpen
  \bibfield  {author} {\bibinfo {author} {\bibfnamefont {C.}~\bibnamefont
  {Benoit}}\ and\ \bibinfo {author} {\bibfnamefont {J.}~\bibnamefont
  {Giordano}},\ }\bibfield  {title} {\bibinfo {title} {Dynamical properties of
  crystals of $\mathrm{MgF}_{2}$, $\mathrm{ZnF}_{2}$ and $\mathrm{FeF}_{2}$.
  {II}. {L}attice dynamics and infrared spectra},\ }\href
  {https://doi.org/10.1088/0022-3719/21/29/016} {\bibfield  {journal} {\bibinfo
   {journal} {Journal of Physics C: Solid State Physics}\ }\textbf {\bibinfo
  {volume} {21}},\ \bibinfo {pages} {5209} (\bibinfo {year}
  {1988})}\BibitemShut {NoStop}%
\bibitem [{\citenamefont {Barker}(1964)}]{berker1964transverse}%
  \BibitemOpen
  \bibfield  {author} {\bibinfo {author} {\bibfnamefont {A.~S.}\ \bibnamefont
  {Barker}},\ }\bibfield  {title} {\bibinfo {title} {{T}ransverse and
  {L}ongitudinal {O}ptic {M}ode {S}tudy in $\mathrm{MgF}_{2}$ and
  $\mathrm{ZnF}_{2}$},\ }\href {https://doi.org/10.1103/PhysRev.136.A1290}
  {\bibfield  {journal} {\bibinfo  {journal} {Phys. Rev.}\ }\textbf {\bibinfo
  {volume} {136}},\ \bibinfo {pages} {A1290} (\bibinfo {year}
  {1964})}\BibitemShut {NoStop}%
\bibitem [{\citenamefont {Hemberger}\ \emph {et~al.}(2006)\citenamefont
  {Hemberger}, \citenamefont {Rudolf}, \citenamefont {Krug~von Nidda},
  \citenamefont {Mayr}, \citenamefont {Pimenov}, \citenamefont {Tsurkan},\ and\
  \citenamefont {Loidl}}]{hemberger2006spin}%
  \BibitemOpen
  \bibfield  {author} {\bibinfo {author} {\bibfnamefont {J.}~\bibnamefont
  {Hemberger}}, \bibinfo {author} {\bibfnamefont {T.}~\bibnamefont {Rudolf}},
  \bibinfo {author} {\bibfnamefont {H.-A.}\ \bibnamefont {Krug~von Nidda}},
  \bibinfo {author} {\bibfnamefont {F.}~\bibnamefont {Mayr}}, \bibinfo {author}
  {\bibfnamefont {A.}~\bibnamefont {Pimenov}}, \bibinfo {author} {\bibfnamefont
  {V.}~\bibnamefont {Tsurkan}},\ and\ \bibinfo {author} {\bibfnamefont
  {A.}~\bibnamefont {Loidl}},\ }\bibfield  {title} {\bibinfo {title}
  {{S}pin-{D}riven {P}honon {S}plitting in {B}ond-{F}rustrated
  $\mathrm{ZnCr}_{2}\mathrm{S}_4$},\ }\href
  {https://doi.org/10.1103/PhysRevLett.97.087204} {\bibfield  {journal}
  {\bibinfo  {journal} {Phys. Rev. Lett.}\ }\textbf {\bibinfo {volume} {97}},\
  \bibinfo {pages} {087204} (\bibinfo {year} {2006})}\BibitemShut {NoStop}%
\bibitem [{\citenamefont {Rudolf}\ \emph {et~al.}(2008)\citenamefont {Rudolf},
  \citenamefont {Kant}, \citenamefont {Mayr},\ and\ \citenamefont
  {Loidl}}]{rudolf2008magnetic}%
  \BibitemOpen
  \bibfield  {author} {\bibinfo {author} {\bibfnamefont {T.}~\bibnamefont
  {Rudolf}}, \bibinfo {author} {\bibfnamefont {C.}~\bibnamefont {Kant}},
  \bibinfo {author} {\bibfnamefont {F.}~\bibnamefont {Mayr}},\ and\ \bibinfo
  {author} {\bibfnamefont {A.}~\bibnamefont {Loidl}},\ }\bibfield  {title}
  {\bibinfo {title} {Magnetic-order induced phonon splitting in $\mathrm{MnO}$
  from far-infrared spectroscopy},\ }\href
  {https://doi.org/10.1103/PhysRevB.77.024421} {\bibfield  {journal} {\bibinfo
  {journal} {Phys. Rev. B}\ }\textbf {\bibinfo {volume} {77}},\ \bibinfo
  {pages} {024421} (\bibinfo {year} {2008})}\BibitemShut {NoStop}%
\bibitem [{\citenamefont {Kant}\ \emph {et~al.}(2009)\citenamefont {Kant},
  \citenamefont {Deisenhofer}, \citenamefont {Rudolf}, \citenamefont {Mayr},
  \citenamefont {Schrettle}, \citenamefont {Loidl}, \citenamefont {Gnezdilov},
  \citenamefont {Wulferding}, \citenamefont {Lemmens},\ and\ \citenamefont
  {Tsurkan}}]{kant2009optical}%
  \BibitemOpen
  \bibfield  {author} {\bibinfo {author} {\bibfnamefont {C.}~\bibnamefont
  {Kant}}, \bibinfo {author} {\bibfnamefont {J.}~\bibnamefont {Deisenhofer}},
  \bibinfo {author} {\bibfnamefont {T.}~\bibnamefont {Rudolf}}, \bibinfo
  {author} {\bibfnamefont {F.}~\bibnamefont {Mayr}}, \bibinfo {author}
  {\bibfnamefont {F.}~\bibnamefont {Schrettle}}, \bibinfo {author}
  {\bibfnamefont {A.}~\bibnamefont {Loidl}}, \bibinfo {author} {\bibfnamefont
  {V.}~\bibnamefont {Gnezdilov}}, \bibinfo {author} {\bibfnamefont
  {D.}~\bibnamefont {Wulferding}}, \bibinfo {author} {\bibfnamefont
  {P.}~\bibnamefont {Lemmens}},\ and\ \bibinfo {author} {\bibfnamefont
  {V.}~\bibnamefont {Tsurkan}},\ }\bibfield  {title} {\bibinfo {title} {Optical
  phonons, spin correlations, and spin-phonon coupling in the frustrated
  pyrochlore magnets $\mathrm{CdCr}_{2}\mathrm{O}_4$ and
  $\mathrm{ZnCr}_{2}\mathrm{O}_4$},\ }\href
  {https://doi.org/10.1103/PhysRevB.80.214417} {\bibfield  {journal} {\bibinfo
  {journal} {Phys. Rev. B}\ }\textbf {\bibinfo {volume} {80}},\ \bibinfo
  {pages} {214417} (\bibinfo {year} {2009})}\BibitemShut {NoStop}%
\bibitem [{\citenamefont {Kant}\ \emph {et~al.}(2012)\citenamefont {Kant},
  \citenamefont {Schmidt}, \citenamefont {Wang}, \citenamefont {Mayr},
  \citenamefont {Tsurkan}, \citenamefont {Deisenhofer},\ and\ \citenamefont
  {Loidl}}]{kant2012universal}%
  \BibitemOpen
  \bibfield  {author} {\bibinfo {author} {\bibfnamefont {C.}~\bibnamefont
  {Kant}}, \bibinfo {author} {\bibfnamefont {M.}~\bibnamefont {Schmidt}},
  \bibinfo {author} {\bibfnamefont {Z.}~\bibnamefont {Wang}}, \bibinfo {author}
  {\bibfnamefont {F.}~\bibnamefont {Mayr}}, \bibinfo {author} {\bibfnamefont
  {V.}~\bibnamefont {Tsurkan}}, \bibinfo {author} {\bibfnamefont
  {J.}~\bibnamefont {Deisenhofer}},\ and\ \bibinfo {author} {\bibfnamefont
  {A.}~\bibnamefont {Loidl}},\ }\bibfield  {title} {\bibinfo {title}
  {{U}niversal {E}xchange-{D}riven {P}honon {S}plitting in
  {A}ntiferromagnets},\ }\href {https://doi.org/10.1103/PhysRevLett.108.177203}
  {\bibfield  {journal} {\bibinfo  {journal} {Phys. Rev. Lett.}\ }\textbf
  {\bibinfo {volume} {108}},\ \bibinfo {pages} {177203} (\bibinfo {year}
  {2012})}\BibitemShut {NoStop}%
\bibitem [{\citenamefont {Wei}\ \emph {et~al.}(2021)\citenamefont {Wei},
  \citenamefont {Sun}, \citenamefont {Li},\ and\ \citenamefont
  {Hong}}]{wei2021phonon}%
  \BibitemOpen
  \bibfield  {author} {\bibinfo {author} {\bibfnamefont {B.}~\bibnamefont
  {Wei}}, \bibinfo {author} {\bibfnamefont {Q.}~\bibnamefont {Sun}}, \bibinfo
  {author} {\bibfnamefont {C.}~\bibnamefont {Li}},\ and\ \bibinfo {author}
  {\bibfnamefont {J.}~\bibnamefont {Hong}},\ }\bibfield  {title} {\bibinfo
  {title} {Phonon anharmonicity: a pertinent review of recent progress and
  perspective},\ }\href {https://doi.org/10.1007/s11433-021-1748-7} {\bibfield
  {journal} {\bibinfo  {journal} {Sci. China Phys. Mech. Astron.}\ }\textbf
  {\bibinfo {volume} {64}},\ \bibinfo {pages} {117001} (\bibinfo {year}
  {2021})}\BibitemShut {NoStop}%
\bibitem [{\citenamefont {Born}\ and\ \citenamefont
  {Huang}(1954)}]{born1954dynamical}%
  \BibitemOpen
  \bibfield  {author} {\bibinfo {author} {\bibfnamefont {M.}~\bibnamefont
  {Born}}\ and\ \bibinfo {author} {\bibfnamefont {K.}~\bibnamefont {Huang}},\
  }\href@noop {} {\emph {\bibinfo {title} {{D}ynamical {T}heory of {C}rystal
  {L}attices}}}\ (\bibinfo  {publisher} {Clarendon press, Oxford},\ \bibinfo
  {year} {1954})\BibitemShut {NoStop}%
\bibitem [{\citenamefont {Balkanski}\ \emph {et~al.}(1983)\citenamefont
  {Balkanski}, \citenamefont {Wallis},\ and\ \citenamefont
  {Haro}}]{balkanski1983anharmonic}%
  \BibitemOpen
  \bibfield  {author} {\bibinfo {author} {\bibfnamefont {M.}~\bibnamefont
  {Balkanski}}, \bibinfo {author} {\bibfnamefont {R.~F.}\ \bibnamefont
  {Wallis}},\ and\ \bibinfo {author} {\bibfnamefont {E.}~\bibnamefont {Haro}},\
  }\bibfield  {title} {\bibinfo {title} {Anharmonic effects in light scattering
  due to optical phonons in silicon},\ }\href
  {https://doi.org/10.1103/PhysRevB.28.1928} {\bibfield  {journal} {\bibinfo
  {journal} {Phys. Rev. B}\ }\textbf {\bibinfo {volume} {28}},\ \bibinfo
  {pages} {1928} (\bibinfo {year} {1983})}\BibitemShut {NoStop}%
\bibitem [{\citenamefont {Lan}\ \emph {et~al.}(2012)\citenamefont {Lan},
  \citenamefont {Tang},\ and\ \citenamefont {Fultz}}]{lan2012phonon}%
  \BibitemOpen
  \bibfield  {author} {\bibinfo {author} {\bibfnamefont {T.}~\bibnamefont
  {Lan}}, \bibinfo {author} {\bibfnamefont {X.}~\bibnamefont {Tang}},\ and\
  \bibinfo {author} {\bibfnamefont {B.}~\bibnamefont {Fultz}},\ }\bibfield
  {title} {\bibinfo {title} {Phonon anharmonicity of rutile $\mathrm{TiO}_{2}$
  studied by {Raman} spectrometry and molecular dynamics simulations},\ }\href
  {https://doi.org/10.1103/PhysRevB.85.094305} {\bibfield  {journal} {\bibinfo
  {journal} {Phys. Rev. B}\ }\textbf {\bibinfo {volume} {85}},\ \bibinfo
  {pages} {094305} (\bibinfo {year} {2012})}\BibitemShut {NoStop}%
\bibitem [{\citenamefont {Cottam}\ and\ \citenamefont
  {Lockwood}(2019)}]{cottam2019spin}%
  \BibitemOpen
  \bibfield  {author} {\bibinfo {author} {\bibfnamefont {M.~G.}\ \bibnamefont
  {Cottam}}\ and\ \bibinfo {author} {\bibfnamefont {D.~J.}\ \bibnamefont
  {Lockwood}},\ }\bibfield  {title} {\bibinfo {title} {Spin-phonon interaction
  in transition-metal difluoride antiferromagnets: {T}heory and experiment},\
  }\href {https://doi.org/10.1063/1.5082316} {\bibfield  {journal} {\bibinfo
  {journal} {Low Temp. Phys.}\ }\textbf {\bibinfo {volume} {45}},\ \bibinfo
  {pages} {78} (\bibinfo {year} {2019})}\BibitemShut {NoStop}%
\bibitem [{\citenamefont {Darby}(1967)}]{darby1967tables}%
  \BibitemOpen
  \bibfield  {author} {\bibinfo {author} {\bibfnamefont {M.~I.}\ \bibnamefont
  {Darby}},\ }\bibfield  {title} {\bibinfo {title} {Tables of the {B}rillouin
  function and of the related function for the spontaneous magnetization},\
  }\href {https://doi.org/10.1088/0508-3443/18/10/307} {\bibfield  {journal}
  {\bibinfo  {journal} {Br. J. Appl. Phys.}\ }\textbf {\bibinfo {volume}
  {18}},\ \bibinfo {pages} {1415} (\bibinfo {year} {1967})}\BibitemShut
  {NoStop}%
\bibitem [{\citenamefont {Moriya}(1966)}]{moriya1966far}%
  \BibitemOpen
  \bibfield  {author} {\bibinfo {author} {\bibfnamefont {T.}~\bibnamefont
  {Moriya}},\ }\bibfield  {title} {\bibinfo {title} {{F}ar {I}nfrared
  {A}bsorption by {T}wo {M}agnon {E}xcitations in {A}niferromagnets},\ }\href
  {https://doi.org/10.1143/JPSJ.21.926} {\bibfield  {journal} {\bibinfo
  {journal} {J. Phys. Soc. Jpn.}\ }\textbf {\bibinfo {volume} {21}},\ \bibinfo
  {pages} {926} (\bibinfo {year} {1966})}\BibitemShut {NoStop}%
\bibitem [{\citenamefont {Martel}\ \emph {et~al.}(1968)\citenamefont {Martel},
  \citenamefont {Cowley},\ and\ \citenamefont
  {Stevenson}}]{martel1968experimental}%
  \BibitemOpen
  \bibfield  {author} {\bibinfo {author} {\bibfnamefont {P.}~\bibnamefont
  {Martel}}, \bibinfo {author} {\bibfnamefont {R.~A.}\ \bibnamefont {Cowley}},\
  and\ \bibinfo {author} {\bibfnamefont {R.~W.~H.}\ \bibnamefont {Stevenson}},\
  }\bibfield  {title} {\bibinfo {title} {Experimental studies of the magnetic
  and phonon excitations in cobalt fluoride},\ }\href
  {https://doi.org/10.1139/p68-456} {\bibfield  {journal} {\bibinfo  {journal}
  {Can. J. Phys.}\ }\textbf {\bibinfo {volume} {46}},\ \bibinfo {pages} {1355}
  (\bibinfo {year} {1968})}\BibitemShut {NoStop}%
\bibitem [{\citenamefont {Allen}\ and\ \citenamefont
  {Guggenheim}(1971{\natexlab{a}})}]{allen1971magnetic1}%
  \BibitemOpen
  \bibfield  {author} {\bibinfo {author} {\bibfnamefont {S.~J.}\ \bibnamefont
  {Allen}}\ and\ \bibinfo {author} {\bibfnamefont {H.~J.}\ \bibnamefont
  {Guggenheim}},\ }\bibfield  {title} {\bibinfo {title} {{M}agnetic
  {E}xcitations in {A}ntiferromagnetic $\mathrm{CoF}_{2}$. {I}.
  {S}pin-{O}ptical-{P}honon {I}nteraction},\ }\href
  {https://doi.org/10.1103/PhysRevB.4.937} {\bibfield  {journal} {\bibinfo
  {journal} {Phys. Rev. B}\ }\textbf {\bibinfo {volume} {4}},\ \bibinfo {pages}
  {937} (\bibinfo {year} {1971}{\natexlab{a}})}\BibitemShut {NoStop}%
\bibitem [{\citenamefont {Allen}\ and\ \citenamefont
  {Guggenheim}(1971{\natexlab{b}})}]{allen1971magnetic2}%
  \BibitemOpen
  \bibfield  {author} {\bibinfo {author} {\bibfnamefont {S.~J.}\ \bibnamefont
  {Allen}}\ and\ \bibinfo {author} {\bibfnamefont {H.~J.}\ \bibnamefont
  {Guggenheim}},\ }\bibfield  {title} {\bibinfo {title} {{M}agnetic
  {E}xcitations in {A}ntiferromagnetic $\mathrm{CoF}_{2}$. {II}. {U}niform
  {M}agnetic {E}xcitations near $t=0^\circ$\,k},\ }\href
  {https://doi.org/10.1103/PhysRevB.4.950} {\bibfield  {journal} {\bibinfo
  {journal} {Phys. Rev. B}\ }\textbf {\bibinfo {volume} {4}},\ \bibinfo {pages}
  {950} (\bibinfo {year} {1971}{\natexlab{b}})}\BibitemShut {NoStop}%
\bibitem [{\citenamefont {Newman}\ and\ \citenamefont
  {Chrenko}(1959)}]{newman1959infrared}%
  \BibitemOpen
  \bibfield  {author} {\bibinfo {author} {\bibfnamefont {R.}~\bibnamefont
  {Newman}}\ and\ \bibinfo {author} {\bibfnamefont {R.~M.}\ \bibnamefont
  {Chrenko}},\ }\bibfield  {title} {\bibinfo {title} {{I}nfrared {A}bsorption
  from ${L} \cdot {S}$ {S}plittings in $\mathrm{Co}^{2+}$ {S}alts},\ }\href
  {https://doi.org/10.1103/PhysRev.115.1147} {\bibfield  {journal} {\bibinfo
  {journal} {Phys. Rev.}\ }\textbf {\bibinfo {volume} {115}},\ \bibinfo {pages}
  {1147} (\bibinfo {year} {1959})}\BibitemShut {NoStop}%
\bibitem [{\citenamefont {Lines}(1965)}]{lines1965magnetic}%
  \BibitemOpen
  \bibfield  {author} {\bibinfo {author} {\bibfnamefont {M.~E.}\ \bibnamefont
  {Lines}},\ }\bibfield  {title} {\bibinfo {title} {Magnetic properties of
  $\mathrm{CoF}_{2}$},\ }\href {https://doi.org/10.1103/PhysRev.137.A982}
  {\bibfield  {journal} {\bibinfo  {journal} {Phys. Rev.}\ }\textbf {\bibinfo
  {volume} {137}},\ \bibinfo {pages} {A982} (\bibinfo {year}
  {1965})}\BibitemShut {NoStop}%
\bibitem [{\citenamefont {H{\"a}ussler}\ \emph {et~al.}(1982)\citenamefont
  {H{\"a}ussler}, \citenamefont {Lehmeyer},\ and\ \citenamefont
  {Merten}}]{haussler1982far}%
  \BibitemOpen
  \bibfield  {author} {\bibinfo {author} {\bibfnamefont {K.}~\bibnamefont
  {H{\"a}ussler}}, \bibinfo {author} {\bibfnamefont {A.}~\bibnamefont
  {Lehmeyer}},\ and\ \bibinfo {author} {\bibfnamefont {L.}~\bibnamefont
  {Merten}},\ }\bibfield  {title} {\bibinfo {title} {{F}ar-{I}nfrared
  {R}eflectivity of {O}ptical {M}agnons in $\mathrm{FeF}_{2}$ and
  $\mathrm{CoF}_{2}$},\ }\href {https://doi.org/10.1002/pssb.2221110213}
  {\bibfield  {journal} {\bibinfo  {journal} {Phys. Status Solidi B}\ }\textbf
  {\bibinfo {volume} {111}},\ \bibinfo {pages} {513} (\bibinfo {year}
  {1982})}\BibitemShut {NoStop}%
\bibitem [{\citenamefont {Johnson}\ and\ \citenamefont
  {Nethercot}(1959)}]{johnson1959antiferromagnetic}%
  \BibitemOpen
  \bibfield  {author} {\bibinfo {author} {\bibfnamefont {F.~M.}\ \bibnamefont
  {Johnson}}\ and\ \bibinfo {author} {\bibfnamefont {A.~H.}\ \bibnamefont
  {Nethercot}},\ }\bibfield  {title} {\bibinfo {title} {{A}ntiferromagnetic
  {R}esonance in $\mathrm{MnF}_{2}$},\ }\href
  {https://doi.org/10.1103/PhysRev.114.705} {\bibfield  {journal} {\bibinfo
  {journal} {Phys. Rev.}\ }\textbf {\bibinfo {volume} {114}},\ \bibinfo {pages}
  {705} (\bibinfo {year} {1959})}\BibitemShut {NoStop}%
\bibitem [{\citenamefont {Duarte}\ \emph {et~al.}(1987)\citenamefont {Duarte},
  \citenamefont {Sanjurjo},\ and\ \citenamefont
  {Katiyar}}]{duarte1987offnormal}%
  \BibitemOpen
  \bibfield  {author} {\bibinfo {author} {\bibfnamefont {J.~L.}\ \bibnamefont
  {Duarte}}, \bibinfo {author} {\bibfnamefont {J.~A.}\ \bibnamefont
  {Sanjurjo}},\ and\ \bibinfo {author} {\bibfnamefont {R.~S.}\ \bibnamefont
  {Katiyar}},\ }\bibfield  {title} {\bibinfo {title} {Off-normal infrared
  reflectivity in uniaxial crystals: $\alpha$-$\mathrm{LiIO}_{3}$ and
  $\alpha$-quartz},\ }\href {https://doi.org/10.1103/PhysRevB.36.3368}
  {\bibfield  {journal} {\bibinfo  {journal} {Phys. Rev. B}\ }\textbf {\bibinfo
  {volume} {36}},\ \bibinfo {pages} {3368} (\bibinfo {year}
  {1987})}\BibitemShut {NoStop}%
\bibitem [{\citenamefont {Allen}\ and\ \citenamefont
  {Guggenheim}(1968)}]{allen1968spin}%
  \BibitemOpen
  \bibfield  {author} {\bibinfo {author} {\bibfnamefont {S.~J.}\ \bibnamefont
  {Allen}}\ and\ \bibinfo {author} {\bibfnamefont {H.~J.}\ \bibnamefont
  {Guggenheim}},\ }\bibfield  {title} {\bibinfo {title}
  {{S}pin-{O}ptical-{P}honon {I}nteraction in {A}ntiferromagnetic
  $\mathrm{CoF}_{2}$},\ }\href {https://doi.org/10.1103/PhysRevLett.21.1807}
  {\bibfield  {journal} {\bibinfo  {journal} {Phys. Rev. Lett.}\ }\textbf
  {\bibinfo {volume} {21}},\ \bibinfo {pages} {1807} (\bibinfo {year}
  {1968})}\BibitemShut {NoStop}%
\bibitem [{\citenamefont {Allen}\ and\ \citenamefont
  {Guggenheim}(1969)}]{allen1969magnon}%
  \BibitemOpen
  \bibfield  {author} {\bibinfo {author} {\bibfnamefont {S.~J.}\ \bibnamefont
  {Allen}}\ and\ \bibinfo {author} {\bibfnamefont {H.~J.}\ \bibnamefont
  {Guggenheim}},\ }\bibfield  {title} {\bibinfo {title} {{M}agnon-{O}ptic
  {P}honon {I}nteraction in {A}ntiferromagnetic $\mathrm{CoF}_{2}$},\ }\href
  {https://doi.org/10.1063/1.1657818} {\bibfield  {journal} {\bibinfo
  {journal} {J. Appl. Phys.}\ }\textbf {\bibinfo {volume} {40}},\ \bibinfo
  {pages} {999} (\bibinfo {year} {1969})}\BibitemShut {NoStop}%
\bibitem [{\citenamefont {Mills}\ and\ \citenamefont
  {Ushioda}(1970)}]{mills1970exciton}%
  \BibitemOpen
  \bibfield  {author} {\bibinfo {author} {\bibfnamefont {D.~L.}\ \bibnamefont
  {Mills}}\ and\ \bibinfo {author} {\bibfnamefont {S.}~\bibnamefont
  {Ushioda}},\ }\bibfield  {title} {\bibinfo {title}
  {{E}xciton---{O}ptical-{P}honon {C}oupling in $\mathrm{CoF}_{2}$},\ }\href
  {https://doi.org/10.1103/PhysRevB.2.3805} {\bibfield  {journal} {\bibinfo
  {journal} {Phys. Rev. B}\ }\textbf {\bibinfo {volume} {2}},\ \bibinfo {pages}
  {3805} (\bibinfo {year} {1970})}\BibitemShut {NoStop}%
\bibitem [{\citenamefont {Allen~{Jr}}\ and\ \citenamefont
  {Guggenheim}(1971)}]{allen1971spin}%
  \BibitemOpen
  \bibfield  {author} {\bibinfo {author} {\bibfnamefont {S.~J.}\ \bibnamefont
  {Allen~{Jr}}}\ and\ \bibinfo {author} {\bibfnamefont {H.~J.}\ \bibnamefont
  {Guggenheim}},\ }\bibfield  {title} {\bibinfo {title} {Spin waves and
  excitons in antiferromagnetic $\mathrm{CoF}_{2}$},\ }\href
  {https://doi.org/10.1063/1.1660387} {\bibfield  {journal} {\bibinfo
  {journal} {J. Appl. Phys.}\ }\textbf {\bibinfo {volume} {42}},\ \bibinfo
  {pages} {1657} (\bibinfo {year} {1971})}\BibitemShut {NoStop}%
\bibitem [{\citenamefont {Gervais}\ and\ \citenamefont
  {Arend}(1983)}]{gervais1983long}%
  \BibitemOpen
  \bibfield  {author} {\bibinfo {author} {\bibfnamefont {F.}~\bibnamefont
  {Gervais}}\ and\ \bibinfo {author} {\bibfnamefont {H.}~\bibnamefont
  {Arend}},\ }\bibfield  {title} {\bibinfo {title} {Long-wavelength phonons in
  the four phases of
  $\{${$\mathrm{N}(\mathrm{CH}_{3})_{4}$}$\}_{2}\mathrm{CuCl}_{4}$ and
  effective charges},\ }\href {https://doi.org/10.1007/BF01307221} {\bibfield
  {journal} {\bibinfo  {journal} {Z. Physik B - Condensed Matter}\ }\textbf
  {\bibinfo {volume} {50}},\ \bibinfo {pages} {17} (\bibinfo {year}
  {1983})}\BibitemShut {NoStop}%
\bibitem [{\citenamefont {Vassiliou}(1986)}]{vassiliou1986pressure}%
  \BibitemOpen
  \bibfield  {author} {\bibinfo {author} {\bibfnamefont {J.~K.}\ \bibnamefont
  {Vassiliou}},\ }\bibfield  {title} {\bibinfo {title} {Pressure and
  temperature dependence of the static dielectric constants and elastic
  anomalies of $\mathrm{ZnF}_{2}$},\ }\href {https://doi.org/10.1063/1.336549}
  {\bibfield  {journal} {\bibinfo  {journal} {J. Appl. Phys.}\ }\textbf
  {\bibinfo {volume} {59}},\ \bibinfo {pages} {1125} (\bibinfo {year}
  {1986})}\BibitemShut {NoStop}%
\bibitem [{\citenamefont {Lowndes}\ and\ \citenamefont
  {Martin}(1970)}]{lowndes1970dielectric}%
  \BibitemOpen
  \bibfield  {author} {\bibinfo {author} {\bibfnamefont {R.}~\bibnamefont
  {Lowndes}}\ and\ \bibinfo {author} {\bibfnamefont {D.}~\bibnamefont
  {Martin}},\ }\bibfield  {title} {\bibinfo {title} {Dielectric constants of
  ionic crystals and their variations with temperature and pressure},\
  }\href@noop {} {\bibfield  {journal} {\bibinfo  {journal} {Proceedings of the
  Royal Society of London. A. Mathematical and Physical Sciences}\ }\textbf
  {\bibinfo {volume} {316}},\ \bibinfo {pages} {351} (\bibinfo {year}
  {1970})}\BibitemShut {NoStop}%
\bibitem [{\citenamefont {Bartels}\ and\ \citenamefont
  {Smith}(1973)}]{bartels1973pressure}%
  \BibitemOpen
  \bibfield  {author} {\bibinfo {author} {\bibfnamefont {R.~A.}\ \bibnamefont
  {Bartels}}\ and\ \bibinfo {author} {\bibfnamefont {P.~A.}\ \bibnamefont
  {Smith}},\ }\bibfield  {title} {\bibinfo {title} {{P}ressure and
  {T}emperature {D}ependence of the {S}tatic {D}ielectric {C}onstants of
  $\mathrm{KCl}$, $\mathrm{NaCl}$, $\mathrm{LiF}$, and $\mathrm{MgO}$},\ }\href
  {https://doi.org/10.1103/PhysRevB.7.3885} {\bibfield  {journal} {\bibinfo
  {journal} {Phys. Rev. B}\ }\textbf {\bibinfo {volume} {7}},\ \bibinfo {pages}
  {3885} (\bibinfo {year} {1973})}\BibitemShut {NoStop}%
\bibitem [{\citenamefont {Wintersgill}\ \emph {et~al.}(1979)\citenamefont
  {Wintersgill}, \citenamefont {Fontanella}, \citenamefont {Andeen},\ and\
  \citenamefont {Schuele}}]{wintersgill1979temperature}%
  \BibitemOpen
  \bibfield  {author} {\bibinfo {author} {\bibfnamefont {M.}~\bibnamefont
  {Wintersgill}}, \bibinfo {author} {\bibfnamefont {J.}~\bibnamefont
  {Fontanella}}, \bibinfo {author} {\bibfnamefont {C.}~\bibnamefont {Andeen}},\
  and\ \bibinfo {author} {\bibfnamefont {D.}~\bibnamefont {Schuele}},\
  }\bibfield  {title} {\bibinfo {title} {The temperature variation of the
  dielectric constant of "pure" $\mathrm{CaF}_{2}$ $\mathrm{CaF}_{2}$,
  $\mathrm{SrF}_{2}$, $\mathrm{BaF}_{2}$, and $\mathrm{MgO}$},\ }\href
  {https://doi.org/10.1063/1.325932} {\bibfield  {journal} {\bibinfo  {journal}
  {J. Appl. Phys.}\ }\textbf {\bibinfo {volume} {50}},\ \bibinfo {pages} {8259}
  (\bibinfo {year} {1979})}\BibitemShut {NoStop}%
\bibitem [{\citenamefont {Fox}\ \emph {et~al.}(1980)\citenamefont {Fox},
  \citenamefont {Tilley}, \citenamefont {Scott},\ and\ \citenamefont
  {Guggenheim}}]{fox1980magnetoelectric}%
  \BibitemOpen
  \bibfield  {author} {\bibinfo {author} {\bibfnamefont {D.~L.}\ \bibnamefont
  {Fox}}, \bibinfo {author} {\bibfnamefont {D.~R.}\ \bibnamefont {Tilley}},
  \bibinfo {author} {\bibfnamefont {J.~F.}\ \bibnamefont {Scott}},\ and\
  \bibinfo {author} {\bibfnamefont {H.~J.}\ \bibnamefont {Guggenheim}},\
  }\bibfield  {title} {\bibinfo {title} {Magnetoelectric phenomena in
  $\mathrm{BaMnF}_{4}$ and
  $\mathrm{BaMn}_{0.99}\mathrm{Co}_{0.01}\mathrm{F}_{4}$},\ }\href
  {https://doi.org/10.1103/PhysRevB.21.2926} {\bibfield  {journal} {\bibinfo
  {journal} {Phys. Rev. B}\ }\textbf {\bibinfo {volume} {21}},\ \bibinfo
  {pages} {2926} (\bibinfo {year} {1980})}\BibitemShut {NoStop}%
\bibitem [{\citenamefont {Seehra}\ and\ \citenamefont
  {Helmick}(1981)}]{seehra1981dielectric}%
  \BibitemOpen
  \bibfield  {author} {\bibinfo {author} {\bibfnamefont {M.~S.}\ \bibnamefont
  {Seehra}}\ and\ \bibinfo {author} {\bibfnamefont {R.~E.}\ \bibnamefont
  {Helmick}},\ }\bibfield  {title} {\bibinfo {title} {Dielectric anomaly in
  $\mathrm{MnO}$ near the magnetic phase transition},\ }\href
  {https://doi.org/10.1103/PhysRevB.24.5098} {\bibfield  {journal} {\bibinfo
  {journal} {Phys. Rev. B}\ }\textbf {\bibinfo {volume} {24}},\ \bibinfo
  {pages} {5098} (\bibinfo {year} {1981})}\BibitemShut {NoStop}%
\bibitem [{\citenamefont {Jahn}(1973)}]{jahn1973linear}%
  \BibitemOpen
  \bibfield  {author} {\bibinfo {author} {\bibfnamefont {I.~R.}\ \bibnamefont
  {Jahn}},\ }\bibfield  {title} {\bibinfo {title} {{L}inear {M}agnetic
  {B}irefringence in the {A}ntiferromagnetic {I}ron {G}roup {D}ifluorides},\
  }\href {https://doi.org/10.1002/pssb.2220570225} {\bibfield  {journal}
  {\bibinfo  {journal} {Phys. Status Solidi B}\ }\textbf {\bibinfo {volume}
  {57}},\ \bibinfo {pages} {681} (\bibinfo {year} {1973})}\BibitemShut
  {NoStop}%
\bibitem [{\citenamefont {Markovin}\ and\ \citenamefont
  {Pisarev}(1979)}]{markovin1979magnetic}%
  \BibitemOpen
  \bibfield  {author} {\bibinfo {author} {\bibfnamefont {P.~A.}\ \bibnamefont
  {Markovin}}\ and\ \bibinfo {author} {\bibfnamefont {R.~V.}\ \bibnamefont
  {Pisarev}},\ }\bibfield  {title} {\bibinfo {title} {Magnetic, thermal, and
  elastic refraction of light in the antiferromagnet $\mathrm{MnF}_{2}$},\
  }\href {http://jetp.ac.ru/cgi-bin/dn/e_050_06_1190.pdf} {\bibfield  {journal}
  {\bibinfo  {journal} {JETP}\ }\textbf {\bibinfo {volume} {50}},\ \bibinfo
  {pages} {1190} (\bibinfo {year} {1979})}\BibitemShut {NoStop}%
\bibitem [{\citenamefont {Dubrovin}\ \emph {et~al.}(2021)\citenamefont
  {Dubrovin}, \citenamefont {Garcia-Castro}, \citenamefont {Siverin},
  \citenamefont {Novikova}, \citenamefont {Boldyrev}, \citenamefont {Romero},\
  and\ \citenamefont {Pisarev}}]{dubrovin2021incipient}%
  \BibitemOpen
  \bibfield  {author} {\bibinfo {author} {\bibfnamefont {R.~M.}\ \bibnamefont
  {Dubrovin}}, \bibinfo {author} {\bibfnamefont {A.~C.}\ \bibnamefont
  {Garcia-Castro}}, \bibinfo {author} {\bibfnamefont {N.~V.}\ \bibnamefont
  {Siverin}}, \bibinfo {author} {\bibfnamefont {N.~N.}\ \bibnamefont
  {Novikova}}, \bibinfo {author} {\bibfnamefont {K.~N.}\ \bibnamefont
  {Boldyrev}}, \bibinfo {author} {\bibfnamefont {A.~H.}\ \bibnamefont
  {Romero}},\ and\ \bibinfo {author} {\bibfnamefont {R.~V.}\ \bibnamefont
  {Pisarev}},\ }\bibfield  {title} {\bibinfo {title} {Incipient geometric
  lattice instability of cubic fluoroperovskites},\ }\href
  {https://doi.org/10.1103/PhysRevB.104.144304} {\bibfield  {journal} {\bibinfo
   {journal} {Phys. Rev. B}\ }\textbf {\bibinfo {volume} {104}},\ \bibinfo
  {pages} {144304} (\bibinfo {year} {2021})}\BibitemShut {NoStop}%
\bibitem [{\citenamefont {Kumar}\ \emph {et~al.}(2012)\citenamefont {Kumar},
  \citenamefont {Fennie},\ and\ \citenamefont {Rabe}}]{kumar2012spin}%
  \BibitemOpen
  \bibfield  {author} {\bibinfo {author} {\bibfnamefont {A.}~\bibnamefont
  {Kumar}}, \bibinfo {author} {\bibfnamefont {C.~J.}\ \bibnamefont {Fennie}},\
  and\ \bibinfo {author} {\bibfnamefont {K.~M.}\ \bibnamefont {Rabe}},\
  }\bibfield  {title} {\bibinfo {title} {Spin-lattice coupling and phonon
  dispersion of $\mathrm{CdCr}_{2}\mathrm{O}_{4}$ from first principles},\
  }\href {https://doi.org/10.1103/PhysRevB.86.184429} {\bibfield  {journal}
  {\bibinfo  {journal} {Phys. Rev. B}\ }\textbf {\bibinfo {volume} {86}},\
  \bibinfo {pages} {184429} (\bibinfo {year} {2012})}\BibitemShut {NoStop}%
\bibitem [{\citenamefont {Momma}\ and\ \citenamefont
  {Izumi}(2011)}]{momma2011vesta}%
  \BibitemOpen
  \bibfield  {author} {\bibinfo {author} {\bibfnamefont {K.}~\bibnamefont
  {Momma}}\ and\ \bibinfo {author} {\bibfnamefont {F.}~\bibnamefont {Izumi}},\
  }\bibfield  {title} {\bibinfo {title} {{\it VESTA 3} for three-dimensional
  visualization of crystal, volumetric and morphology data},\ }\href
  {https://doi.org/10.1107/S0021889811038970} {\bibfield  {journal} {\bibinfo
  {journal} {J. Appl. Crystallogr.}\ }\textbf {\bibinfo {volume} {44}},\
  \bibinfo {pages} {1272} (\bibinfo {year} {2011})}\BibitemShut {NoStop}%
\bibitem [{\citenamefont {Barreda-Arg\"ueso}\ \emph {et~al.}(2013)\citenamefont
  {Barreda-Arg\"ueso}, \citenamefont {L\'opez-Moreno}, \citenamefont
  {Sanz-Ortiz}, \citenamefont {Aguado}, \citenamefont {Valiente}, \citenamefont
  {Gonz\'alez}, \citenamefont {Rodr\'{\i}guez}, \citenamefont {Romero},
  \citenamefont {Mu\~noz}, \citenamefont {Nataf},\ and\ \citenamefont
  {Baudelet}}]{barreda2013pressure}%
  \BibitemOpen
  \bibfield  {author} {\bibinfo {author} {\bibfnamefont {J.~A.}\ \bibnamefont
  {Barreda-Arg\"ueso}}, \bibinfo {author} {\bibfnamefont {S.}~\bibnamefont
  {L\'opez-Moreno}}, \bibinfo {author} {\bibfnamefont {M.~N.}\ \bibnamefont
  {Sanz-Ortiz}}, \bibinfo {author} {\bibfnamefont {F.}~\bibnamefont {Aguado}},
  \bibinfo {author} {\bibfnamefont {R.}~\bibnamefont {Valiente}}, \bibinfo
  {author} {\bibfnamefont {J.}~\bibnamefont {Gonz\'alez}}, \bibinfo {author}
  {\bibfnamefont {F.}~\bibnamefont {Rodr\'{\i}guez}}, \bibinfo {author}
  {\bibfnamefont {A.~H.}\ \bibnamefont {Romero}}, \bibinfo {author}
  {\bibfnamefont {A.}~\bibnamefont {Mu\~noz}}, \bibinfo {author} {\bibfnamefont
  {L.}~\bibnamefont {Nataf}},\ and\ \bibinfo {author} {\bibfnamefont
  {F.}~\bibnamefont {Baudelet}},\ }\bibfield  {title} {\bibinfo {title}
  {Pressure-induced phase-transition sequence in $\mathrm{CoF}_{2}$: An
  experimental and first-principles study on the crystal, vibrational, and
  electronic properties},\ }\href {https://doi.org/10.1103/PhysRevB.88.214108}
  {\bibfield  {journal} {\bibinfo  {journal} {Phys. Rev. B}\ }\textbf {\bibinfo
  {volume} {88}},\ \bibinfo {pages} {214108} (\bibinfo {year}
  {2013})}\BibitemShut {NoStop}%
\bibitem [{\citenamefont {Lee}\ \emph {et~al.}(1994)\citenamefont {Lee},
  \citenamefont {Ghosez},\ and\ \citenamefont {Gonze}}]{lee1994lattice}%
  \BibitemOpen
  \bibfield  {author} {\bibinfo {author} {\bibfnamefont {C.}~\bibnamefont
  {Lee}}, \bibinfo {author} {\bibfnamefont {P.}~\bibnamefont {Ghosez}},\ and\
  \bibinfo {author} {\bibfnamefont {X.}~\bibnamefont {Gonze}},\ }\bibfield
  {title} {\bibinfo {title} {Lattice dynamics and dielectric properties of
  incipient ferroelectric $\mathrm{TiO}_{2}$ rutile},\ }\href
  {https://doi.org/10.1103/PhysRevB.50.13379} {\bibfield  {journal} {\bibinfo
  {journal} {Phys. Rev. B}\ }\textbf {\bibinfo {volume} {50}},\ \bibinfo
  {pages} {13379} (\bibinfo {year} {1994})}\BibitemShut {NoStop}%
\bibitem [{\citenamefont {Raeliarijaona}\ and\ \citenamefont
  {Fu}(2015)}]{raeliarijaona2015mode}%
  \BibitemOpen
  \bibfield  {author} {\bibinfo {author} {\bibfnamefont {A.}~\bibnamefont
  {Raeliarijaona}}\ and\ \bibinfo {author} {\bibfnamefont {H.}~\bibnamefont
  {Fu}},\ }\bibfield  {title} {\bibinfo {title} {Mode sequence, frequency
  change of nonsoft phonons, and {LO}-{TO} splitting in strained tetragonal
  $\mathrm{BaTiO}_{3}$},\ }\href {https://doi.org/10.1103/PhysRevB.92.094303}
  {\bibfield  {journal} {\bibinfo  {journal} {Phys. Rev. B}\ }\textbf {\bibinfo
  {volume} {92}},\ \bibinfo {pages} {094303} (\bibinfo {year}
  {2015})}\BibitemShut {NoStop}%
\bibitem [{\citenamefont {Fredrickson}\ \emph {et~al.}(2016)\citenamefont
  {Fredrickson}, \citenamefont {Lin}, \citenamefont {Zollner},\ and\
  \citenamefont {Demkov}}]{fredrickson2016theoretical}%
  \BibitemOpen
  \bibfield  {author} {\bibinfo {author} {\bibfnamefont {K.~D.}\ \bibnamefont
  {Fredrickson}}, \bibinfo {author} {\bibfnamefont {C.}~\bibnamefont {Lin}},
  \bibinfo {author} {\bibfnamefont {S.}~\bibnamefont {Zollner}},\ and\ \bibinfo
  {author} {\bibfnamefont {A.~A.}\ \bibnamefont {Demkov}},\ }\bibfield  {title}
  {\bibinfo {title} {Theoretical study of negative optical mode splitting in
  $\mathrm{LaAlO}_{3}$},\ }\href {https://doi.org/10.1103/PhysRevB.93.134301}
  {\bibfield  {journal} {\bibinfo  {journal} {Phys. Rev. B}\ }\textbf {\bibinfo
  {volume} {93}},\ \bibinfo {pages} {134301} (\bibinfo {year}
  {2016})}\BibitemShut {NoStop}%
\bibitem [{\citenamefont {Schubert}\ \emph {et~al.}(2019)\citenamefont
  {Schubert}, \citenamefont {Mock}, \citenamefont {Korlacki},\ and\
  \citenamefont {Darakchieva}}]{schubert2019phonon}%
  \BibitemOpen
  \bibfield  {author} {\bibinfo {author} {\bibfnamefont {M.}~\bibnamefont
  {Schubert}}, \bibinfo {author} {\bibfnamefont {A.}~\bibnamefont {Mock}},
  \bibinfo {author} {\bibfnamefont {R.}~\bibnamefont {Korlacki}},\ and\
  \bibinfo {author} {\bibfnamefont {V.}~\bibnamefont {Darakchieva}},\
  }\bibfield  {title} {\bibinfo {title} {Phonon order and reststrahlen bands of
  polar vibrations in crystals with monoclinic symmetry},\ }\href
  {https://doi.org/10.1103/PhysRevB.99.041201} {\bibfield  {journal} {\bibinfo
  {journal} {Phys. Rev. B}\ }\textbf {\bibinfo {volume} {99}},\ \bibinfo
  {pages} {041201} (\bibinfo {year} {2019})}\BibitemShut {NoStop}%
\bibitem [{\citenamefont {Granado}\ \emph {et~al.}(1999)\citenamefont
  {Granado}, \citenamefont {Garc\'{\i}a}, \citenamefont {Sanjurjo},
  \citenamefont {Rettori}, \citenamefont {Torriani}, \citenamefont {Prado},
  \citenamefont {S\'anchez}, \citenamefont {Caneiro},\ and\ \citenamefont
  {Oseroff}}]{granado1999magnetic}%
  \BibitemOpen
  \bibfield  {author} {\bibinfo {author} {\bibfnamefont {E.}~\bibnamefont
  {Granado}}, \bibinfo {author} {\bibfnamefont {A.}~\bibnamefont
  {Garc\'{\i}a}}, \bibinfo {author} {\bibfnamefont {J.~A.}\ \bibnamefont
  {Sanjurjo}}, \bibinfo {author} {\bibfnamefont {C.}~\bibnamefont {Rettori}},
  \bibinfo {author} {\bibfnamefont {I.}~\bibnamefont {Torriani}}, \bibinfo
  {author} {\bibfnamefont {F.}~\bibnamefont {Prado}}, \bibinfo {author}
  {\bibfnamefont {R.~D.}\ \bibnamefont {S\'anchez}}, \bibinfo {author}
  {\bibfnamefont {A.}~\bibnamefont {Caneiro}},\ and\ \bibinfo {author}
  {\bibfnamefont {S.~B.}\ \bibnamefont {Oseroff}},\ }\bibfield  {title}
  {\bibinfo {title} {Magnetic ordering effects in the raman spectra of
  $\mathrm{La}_{1-x}\mathrm{Mn}_{1-x}\mathrm{O}_{3}$},\ }\href
  {https://doi.org/10.1103/PhysRevB.60.11879} {\bibfield  {journal} {\bibinfo
  {journal} {Phys. Rev. B}\ }\textbf {\bibinfo {volume} {60}},\ \bibinfo
  {pages} {11879} (\bibinfo {year} {1999})}\BibitemShut {NoStop}%
\bibitem [{\citenamefont {Zhong}\ \emph {et~al.}(1994)\citenamefont {Zhong},
  \citenamefont {King-Smith},\ and\ \citenamefont
  {Vanderbilt}}]{zhong1994giant}%
  \BibitemOpen
  \bibfield  {author} {\bibinfo {author} {\bibfnamefont {W.}~\bibnamefont
  {Zhong}}, \bibinfo {author} {\bibfnamefont {R.~D.}\ \bibnamefont
  {King-Smith}},\ and\ \bibinfo {author} {\bibfnamefont {D.}~\bibnamefont
  {Vanderbilt}},\ }\bibfield  {title} {\bibinfo {title} {{G}iant {LO}-{TO}
  {S}plittings in {P}erovskite {F}erroelectrics},\ }\href
  {https://doi.org/10.1103/PhysRevLett.72.3618} {\bibfield  {journal} {\bibinfo
   {journal} {Phys. Rev. Lett.}\ }\textbf {\bibinfo {volume} {72}},\ \bibinfo
  {pages} {3618} (\bibinfo {year} {1994})}\BibitemShut {NoStop}%
\bibitem [{\citenamefont {Khedidji}\ \emph {et~al.}(2021)\citenamefont
  {Khedidji}, \citenamefont {Amoroso},\ and\ \citenamefont
  {Djani}}]{khedidji2021microscopic}%
  \BibitemOpen
  \bibfield  {author} {\bibinfo {author} {\bibfnamefont {M.}~\bibnamefont
  {Khedidji}}, \bibinfo {author} {\bibfnamefont {D.}~\bibnamefont {Amoroso}},\
  and\ \bibinfo {author} {\bibfnamefont {H.}~\bibnamefont {Djani}},\ }\bibfield
   {title} {\bibinfo {title} {Microscopic mechanisms behind
  hyperferroelectricity},\ }\href {https://doi.org/10.1103/PhysRevB.103.014116}
  {\bibfield  {journal} {\bibinfo  {journal} {Phys. Rev. B}\ }\textbf {\bibinfo
  {volume} {103}},\ \bibinfo {pages} {014116} (\bibinfo {year}
  {2021})}\BibitemShut {NoStop}%
\bibitem [{\citenamefont {Cowley}\ \emph {et~al.}(1975)\citenamefont {Cowley},
  \citenamefont {Dietrich},\ and\ \citenamefont {Jones}}]{cowley1975magnetic}%
  \BibitemOpen
  \bibfield  {author} {\bibinfo {author} {\bibfnamefont {R.~A.}\ \bibnamefont
  {Cowley}}, \bibinfo {author} {\bibfnamefont {O.~W.}\ \bibnamefont
  {Dietrich}},\ and\ \bibinfo {author} {\bibfnamefont {D.~A.}\ \bibnamefont
  {Jones}},\ }\bibfield  {title} {\bibinfo {title} {Magnetic excitations of
  mixed $\mathrm{CoF}_{2}$/$\mathrm{ZnF}_{2}$ crystals},\ }\href
  {https://doi.org/10.1088/0022-3719/8/18/022} {\bibfield  {journal} {\bibinfo
  {journal} {J. Phys. C: Solid State Phys.}\ }\textbf {\bibinfo {volume} {8}},\
  \bibinfo {pages} {3023} (\bibinfo {year} {1975})}\BibitemShut {NoStop}%
\bibitem [{\citenamefont {L{\'o}pez-Moreno}\ \emph {et~al.}(2016)\citenamefont
  {L{\'o}pez-Moreno}, \citenamefont {Romero}, \citenamefont
  {Mej{\'\i}a-L{\'o}pez},\ and\ \citenamefont {Mu{\~n}oz}}]{lopez2016first}%
  \BibitemOpen
  \bibfield  {author} {\bibinfo {author} {\bibfnamefont {S.}~\bibnamefont
  {L{\'o}pez-Moreno}}, \bibinfo {author} {\bibfnamefont {A.}~\bibnamefont
  {Romero}}, \bibinfo {author} {\bibfnamefont {J.}~\bibnamefont
  {Mej{\'\i}a-L{\'o}pez}},\ and\ \bibinfo {author} {\bibfnamefont
  {A.}~\bibnamefont {Mu{\~n}oz}},\ }\bibfield  {title} {\bibinfo {title}
  {First-principles study of pressure-induced structural phase transitions in
  $\mathrm{MnF}_{2}$},\ }\href {https://doi.org/10.1039/C6CP05467F} {\bibfield
  {journal} {\bibinfo  {journal} {Phys. Chem. Chem. Phys.}\ }\textbf {\bibinfo
  {volume} {18}},\ \bibinfo {pages} {33250} (\bibinfo {year}
  {2016})}\BibitemShut {NoStop}%
\bibitem [{\citenamefont {Fechner}\ and\ \citenamefont
  {Spaldin}(2016)}]{fechner2016effects}%
  \BibitemOpen
  \bibfield  {author} {\bibinfo {author} {\bibfnamefont {M.}~\bibnamefont
  {Fechner}}\ and\ \bibinfo {author} {\bibfnamefont {N.~A.}\ \bibnamefont
  {Spaldin}},\ }\bibfield  {title} {\bibinfo {title} {Effects of intense
  optical phonon pumping on the structure and electronic properties of yttrium
  barium copper oxide},\ }\href {https://doi.org/10.1103/PhysRevB.94.134307}
  {\bibfield  {journal} {\bibinfo  {journal} {Phys. Rev. B}\ }\textbf {\bibinfo
  {volume} {94}},\ \bibinfo {pages} {134307} (\bibinfo {year}
  {2016})}\BibitemShut {NoStop}%
\bibitem [{\citenamefont {Fechner}\ \emph {et~al.}(2018)\citenamefont
  {Fechner}, \citenamefont {Sukhov}, \citenamefont {Chotorlishvili},
  \citenamefont {Kenel}, \citenamefont {Berakdar},\ and\ \citenamefont
  {Spaldin}}]{fechner2018magnetophotonics}%
  \BibitemOpen
  \bibfield  {author} {\bibinfo {author} {\bibfnamefont {M.}~\bibnamefont
  {Fechner}}, \bibinfo {author} {\bibfnamefont {A.}~\bibnamefont {Sukhov}},
  \bibinfo {author} {\bibfnamefont {L.}~\bibnamefont {Chotorlishvili}},
  \bibinfo {author} {\bibfnamefont {C.}~\bibnamefont {Kenel}}, \bibinfo
  {author} {\bibfnamefont {J.}~\bibnamefont {Berakdar}},\ and\ \bibinfo
  {author} {\bibfnamefont {N.~A.}\ \bibnamefont {Spaldin}},\ }\bibfield
  {title} {\bibinfo {title} {{M}agnetophononics: {U}ltrafast spin control
  through the lattice},\ }\href
  {https://doi.org/10.1103/PhysRevMaterials.2.064401} {\bibfield  {journal}
  {\bibinfo  {journal} {Phys. Rev. Mater.}\ }\textbf {\bibinfo {volume} {2}},\
  \bibinfo {pages} {064401} (\bibinfo {year} {2018})}\BibitemShut {NoStop}%
\bibitem [{\citenamefont {Disa}\ \emph {et~al.}(2023)\citenamefont {Disa},
  \citenamefont {Curtis}, \citenamefont {Fechner}, \citenamefont {Liu},
  \citenamefont {von Hoegen}, \citenamefont {F\"{o}rst}, \citenamefont {Nova},
  \citenamefont {Narang}, \citenamefont {Maljuk}, \citenamefont {Boris},
  \citenamefont {Keimer},\ and\ \citenamefont {Cavalleri}}]{disa2023photo}%
  \BibitemOpen
  \bibfield  {author} {\bibinfo {author} {\bibfnamefont {A.~S.}\ \bibnamefont
  {Disa}}, \bibinfo {author} {\bibfnamefont {J.}~\bibnamefont {Curtis}},
  \bibinfo {author} {\bibfnamefont {M.}~\bibnamefont {Fechner}}, \bibinfo
  {author} {\bibfnamefont {A.}~\bibnamefont {Liu}}, \bibinfo {author}
  {\bibfnamefont {A.}~\bibnamefont {von Hoegen}}, \bibinfo {author}
  {\bibfnamefont {M.}~\bibnamefont {F\"{o}rst}}, \bibinfo {author}
  {\bibfnamefont {T.~F.}\ \bibnamefont {Nova}}, \bibinfo {author}
  {\bibfnamefont {P.}~\bibnamefont {Narang}}, \bibinfo {author} {\bibfnamefont
  {A.}~\bibnamefont {Maljuk}}, \bibinfo {author} {\bibfnamefont {A.~V.}\
  \bibnamefont {Boris}}, \bibinfo {author} {\bibfnamefont {B.}~\bibnamefont
  {Keimer}},\ and\ \bibinfo {author} {\bibfnamefont {A.}~\bibnamefont
  {Cavalleri}},\ }\bibfield  {title} {\bibinfo {title} {Photo-induced
  high-temperature ferromagnetism in $\mathrm{YTiO}_{3}$},\ }\href
  {https://doi.org/10.1038/s41586-023-05853-8} {\bibfield  {journal} {\bibinfo
  {journal} {Nature}\ }\textbf {\bibinfo {volume} {617}},\ \bibinfo {pages}
  {73} (\bibinfo {year} {2023})}\BibitemShut {NoStop}%
\bibitem [{\citenamefont {Gurevich}\ and\ \citenamefont
  {Melkov}(1996)}]{gurevich1996magnetization}%
  \BibitemOpen
  \bibfield  {author} {\bibinfo {author} {\bibfnamefont {A.~G.}\ \bibnamefont
  {Gurevich}}\ and\ \bibinfo {author} {\bibfnamefont {G.~A.}\ \bibnamefont
  {Melkov}},\ }\href@noop {} {\emph {\bibinfo {title} {Magnetization
  oscillations and waves}}}\ (\bibinfo  {publisher} {CRC},\ \bibinfo {year}
  {1996})\BibitemShut {NoStop}%
\bibitem [{\citenamefont {Chen}\ and\ \citenamefont
  {Wang}(2021)}]{chen2021one}%
  \BibitemOpen
  \bibfield  {author} {\bibinfo {author} {\bibfnamefont {X.-Y.}\ \bibnamefont
  {Chen}}\ and\ \bibinfo {author} {\bibfnamefont {Y.-P.}\ \bibnamefont
  {Wang}},\ }\bibfield  {title} {\bibinfo {title} {One-dimensional model for
  coupling between magnon and optical phonon},\ }\href
  {https://doi.org/10.1103/PhysRevB.104.155132} {\bibfield  {journal} {\bibinfo
   {journal} {Phys. Rev. B}\ }\textbf {\bibinfo {volume} {104}},\ \bibinfo
  {pages} {155132} (\bibinfo {year} {2021})}\BibitemShut {NoStop}%
\bibitem [{\citenamefont {Wang}\ \emph {et~al.}(2020)\citenamefont {Wang},
  \citenamefont {Xie}, \citenamefont {Xu},\ and\ \citenamefont
  {Xia}}]{wang2020first}%
  \BibitemOpen
  \bibfield  {author} {\bibinfo {author} {\bibfnamefont {L.-W.}\ \bibnamefont
  {Wang}}, \bibinfo {author} {\bibfnamefont {L.-S.}\ \bibnamefont {Xie}},
  \bibinfo {author} {\bibfnamefont {P.-X.}\ \bibnamefont {Xu}},\ and\ \bibinfo
  {author} {\bibfnamefont {K.}~\bibnamefont {Xia}},\ }\bibfield  {title}
  {\bibinfo {title} {First-principles study of magnon-phonon interactions in
  gadolinium iron garnet},\ }\href
  {https://doi.org/10.1103/PhysRevB.101.165137} {\bibfield  {journal} {\bibinfo
   {journal} {Phys. Rev. B}\ }\textbf {\bibinfo {volume} {101}},\ \bibinfo
  {pages} {165137} (\bibinfo {year} {2020})}\BibitemShut {NoStop}%
\bibitem [{\citenamefont {Wang}\ \emph {et~al.}(2023)\citenamefont {Wang},
  \citenamefont {Ren}, \citenamefont {Hou}, \citenamefont {Cheng},\ and\
  \citenamefont {Zhang}}]{wang2023magnon}%
  \BibitemOpen
  \bibfield  {author} {\bibinfo {author} {\bibfnamefont {K.}~\bibnamefont
  {Wang}}, \bibinfo {author} {\bibfnamefont {K.}~\bibnamefont {Ren}}, \bibinfo
  {author} {\bibfnamefont {Y.}~\bibnamefont {Hou}}, \bibinfo {author}
  {\bibfnamefont {Y.}~\bibnamefont {Cheng}},\ and\ \bibinfo {author}
  {\bibfnamefont {G.}~\bibnamefont {Zhang}},\ }\bibfield  {title} {\bibinfo
  {title} {Magnon--phonon coupling: from fundamental physics to applications},\
  }\href {https://doi.org/10.1039/D3CP02683C} {\bibfield  {journal} {\bibinfo
  {journal} {Phys. Chem. Chem. Phys.}\ }\textbf {\bibinfo {volume} {25}},\
  \bibinfo {pages} {21802} (\bibinfo {year} {2023})}\BibitemShut {NoStop}%
\bibitem [{\citenamefont {Liu}\ \emph {et~al.}(2020)\citenamefont {Liu},
  \citenamefont {Chaloupka},\ and\ \citenamefont {Khaliullin}}]{liu2020kitaev}%
  \BibitemOpen
  \bibfield  {author} {\bibinfo {author} {\bibfnamefont {H.}~\bibnamefont
  {Liu}}, \bibinfo {author} {\bibfnamefont {J.}~\bibnamefont {Chaloupka}},\
  and\ \bibinfo {author} {\bibfnamefont {G.}~\bibnamefont {Khaliullin}},\
  }\bibfield  {title} {\bibinfo {title} {{K}itaev {S}pin {L}iquid in $3d$
  {T}ransition {M}etal {C}ompounds},\ }\href
  {https://doi.org/10.1103/PhysRevLett.125.047201} {\bibfield  {journal}
  {\bibinfo  {journal} {Phys. Rev. Lett.}\ }\textbf {\bibinfo {volume} {125}},\
  \bibinfo {pages} {047201} (\bibinfo {year} {2020})}\BibitemShut {NoStop}%
\bibitem [{\citenamefont {Hong}\ \emph {et~al.}(2021)\citenamefont {Hong},
  \citenamefont {Gillig}, \citenamefont {Hentrich}, \citenamefont {Yao},
  \citenamefont {Kocsis}, \citenamefont {Witte}, \citenamefont {Schreiner},
  \citenamefont {Baumann}, \citenamefont {P\'erez}, \citenamefont {Wolter},
  \citenamefont {Li}, \citenamefont {B\"uchner},\ and\ \citenamefont
  {Hess}}]{hong2021strongly}%
  \BibitemOpen
  \bibfield  {author} {\bibinfo {author} {\bibfnamefont {X.}~\bibnamefont
  {Hong}}, \bibinfo {author} {\bibfnamefont {M.}~\bibnamefont {Gillig}},
  \bibinfo {author} {\bibfnamefont {R.}~\bibnamefont {Hentrich}}, \bibinfo
  {author} {\bibfnamefont {W.}~\bibnamefont {Yao}}, \bibinfo {author}
  {\bibfnamefont {V.}~\bibnamefont {Kocsis}}, \bibinfo {author} {\bibfnamefont
  {A.~R.}\ \bibnamefont {Witte}}, \bibinfo {author} {\bibfnamefont
  {T.}~\bibnamefont {Schreiner}}, \bibinfo {author} {\bibfnamefont
  {D.}~\bibnamefont {Baumann}}, \bibinfo {author} {\bibfnamefont
  {N.}~\bibnamefont {P\'erez}}, \bibinfo {author} {\bibfnamefont {A.~U.~B.}\
  \bibnamefont {Wolter}}, \bibinfo {author} {\bibfnamefont {Y.}~\bibnamefont
  {Li}}, \bibinfo {author} {\bibfnamefont {B.}~\bibnamefont {B\"uchner}},\ and\
  \bibinfo {author} {\bibfnamefont {C.}~\bibnamefont {Hess}},\ }\bibfield
  {title} {\bibinfo {title} {Strongly scattered phonon heat transport of the
  candidate {K}itaev material
  $\mathrm{Na}_{2}\mathrm{Co}_{2}\mathrm{TeO}_{6}$},\ }\href
  {https://doi.org/10.1103/PhysRevB.104.144426} {\bibfield  {journal} {\bibinfo
   {journal} {Phys. Rev. B}\ }\textbf {\bibinfo {volume} {104}},\ \bibinfo
  {pages} {144426} (\bibinfo {year} {2021})}\BibitemShut {NoStop}%
\bibitem [{\citenamefont {Li}\ \emph {et~al.}(2022)\citenamefont {Li},
  \citenamefont {Gu}, \citenamefont {Chen}, \citenamefont {Garlea},
  \citenamefont {Iida}, \citenamefont {Kamazawa}, \citenamefont {Li},
  \citenamefont {Deng}, \citenamefont {Xiao}, \citenamefont {Zheng},
  \citenamefont {Ye}, \citenamefont {Peng}, \citenamefont {Zaliznyak},
  \citenamefont {Tranquada},\ and\ \citenamefont {Li}}]{li2022giant}%
  \BibitemOpen
  \bibfield  {author} {\bibinfo {author} {\bibfnamefont {X.}~\bibnamefont
  {Li}}, \bibinfo {author} {\bibfnamefont {Y.}~\bibnamefont {Gu}}, \bibinfo
  {author} {\bibfnamefont {Y.}~\bibnamefont {Chen}}, \bibinfo {author}
  {\bibfnamefont {V.~O.}\ \bibnamefont {Garlea}}, \bibinfo {author}
  {\bibfnamefont {K.}~\bibnamefont {Iida}}, \bibinfo {author} {\bibfnamefont
  {K.}~\bibnamefont {Kamazawa}}, \bibinfo {author} {\bibfnamefont
  {Y.}~\bibnamefont {Li}}, \bibinfo {author} {\bibfnamefont {G.}~\bibnamefont
  {Deng}}, \bibinfo {author} {\bibfnamefont {Q.}~\bibnamefont {Xiao}}, \bibinfo
  {author} {\bibfnamefont {X.}~\bibnamefont {Zheng}}, \bibinfo {author}
  {\bibfnamefont {Z.}~\bibnamefont {Ye}}, \bibinfo {author} {\bibfnamefont
  {Y.}~\bibnamefont {Peng}}, \bibinfo {author} {\bibfnamefont {I.~A.}\
  \bibnamefont {Zaliznyak}}, \bibinfo {author} {\bibfnamefont {J.~M.}\
  \bibnamefont {Tranquada}},\ and\ \bibinfo {author} {\bibfnamefont
  {Y.}~\bibnamefont {Li}},\ }\bibfield  {title} {\bibinfo {title} {{G}iant
  {M}agnetic {I}n-{P}lane {A}nisotropy and {C}ompeting {I}nstabilities in
  $\mathrm{Na}_{3}\mathrm{Co}_{2}\mathrm{SbO}_{6}$},\ }\href
  {https://doi.org/10.1103/PhysRevX.12.041024} {\bibfield  {journal} {\bibinfo
  {journal} {Phys. Rev. X}\ }\textbf {\bibinfo {volume} {12}},\ \bibinfo
  {pages} {041024} (\bibinfo {year} {2022})}\BibitemShut {NoStop}%
\bibitem [{\citenamefont {Feng}\ \emph {et~al.}(2022)\citenamefont {Feng},
  \citenamefont {Swarup},\ and\ \citenamefont {Perkins}}]{feng2022footprints}%
  \BibitemOpen
  \bibfield  {author} {\bibinfo {author} {\bibfnamefont {K.}~\bibnamefont
  {Feng}}, \bibinfo {author} {\bibfnamefont {S.}~\bibnamefont {Swarup}},\ and\
  \bibinfo {author} {\bibfnamefont {N.~B.}\ \bibnamefont {Perkins}},\
  }\bibfield  {title} {\bibinfo {title} {{F}ootprints of {K}itaev spin liquid
  in the {F}ano lineshape of {R}aman-active optical phonons},\ }\href
  {https://doi.org/10.1103/PhysRevB.105.L121108} {\bibfield  {journal}
  {\bibinfo  {journal} {Phys. Rev. B}\ }\textbf {\bibinfo {volume} {105}},\
  \bibinfo {pages} {L121108} (\bibinfo {year} {2022})}\BibitemShut {NoStop}%
\bibitem [{\citenamefont {Li}\ \emph {et~al.}(2023)\citenamefont {Li},
  \citenamefont {Neumann}, \citenamefont {Guang}, \citenamefont {Huang},
  \citenamefont {Liu}, \citenamefont {Xia}, \citenamefont {Yue}, \citenamefont
  {Sun}, \citenamefont {Wang}, \citenamefont {Li}, \citenamefont {Jiang},
  \citenamefont {Fang}, \citenamefont {Jiang}, \citenamefont {Zhao},
  \citenamefont {Mook}, \citenamefont {Henk}, \citenamefont {Mertig},
  \citenamefont {Zhou},\ and\ \citenamefont {Sun}}]{li2023magnon}%
  \BibitemOpen
  \bibfield  {author} {\bibinfo {author} {\bibfnamefont {N.}~\bibnamefont
  {Li}}, \bibinfo {author} {\bibfnamefont {R.~R.}\ \bibnamefont {Neumann}},
  \bibinfo {author} {\bibfnamefont {S.~K.}\ \bibnamefont {Guang}}, \bibinfo
  {author} {\bibfnamefont {Q.}~\bibnamefont {Huang}}, \bibinfo {author}
  {\bibfnamefont {J.}~\bibnamefont {Liu}}, \bibinfo {author} {\bibfnamefont
  {K.}~\bibnamefont {Xia}}, \bibinfo {author} {\bibfnamefont {X.~Y.}\
  \bibnamefont {Yue}}, \bibinfo {author} {\bibfnamefont {Y.}~\bibnamefont
  {Sun}}, \bibinfo {author} {\bibfnamefont {Y.~Y.}\ \bibnamefont {Wang}},
  \bibinfo {author} {\bibfnamefont {Q.~J.}\ \bibnamefont {Li}}, \bibinfo
  {author} {\bibfnamefont {Y.}~\bibnamefont {Jiang}}, \bibinfo {author}
  {\bibfnamefont {J.}~\bibnamefont {Fang}}, \bibinfo {author} {\bibfnamefont
  {Z.}~\bibnamefont {Jiang}}, \bibinfo {author} {\bibfnamefont
  {X.}~\bibnamefont {Zhao}}, \bibinfo {author} {\bibfnamefont {A.}~\bibnamefont
  {Mook}}, \bibinfo {author} {\bibfnamefont {J.}~\bibnamefont {Henk}}, \bibinfo
  {author} {\bibfnamefont {I.}~\bibnamefont {Mertig}}, \bibinfo {author}
  {\bibfnamefont {H.~D.}\ \bibnamefont {Zhou}},\ and\ \bibinfo {author}
  {\bibfnamefont {X.~F.}\ \bibnamefont {Sun}},\ }\bibfield  {title} {\bibinfo
  {title} {Magnon-polaron driven thermal {H}all effect in a
  {H}eisenberg-{K}itaev antiferromagnet},\ }\href
  {https://doi.org/10.1103/PhysRevB.108.L140402} {\bibfield  {journal}
  {\bibinfo  {journal} {Phys. Rev. B}\ }\textbf {\bibinfo {volume} {108}},\
  \bibinfo {pages} {L140402} (\bibinfo {year} {2023})}\BibitemShut {NoStop}%
\bibitem [{\citenamefont {Zhang}\ \emph {et~al.}(2023)\citenamefont {Zhang},
  \citenamefont {Lee}, \citenamefont {Woods}, \citenamefont {Peria},
  \citenamefont {Thomas}, \citenamefont {Movshovich}, \citenamefont {Brosha},
  \citenamefont {Huang}, \citenamefont {Zhou}, \citenamefont {Zapf},\ and\
  \citenamefont {Lee}}]{zhang2023electronic}%
  \BibitemOpen
  \bibfield  {author} {\bibinfo {author} {\bibfnamefont {S.}~\bibnamefont
  {Zhang}}, \bibinfo {author} {\bibfnamefont {S.}~\bibnamefont {Lee}}, \bibinfo
  {author} {\bibfnamefont {A.~J.}\ \bibnamefont {Woods}}, \bibinfo {author}
  {\bibfnamefont {W.~K.}\ \bibnamefont {Peria}}, \bibinfo {author}
  {\bibfnamefont {S.~M.}\ \bibnamefont {Thomas}}, \bibinfo {author}
  {\bibfnamefont {R.}~\bibnamefont {Movshovich}}, \bibinfo {author}
  {\bibfnamefont {E.}~\bibnamefont {Brosha}}, \bibinfo {author} {\bibfnamefont
  {Q.}~\bibnamefont {Huang}}, \bibinfo {author} {\bibfnamefont
  {H.}~\bibnamefont {Zhou}}, \bibinfo {author} {\bibfnamefont {V.~S.}\
  \bibnamefont {Zapf}},\ and\ \bibinfo {author} {\bibfnamefont
  {M.}~\bibnamefont {Lee}},\ }\bibfield  {title} {\bibinfo {title} {Electronic
  and magnetic phase diagrams of the {K}itaev quantum spin liquid candidate
  $\mathrm{Na}_{2}\mathrm{Co}_{2}\mathrm{TeO}_{6}$},\ }\href
  {https://doi.org/10.1103/PhysRevB.108.064421} {\bibfield  {journal} {\bibinfo
   {journal} {Phys. Rev. B}\ }\textbf {\bibinfo {volume} {108}},\ \bibinfo
  {pages} {064421} (\bibinfo {year} {2023})}\BibitemShut {NoStop}%
\end{thebibliography}%

\end{document}